\documentclass[12pt,preprint]{aastex}
\usepackage{natbib}
\usepackage[caption=false]{subfig}
\usepackage{float}
\begin{document}

\title{Constraining the Sub-AU-Scale Distribution of Hydrogen and
  Carbon Monoxide Gas around Young Stars with the Keck Interferometer}

\author{J.A. Eisner\altaffilmark{1},
  L. A. Hillenbrand\altaffilmark{2}, \& Jordan M. Stone\altaffilmark{1}}

\altaffiltext{1}{Steward Observatory, University of Arizona, Tucson, AZ 85721}
\altaffiltext{2}{Astrophysics Department, California Institute of Technology, Pasadena, CA 91125}


\begin{abstract}
We present Keck Interferometer observations of T Tauri and Herbig
Ae/Be stars with a
spatial resolution of a few milliarcseconds and a spectral 
resolution of $\sim 2000$.  Our observations span the $K$-band, and
include the Br$\gamma$ transition of Hydrogen and the $v=2
\rightarrow 0$ and $v=3 \rightarrow 1$ transitions of carbon
monoxide.  For several targets we also present data from Keck/NIRSPEC
that provide higher spectral resolution, but a seeing-limited spatial resolution,
of the same spectral features.  We analyze the Br$\gamma$ emission in
the context of both disk and infall/outflow models, and conclude that
the Br$\gamma$ emission traces gas at very small stellocentric radii,
consistent with the magnetospheric scale.  However some Br$\gamma$-emitting
gas also seems to be located at radii of $\ga 0.1$ AU, perhaps tracing
the inner regions of magnetically launched outflows.   CO emission is
detected from several objects, and we generate disk models that
reproduce both the KI and NIRSPEC data well.   
We infer the CO spatial distribution to be
coincident with the distribution of continuum emission in most cases.
Furthermore the Br$\gamma$ emission in these objects is
roughly coincident with both the CO and continuum emission.
We present potential explanations for the spatial coincidence of continuum,
Br$\gamma$, and CO overtone emission, and explore the implications for
the low occurrence rate of CO overtone emission in young stars. 
Finally, we provide additional discussion of V1685 Cyg, which is
unusual among our sample in showing large differences in emitting
region size and spatial position as a function of wavelength. 
\end{abstract}

\section{Introduction \label{sec:intro}}
Protoplanetary disks are an integral part of the star and planet
formation process.  Angular momentum conservation demands disk
creation during the protostellar collapse process, and these disks
then provide a reservoir from which stars and planets accrete
material.  Gas within 1 AU of young stars may reside in
protoplanetary disks, infalling streams from inner disks onto the
central stars, or in outflows.  Observations of gas on sub-AU scales
can thus constrain the composition and dynamics of inner disks,
accretion flows, and outflows.

To probe inner disk gas requires observations with high spatial and
spectral resolution.  At a distance of 140 pc, 1 AU subtends 7 mas.
Gas in Keplerian rotation around a solar-mass star 
at this radius would produce emission lines with velocity widths of
30 km s$^{-1}$; in the near-IR this corresponds to a spectral
dispersion $\lambda / \Delta \lambda \approx 10,000$.  At smaller
radii, or for gas that is infalling or outflowing, linewidths may be
higher and the required spectral dispersion somewhat lower.

Spatially resolved spectroscopy at these resolutions is challenging,
but has been recently enabled by near-IR interferometers.  Most
studies have focused on the Br$\gamma$ transition of hydrogen
\citep{EISNER07,MALBET+07,TATULLI+07,KRAUS+08,EISNER+10,
WEIGELT+11}.   The Br$\gamma$ transition, from the $n=7\rightarrow 4$
electronic states, produces a  spectral line at 2.1662 $\mu$m.
This line has been shown to be strongly correlated with accretion onto
young stars \citep{MHC98}. While Balmer series hydrogen
lines often show profiles associated with winds \citep[or a
combination of winds and infall; e.g.,][]{KHS06}, Br$\gamma$
line profiles are often more consistent with infall kinematics
\citep[e.g.,][]{NCT96}.

Br$\gamma$ is a prime target for spatially resolved spectroscopy
because the line generally has a high flux and large linewidth, and is
very common around young stars \citep[e.g.,][]{FE01}.  For most
objects, spatially resolved data indicates Br$\gamma$ emission more
compact than the continuum emission; in these cases the gas probably
arises in accretion inflows \citep[e.g.,][]{KRAUS+08,EISNER+10}.  However there
are exceptions, where extended Br$\gamma$ emission suggests a wind
origin \citep[e.g.,][]{MALBET+07,TATULLI+07,WEIGELT+11}.

Other studies have also targeted or included the CO overtone bandheads
\citep{TATULLI+08, EISNER+09,EH11}.  
The CO overtone bandheads are made up of numerous lines tracing
rovibrational transitions with $\Delta v = 2$. Overtone transitions
are rare compared to the occurrence rate of Br$\gamma$ emission, but
are found toward a number of young stars
\citep{CARR89,NAJITA+96,NAJITA+00,NAJITA+06,NAJITA+09,
BISCAYA+97,THI+05,BRITTAIN+07,BERTHOUD+07,BERTHOUD08,EISNER+13}.

Analyses of spectrally resolved line profiles indicate
that CO overtone emission generally arises in the inner $\sim 0.15$ AU
of Keplerian disks \citep[e.g.,][]{NAJITA+96,NAJITA+09,THI+05},
consistent with expectations based on the high excitation temperatures
of these transitions.  Furthermore, the size of the CO
emitting region has been directly measured (interferometrically) for
the young star 51 Oph
to be $\sim 0.15$ AU
\citep{TATULLI+08}. Spatially resolved observations of this object and
others can remove ambiguities inherent in previous modeling of spatially
unresolved observations.

Finally, a number of studies have constrained the gas content of inner
disk regions
indirectly from low-dispersion spectroscopy or continuum data
\citep{EISNER+07a,TANNIRKULAM+08,ISELLA+08,EISNER+09}.  The results of
these studies suggest the presence of gas at stellocentric radii
smaller than the dust sublimation radius.
This compact matter emits as  a
(pseudo-)continuum, perhaps tracing free-free emission
\citep[e.g.,][]{EISNER+09} or emission from highly refractory dust grains
\citep[e.g.,][]{BENISTY+10}.

The Br$\gamma$, CO, and continuum emission probably trace different
physical components of star+disk systems.  Br$\gamma$ data constrains accretion
and outflow processes on sub-AU scales, and can distinguish between
various accretion and wind-launching models
\citep[e.g.,][]{LP74,KONIGL91,SHU+94,KP00}.  The CO emission probably
traces Keplerian inner disks \citep[e.g.,][]{NAJITA+96} on sub-AU
scales. Continuum
emission is dominated by dust near the sublimation radius
\citep[e.g.,][and references therein]{DM10}.  Observing dust and gas
can provide a complete picture of disk inner regions from the stellar
surface to the dust sublimation radius.

Here we use the Keck Interferometer (KI) to spatially and spectrally resolve gas
within 1 AU of a sample of 21 young stars spanning a mass range from
$\sim 0.5$--10 M$_{\odot}$.  15 of these objects were included in
\citet{EISNER+10}, but we include an expanded wavelength coverage
here, as well as additional data for many sources.  The observations
presented here target the Br$\gamma$ line as well as the spectral
region containing the $v=2 \rightarrow 0$ and $v=3 \rightarrow 1$ CO
overtone bandheads.

\section{Observations and Data Reduction \label{sec:odr}}

\subsection{Sample \label{sec:sample}}
We selected a sample of young stars (Table \ref{tab:sample})
known to be surrounded by protoplanetary disks, most 
of which have been observed previously at near-IR
wavelengths with long-baseline interferometers.
Our sample
includes: 11 T Tauri stars, pre-main-sequence analogs of solar-type
stars like our own sun; 8 Herbig Ae/Be stars, 2--10 M$_{\odot}$ 
pre-main-sequence stars; and 2 stars (AS 353 A and V1331 Cyg)
with heavily veiled stellar photospheres whose spectral types are uncertain.  
Three FU Ori objects, FU Ori, V1057 Cyg, and V1515 Cyg, are included
in a separate publication \citep{EH11}.


Among the T Tauri stars in our sample, RY Tau, T Tau A, DG Tau, DK
Tau A, GK Tau, DO Tau, DR Tau, SU Aur, and RW Aur A are all in the
Taurus/Auriga complex, at an assumed distance of 140 pc
\citep[e.g.,][]{KDH94}.   AS 205 A and
V2508 Oph reside in the $\rho$ Oph cloud, at an asumed distance of 160
pc \citep{CHINI81}.  
These T Tauri stars span
spectral types between M0 and G2, corresponding to stellar masses
between $\sim 0.5$ and 1.5 M$_{\odot}$
\citep[e.g.,][]{HEG95,WG01,EISNER+05}.  The sample also includes objects
with a broad range of veilings, including sources like DK Tau A where the
circumstellar flux is a small fraction of the stellar flux, and
objects like AS 205 A where the circumstellar flux dominates
\citep[e.g.,][]{EISNER+05,EISNER+09,HH14}.  These stars span a
range of $>10$ in accretion luminosity \citep[e.g.,][]{FE01,EISNER+10},
and in outflow properties \citep[e.g.,][]{HAMANN94,HEG95}.  
Among our sample,
DG Tau, DO Tau,  DR Tau,  and AS 205 A have particularly high accretion
luminosities \citep{NCT96,FE01,EISNER+10}.  DG Tau, DO Tau,
and DR Tau  also  have large forbidden line equivalent widths
indicative of powerful outflows \citep{HAMANN94,HEG95}.  DG Tau
 in particular, stands out because it is a known
CO overtone emission source \citep[e.g.,][]{CARR89}.

The Herbig Ae/Be stars in our sample are distributed across several
star forming regions (see Table \ref{tab:sample} for distances).
These stars span a range of spectral type from A3 to B0, corresponding
to stellar masses between $\sim 2$ and 10 M$_{\odot}$
\citep[e.g.,][]{PS93}.   The
circumstellar emission typically dominates the total flux for these
systems, although there is a small range of circumstellar-to-stellar
flux ratios across the sample.  The selected Herbig Ae/Be stars span a
broad range of accretion luminosities \citep[e.g.,][]{EISNER+10}, and
show a range of outflow strengths \citep[as traced by forbidden line
emission; e.g.,][]{CR97}. 

While the spectral types,
and hence masses, of AS 353 A and V1331 Cyg are uncertain, they are
known to show strong CO overtone emission \citep[e.g.,][]{CARR89} and
the highest equivalent width Br$\gamma$ emission---suggesting the
highest accretion luminosities---of any objects in our
sample \citep[e.g.,][]{NCT96,EISNER+10}.  These objects all also known
to drive powerful outflows \citep[e.g.,][]{HJ83,HMS86,KUHI64,ME98}.


\subsection{Keck Interferometer Data}
We briefly summarize the experimental setup and data calibration
procedures employed in this work.  For additional details we refer to
\citet{EISNER+10}, which used the same setup as the work described here.

\subsubsection{Experimental Setup \label{sec:setup}}

KI was a fringe-tracking long baseline near-IR Michelson
interferometer that combined light from the two 10-m Keck apertures 
\citep{CW03,COLAVITA+03,COLAVITA+13}.  Each of the 10-m apertures was
equipped with a natural guide star adaptive optics (NGS-AO) system
that corrected phase errors caused by
atmospheric turbulence across each telescope pupil, and thereby maintained 
spatial coherence of the light from the source across each aperture.  
We used the ``self phase referencing'' (SPR)  mode of KI
\citep{WOILLEZ+12}, which was implemented as a precursor to the
dual-field phase referencing mode \citep{WOILLEZ+14}.  
In SPR a servo loop used the phase information
measured on the primary (continuum) channel to stabilize the
atmospheric phase motions, and enabled longer integration times on the
secondary (spectrally-dispersed) channel.  For the spectrally-dispersed 
channel we
used integration times between 0.5 and 2 s, approximately 1,000 times
longer than possible with the uncorrected primary side.  

The spectrally-dispersed channel included a grism that, used in first-order, 
passed the entire $K$-band
with a dispersion of $\lambda/\Delta \lambda \approx 2000$.  
This spectral resolution was confirmed with measurements of a neon lamp
spectrum.  Note, however, that the lines were not fully Nyquist sampled with
our detector; spectra were Nyquist sampled at a resolution of $\sim
1000$.   Neon lamp spectra and/or Fourier Transform Spectroscopy were
also used to determine the wavelength scale for each night of observed data.

While the entire $K$-band
fell on the detector, vignetting in the camera led to lower throughput toward the band 
edges.  The effective bandpass of our observations is approximately
2.05 to 2.35 $\mu$m.  In this paper we focus on two sub-regions of the
$K$-band: one centered on the Br$\gamma$ feature at 2.1662 $\mu$m, 
and one including the first two CO overtone bandheads, at 2.2936
$\mu$m and 2.3227 $\mu$m.  The third CO overtone bandhead
at 2.3527 $\mu$m is included in the KI bandpass, but in the
lower-throughput edge region.  We therefore exclude it from the
analysis below.

\subsubsection{Observations and Data Calibration  \label{sec:obs}}
We obtained Keck Interferometer (KI) observations of our sample during 
9 nights between April 2008 and November 2011 (see Table \ref{tab:obs}).
When observing our sample, targets were interleaved with calibrators
every 10--15 minutes.

We used the count rates in each spectral channel observed during 
``foreground integrations'', corrected for biases 
\citep{COLAVITA99}, to recover crude
spectra for our targets.  These spectra were measured when no fringes
were present, and include all flux measured
within the $\sim 50$ mas diameter of the instrumental field of view.
We divided the measured flux versus wavelength
for our targets by the observed fluxes from the calibrator stars, using
calibrator scans nearest in time to given target scans, and then multiplied
the results by Nextgen template spectra \citep{HAB99} suitable for the 
spectral types of the calibrators.  All calibrators have spectral
types earlier than K2 (Table \ref{tab:sample}), where Nextgen models
are well-matched to empirical stellar spectra.
For our calibration procedure we prefer Nextgen templates to observed
stellar spectra, because it is easier to compile a
grid of photospheric spectra that is densely sampled in spectral type.

\begin{figure}
\plotone{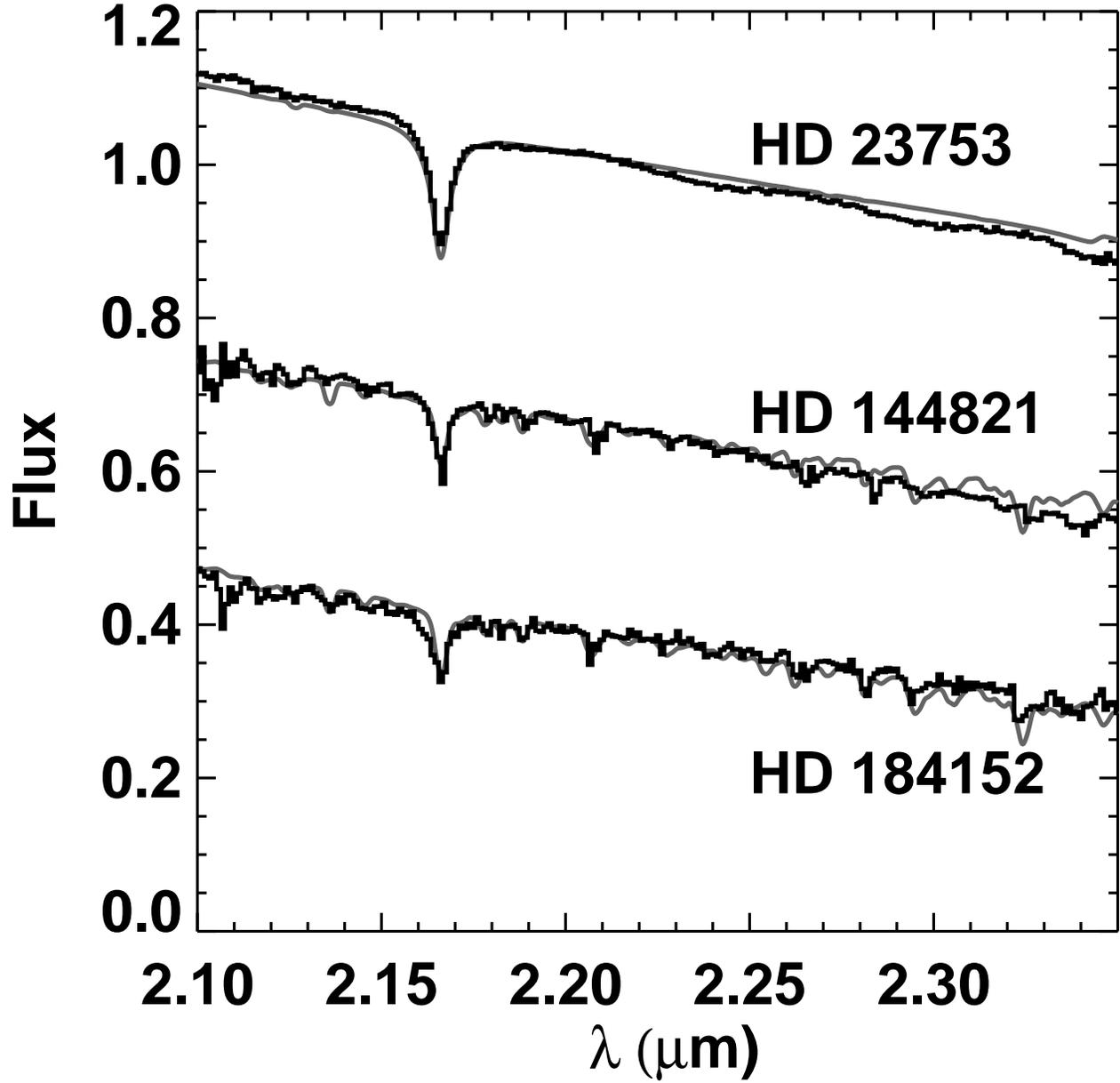}
\caption{Spectra for three calibrators used to check the flux
  calibration procedure for KI (Section \ref{sec:obs}).  The
  calibrated spectra are shown as black histograms.  Nextgen spectra
  computed at the KI spectral resolution, using the source spectral
  types (see Table \ref{tab:sample}), are shown as gray curves.
\label{fig:checkstar}}
\end{figure}

Instrumental scattered or thermal
background light could add continuum emission to the KI data, leading
to observed line-to-continuum ratios lower than intrinsic spectral
features.  To test for such effects, we compared observed and
theoretical spectra for several of our main-sequence calibrator
stars.  
Figure \ref{fig:checkstar} demonstrates that the observed
photospheric Br$\gamma$ absorption in these ``check stars'' 
is consistent with
theoretical expectations, across a range of source colors and
brightnesses (see Table \ref{tab:sample} for spectral types and
magnitudes).   Thus the Br$\gamma$ spectra derived from the KI data
do not seem to be affected substantially by  instrumental
scattering or background emission.

Since CO overtone emission occurs in the red end of the $K$-band,
where thermal emission is stronger, uncorrected background emission
may be more significant for CO than for the Br$\gamma$ region.  
Two of the check stars, HD 144841 and HD 184152, are expected to
show some CO overtone absorption.   The observed absorption features 
appear weaker than expected from the synthetic spectra (Figure
\ref{fig:checkstar}).  The largest discrepancy is seen for HD 184152,
one of the faintest objects observed with KI.  Uncorrected background
emission would be relatively stronger for fainter objects, consistent
with this observation.  Thus, some caution is required when
interpreting the CO overtone emission or absorption features in the KI
spectra, particularly for fainter targets.

We measured squared visibilities ($V^2$) for our targets and calibrator
stars in each of the 330 spectral channels across the $K$-band
provided by the grism. The calibrator stars are main
sequence stars, with known parallaxes,
whose $K$ magnitudes are within 0.5 mags of the target $K$ magnitudes
(Table \ref{tab:sample}).
We calculated the system visibility appropriate to each target scan
by weighting the calibrator data by the internal scatter and the temporal and 
angular proximity to the target data \citep{BODEN+98}.  For comparison, we 
also computed the straight average of the $V^2$ for all calibrators used for a 
given source, and the system visibility for the calibrator observations 
closest in time.  These methods all produce results consistent within the 
measurement uncertainties.  We adopt the first method in our analysis.

Phases are measured in each spectral channel using the same ``ABCD''
procedure \citep{COLAVITA99} used to determine $V^2$.  
These phases are then de-rotated so 
that the average phase of all channels is zero.  Next,
the phases versus wavelength are unwrapped to 
eliminate any jumps between -180$^{\circ}$ and 180$^{\circ}$.  
After computing weighted average
differential phases for each target and calibrator scan,
we determine a ``system differential phase'' using similar weighting
employed above to calculate the system visibility.

The system differential phase is subtracted from the target
differential phase.  Since targets
and calibrators are observed at similar airmasses, this
calibration procedure removes most atmospheric and
instrumental refraction effects.  Finally, we remove any
residual slope in the differential phase spectrum, since
we can not distinguish instrumental slopes from those
intrinsic to the target signal \citep[e.g.,][]{WOILLEZ+12}.  Since we
are focused
on small spectral regions around individual spectral features,
we are largely insensitive to errors in these calibrations.

The measured fluxes, $V^2$, and differential phases contain
contributions from the circumstellar material and from the central
stars.  To remove the central star, we first estimate its flux in each
observed spectral channel using stellar parameters from the literature
(see Table \ref{tab:sample}) and a suitable Nextgen synthetic
spectrum.  The synthetic spectra are subtracted from the observed KI
spectra to produce circumstellar fluxes for each spectral
channel\footnote{For AS 353 A and V1331 Cyg, which appear to be heavily
  veiled, we assume 100\% of the
  observed flux comes from circumstellar matter.}.  

Circumstellar fluxes ($F_{\rm circ}$), visibilities ($V_{\rm circ}$), and differential
phases ($\Delta \phi_{\rm circ}$) are given by the following equations
\citep[e.g.,][]{EISNER+10}:
\begin{equation}
F_{\rm circ} = F_{\rm meas} - F_{\ast},
\label{eq:fcirc}
\end{equation}
\begin{equation}
  \Delta \phi_{\rm circ} = \tan^{-1} \left[ \frac{V_{\rm meas} \sin
      (\Delta \phi_{\rm meas}) (F_{\ast} + F_{\rm circ})}{V_{\rm meas}
      \cos (\Delta \phi_{\rm meas}) (F_{\ast}+F_{\rm circ}) - F_{\ast}}\right],
  \label{eq:dpcirc}
\end{equation}
\begin{equation}
  V_{\rm circ} = V_{\rm meas} \frac{\sin (\Delta \phi_{\rm meas})}{\sin
    (\Delta \phi_{\rm circ})} \left(\frac{F_{\ast}+F_{\rm circ}}{F_{\rm circ}}\right).
  \label{eq:vdisk}
\end{equation}
These equations imply that even if a spectral feature is not apparent
in the observed spectrum, its presence may be implied in the
circumstellar emission.  For example, late-type stars have
photospheric CO overtone absorption.  If no absorption
is seen in the observed spectra of such objects, this 
implies that the photospheric
absorption is filled in by circumstellar emission.

One potential pitfall with this procedure is that the Nextgen
templates only cover stars up to 10,000 K, while the Herbig Be stars
in our sample (V1685 Cyg, AS 442, and MWC 1080) have higher effective
temperatures.   Such hot stars typically have shallower Br$\gamma$
absorption than cooler A stars, since the H ionization fraction in the
stellar photosphere is higher. Thus the circumstellar Br$\gamma$
spectra for these objects may underestimate the true line-to-continuum
ratios.

\subsection{NIRSPEC Data}
For four of our sample objects, VV Ser, AS 353 A, V1685 Cyg, and AS 442,
we obtained NIRSPEC data on 2011 September 15.  
We used the high-dispersion mode with the 3 pixel
slit, which provides a resolving power of $R \approx 24,000$.
The NIRSPEC-7 filter was used with an echelle position of 62.25 and a
cross-disperser position of 35.45. This provided seven spectral orders
covering portions of the wavelength range between 1.99 and 2.39
$\mu$m. Included in these orders are the Br$\gamma$ line and the
$v=2\rightarrow 0$ and $v=4 \rightarrow 2$ CO bandheads.

Spectra were calibrated and extracted using the REDSPEC
package \citep[e.g.,][]{MCLEAN+03}.  Reduction included mapping of
spatial distortions, spectral extraction, wavelength calibration,
heliocentric radial velocity corrections, bias correction, flat 
fielding, and sky subtraction. We divided our target spectra by the
observed spectrum of HD 201320, an A0V star.  We interpolated the A0V
spectrum over the broad Br$\gamma$ absorption feature.  Finally we
multiplied the divided spectra by an appropriate blackbody template to
calibrate the bandpass of the instrument.
We did not attempt to flux calibrate the spectra, but
instead applied scaling factors so that the NIRSPEC data had the same
mean flux level (in a given spectral region) as the KI spectra.

\section{Basic Results and Analysis \label{sec:res}}

In this section we present the calibrated KI and NIRSPEC data, and
discuss physical implications for the distribution of gaseous emission
around our sample objects.  We begin by comparing the KI spectra with
our NIRSPEC observations (where available) and previous data from the
literature.  We then use the observed $V^2$ and $\Delta \phi$ for our
sample to derive estimates of the spatial distributions and centroid
positions of the emission as a function of wavelength.  As we discuss
below, Br$\gamma$ emission
is prevalent from the sample, while CO overtone emission is relatively
rare. We therefore present the results for the Br$\gamma$ and CO
spectral regions separately in Sections \ref{sec:brgres} and
\ref{sec:cores}.

\subsection{Spectra and Spatial Distributions of Br $\gamma$
  Emission \label{sec:brgres}}
\begin{figure}
\plotone{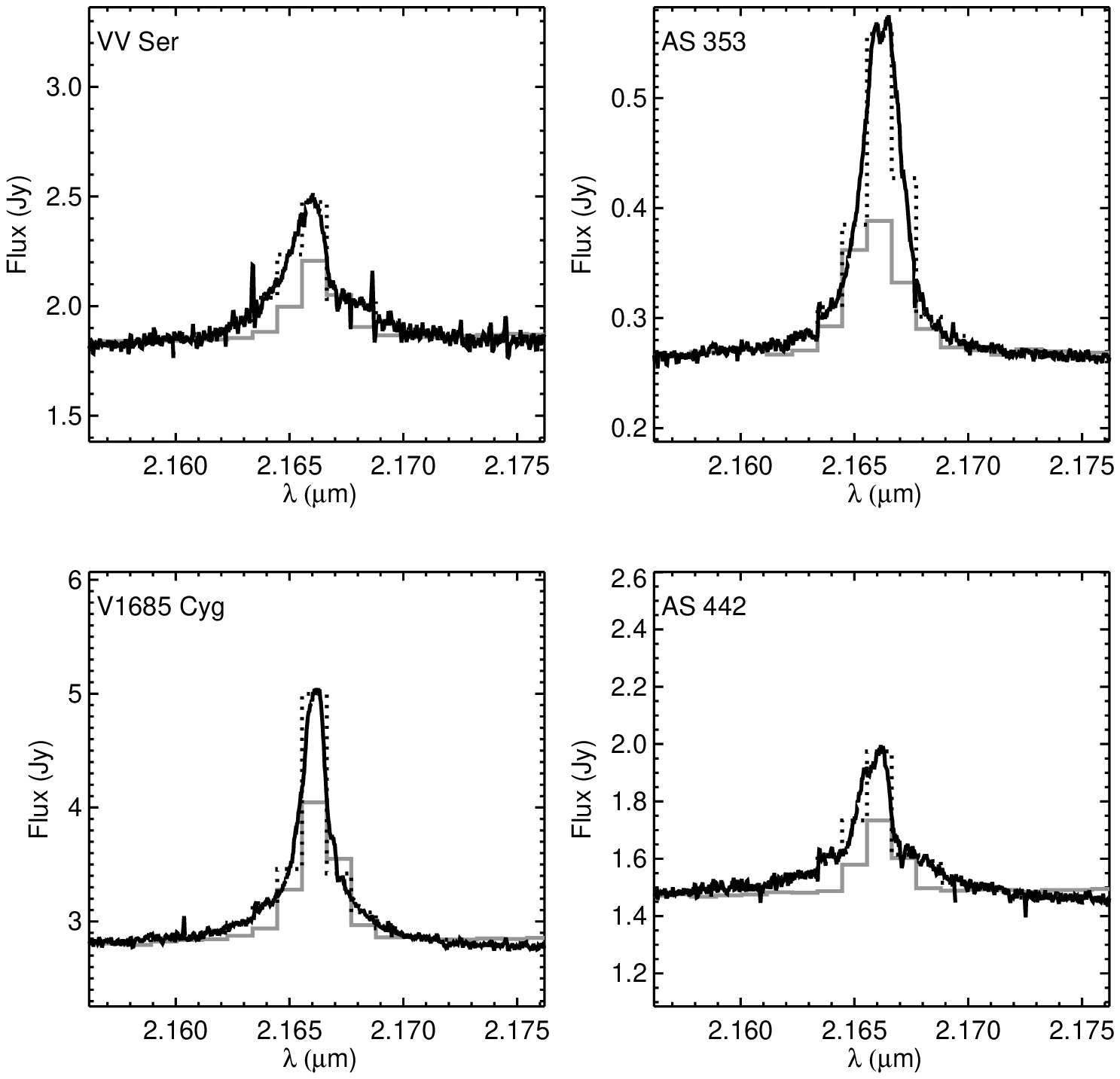}
\caption{Spectra of Br$\gamma$ emission for 
four of our sample objects obtained with NIRSPEC
  (black curves).  We also plot the KI spectra as gray histograms, and
  the NIRSPEC data smoothed to the KI spectral resolution as dotted
  black hitograms.
\label{fig:nspec_brg}}
\end{figure}

Calibrated spectra from KI, observed and with the stellar components
subtracted (Equation \ref{eq:fcirc}), are shown for the Br$\gamma$ 
spectral region in Figure \ref{fig:spectra}.  Calibrated and scaled
NIRSPEC spectra are plotted in Figure \ref{fig:nspec_brg}.   
While the NIRSPEC data have approximately 10 times higher spectral
resolution than the KI data, the spatial resolution of the
seeing-limited NIRSPEC observations is $\sim 1000$ times coarser.

The NIRSPEC data show higher line-to-continuum ratios for the
Br$\gamma$ emission than do the KI data (Figure \ref{fig:nspec_brg}).
Previous near-IR spectroscopy \citep{FE01}, with a spectral dispersion
similar to the NIRSPEC data presented here, also found
line-to-continuum ratios of the Br$\gamma$ emission higher than seen
in our KI data.  The lower line-to-continuum ratios in the KI data are 
due, in part, to the lower spectral dispersion.  With $>10$ times higher
dispersion, the NIRSPEC data suffer less dilution of
strong spectral features.  However the Br$\gamma$ lines are
well-resolved, and dilution is only a minor effect (Figure
\ref{fig:nspec_brg}).

The different fields of view of the
two instruments may also play a role, with the larger field of view of
NIRSPEC potentially sensitive to spatially extended Br$\gamma$
emission outside of the 50 mas field of view of KI.  Br$\gamma$
emission is seen on $\ga 1''$ scales in some young stars, comprising
up to 10\% of the total Br$\gamma$ flux \citep[e.g.,][]{BBM10}.
Extrapolating this finding to scales between 50 mas and $1''$ may
explain the difference between the KI and NIRSPEC spectra.  
We explore this issue quantitatively in Section \ref{sec:modeling}.

From the KI spectra, we compute the equivalent width of the
Br$\gamma$ line for each source.  
Following \citet{EISNER+07c}, we use these equivalent
widths in conjunction with literature photometry and
extinction estimates to determine the Br$\gamma$
line luminosities.   These line luminosities are then converted into
accretion luminosities using empirically-determined relationships for
stars with masses $<1$ M$_{\odot}$ \citep{MHC98} and 
$>1$ M$_{\odot}$ \citep{MENDIGUTIA+11b}.   These
relationships have not been calibrated beyond 6 M$_{\odot}$, and may
not hold for the most massive Herbig Be stars in our sample,
V1685 Cyg and MWC 1080.  Equivalent widths, Br$\gamma$ line
luminosities, and accretion luminosities for our sample are listed in
Table \ref{tab:brg}.

In addition to spectra, we measured squared visibilities for our
targets, which constrain the relative spatial distributions of
continuum and line emission components.
We fit the $V^2$ data for each source, in
each channel, with a simple uniform ring model
\citep[e.g.,][]{EISNER+04} to produce a ``spectral size distribution'' of the
emission.  The spectral size distributions based
on the observed data, and for the circumstellar emission components
(Equation \ref{eq:vdisk}), are shown in Figure \ref{fig:v2}.

For the majority of the observed targets, we infer a distinct spatial
distribution for Br$\gamma$ emission and continuum emission (Figure
\ref{fig:v2}). 
To quantify the relative distributions, we separate the complex visibility
components due to the Br$\gamma$ line and the continuum, and fit each
with uniform ring models \citep[as in][]{EISNER+10}.  The estimated
sizes for the two components are listed for our sample in Table
\ref{tab:brgsizes}. 

The Br$\gamma$ emission is generally found in a significantly more
compact distribution.  However for AS 353 A, V1685 Cyg, and V1331 Cyg,
the inferred sizes for the line and continuum emission components are
comparable.  A few other objects (including DG Tau and DO Tau) show
Br$\gamma$ emission sizes that are a significant fraction of the
continuum emission sizes.

Differential phase signatures in the KI data indicate centroid offsets
of various line emission components.
We convert the measured $\Delta \phi$ values into centroid offsets
as follows:
\begin{equation}
\Delta \theta = \Delta \phi \frac{\lambda}{2\pi B},
\label{eq:dp}
\end{equation}
where $\lambda$ is the observed wavelength and $B$
is the projected baseline length.  
We also compute the centroid offsets for the circumstellar emission 
components (Equation \ref{eq:dpcirc}).   Centroid offsets are plotted
in Figure \ref{fig:dp} for the Br$\gamma$ order.

A number of targets show evidence for position offsets of different
velocity components across the spectrally-resolved Br$\gamma$ line
(Figure \ref{fig:dp}).   For example,
in RY Tau the red- and blue-shifted components of the emission have
opposite (and approximately equal) position offsets.  Such a signature
resembles expectations for a spatially resolved Keplerian disk of
emission.  In contrast, MWC 275 shows a significant positional offset
of only the red-shifted emission component, perhaps arguing for a
bipolar infall where one lobe is (partially) obscured from view.  We
model these signatures quantitatively below in Section
\ref{sec:modeling}.

\subsection{Spectra and Spatial Distributions of CO Overtone
  Emission \label{sec:cores}}
\begin{figure}
\plotone{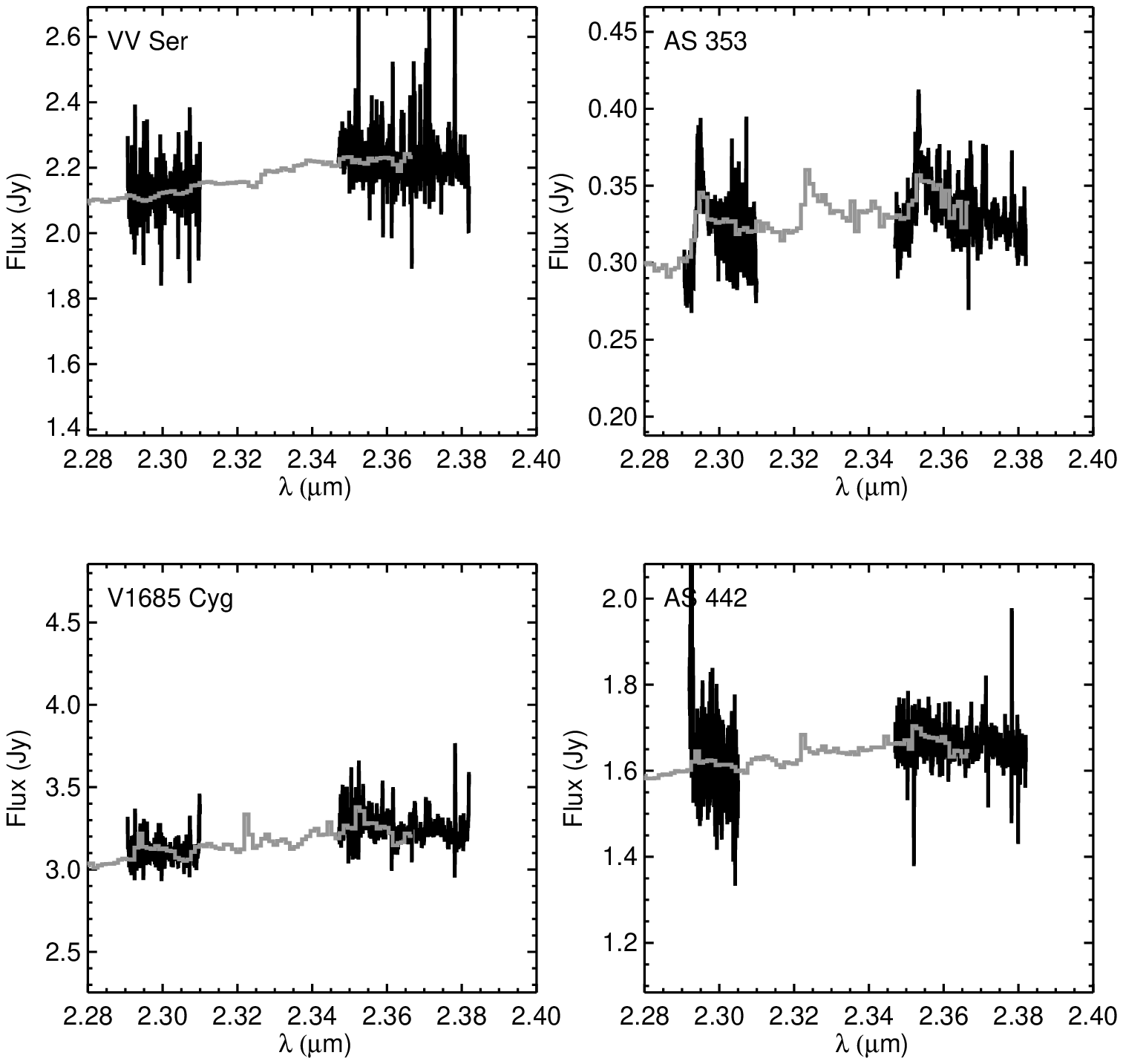}
\caption{Spectra of CO overtone emission from 
  four of our sample objects obtained with NIRSPEC
  (black curves).  These spectra, obtained in two separate orders,
  cover the $v=2\rightarrow 0$ and $v=4\rightarrow2$ bandheads.
  We also plot the KI spectra as gray histograms.
\label{fig:nspec_co}}
\end{figure}

KI and NIRSPEC spectra of our sample, covering the spectral region
including the CO overtone bandheads, are shown in Figures
\ref{fig:cospectra} and \ref{fig:nspec_co}.  Spectral size
distributions, and centroid offsets across the CO bandheads, are shown
in Figures \ref{fig:cov2} and \ref{fig:codp}, respectively.

Of the 21 objects in our sample, only two---AS 353 A and V1331
Cyg---show strong CO overtone emission.  CO emission has been
observed previously for both of these
\citep[e.g.,][]{CARR89,BISCAYA+97,PGS03}.
The line-to-continuum ratio observed here for V1331 Cyg is comparable
to that seen in previous observations, including ones with
similar resolution to the KI data \citep{BISCAYA+97,EISNER+13}.  The
line-to-continuum ratio of AS 353 A in our KI and NIRSPEC data is
somewhat higher than seen in previous observations
\citep{CARR89,PGS03}, although there does seem to be some variability
in the CO emission strength \citep{BISCAYA+97}.

In addition to AS 353 A and V1331 Cyg, a number of targets show
tentative evidence of CO overtone emission (Figure
\ref{fig:cospectra}).  DG Tau and DO Tau both exhibit
circumstellar CO emission features, although such features are absent
in the calibrated (star+circumstellar) spectra.  There are also hints
of circumstellar CO emission in DK Tau A and GK Tau, although the
calibrated spectra show CO absorption.  Thus, any inferred CO emission
in these objects depends on the accuracy of the removal of stellar CO
absorption.

In Section \ref{sec:obs} we suggested that CO overtone absorption
features in faint objects might appear weaker in observed spectra
 than their intrinsic values, perhaps because of uncorrected thermal
background emission.  
DG Tau, DK Tau A, GK Tau, and DO Tau are all among the
faintest objects in the sample, where uncorrected background emission
would have a larger effect.  The small apparent CO emission in DK Tau
A and GK Tau is likely an artifact of the calibration process.
However, DG Tau is known to show CO
overtone emission \citep[e.g.,][]{CARR89}, and the signal in DO Tau is
fairly large. 

Given that AS 353 A and V1331 Cyg exhibit CO emission at levels similar
to (or higher than) previous observations, it seems improbable that
the CO features in DO Tau (which is similarly faint to those targets)
could be completely
spurious.  However, any line enhancement arising from instrumental
errors will affect the data strongly for DO Tau, because the ratio of
stellar to circumstellar flux is high (see Equations
\ref{eq:fcirc}--\ref{eq:vdisk}).  As seen in Figure
\ref{fig:cospectra}, the stellar component is approximately the same
brightness as the circumstellar component in the CO spectral region.
In contrast, errors for DG Tau will be less severe, since its stellar
flux is only a small fraction of its circumstellar flux (Figure
\ref{fig:cospectra}).

V1685 Cyg and AS 442 also show weak CO emission features, while MWC
275 shows apparent CO overtone absorption in its
circumstellar spectrum (Figure \ref{fig:cospectra}).  These are
hot stars, and these tentative detections do not rely on accurate
subtraction of the stellar photospheric spectra.  
However calibration artifacts (e.g., due to mismatches between
observed and template spectra for late-type calibrators) 
may be responsible for the apparent features.  
Recent CO overtone spectra of these
sources show no features in AS 442 or MWC 275, but possible evidence for
weak CO emission in V1685 Cyg \citep{EISNER+13}.  
Strong features seen in the $V^2$ and differential phase data for V1685 Cyg
(Figures \ref{fig:cov2} and \ref{fig:codp})  strengthen the case for
real CO emission in this source.  We therefore discount the apparent
CO features in MWC 275 and AS 442, but include V1685 Cyg in our
further analysis of the CO spectral region.

For our analysis we select those objects with the most
significant detections of CO emission features.
This sub-sample includes DG Tau, DO Tau, AS 353 A, V1685 Cyg, and V1331
Cyg.  Since the CO detections for DG Tau and DO Tau depend on the
removal of the stellar photospheric spectrum, these should be treated
with some caution.  The detection rate in our sample, 5/21, is similar
to the detection rate in previous surveys for CO overtone emission
\citep[e.g., 9/40 in][]{CARR89}.  

\begin{figure}
\plotone{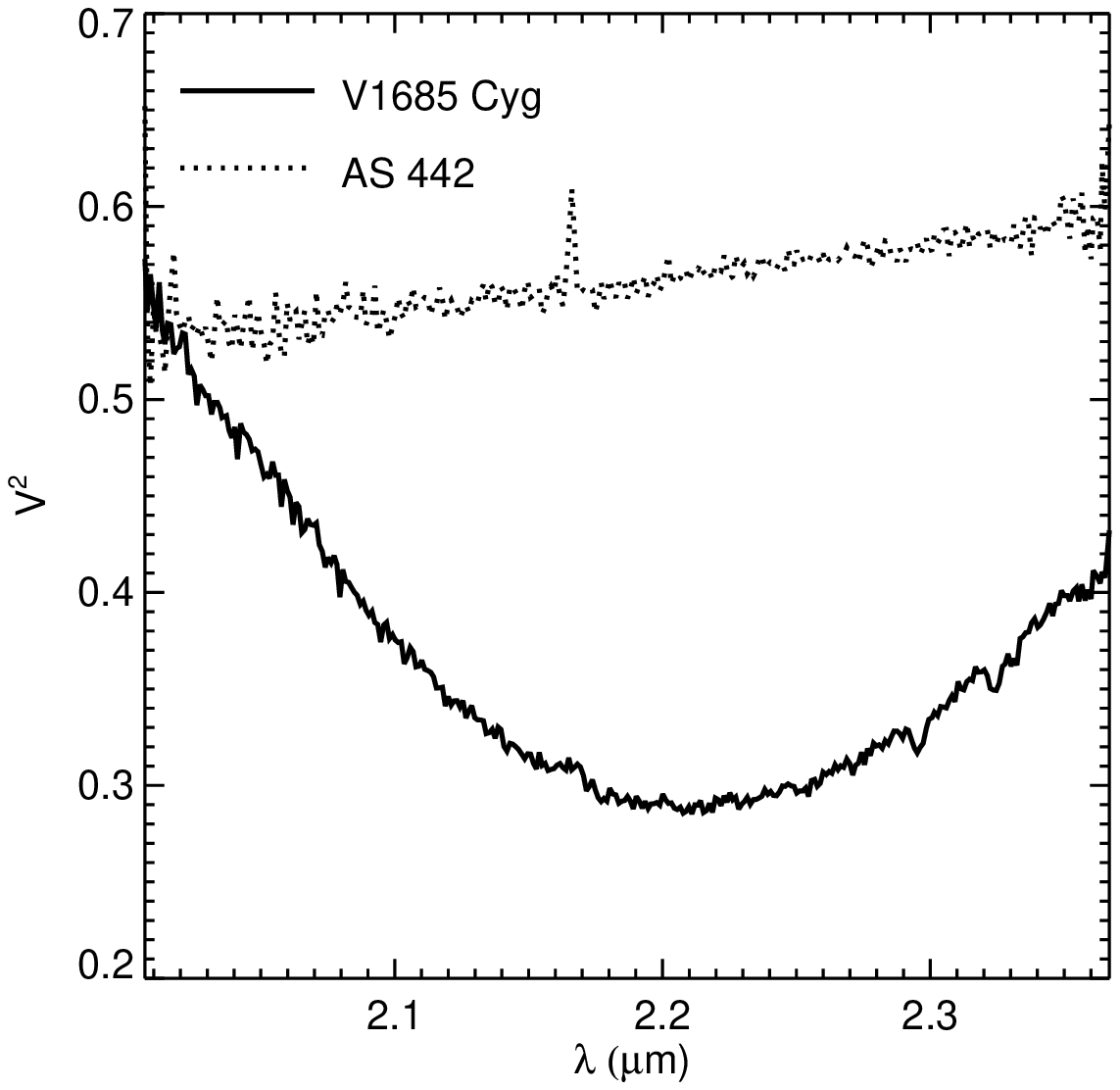}
\caption{$V^2$ observed with KI across the entire $K$-band for V1685
  Cyg (solid curve) and AS 442 (dotted curve), illustrating the
  unusual spectral behavior of the V1685 Cyg visibilities.  AS 442 is
  representative of other objects in our sample, and is chosen for
  this plot because it shared the same calibration as V1685 Cyg (Table
  \ref{tab:sample}).
\label{fig:v1685cyg}}
\end{figure}

For this sub-sample of CO overtone emitters, we estimate the spectral
size distribution in the CO region.  The continuum sizes in
the CO region (Table
\ref{tab:cosizes}) may differ from those in the Br$\gamma$ region
(Table \ref{tab:brgsizes}).  Small, monotonic, changes with wavelength
are expected (for all targets) if continuum emission arises in
extended distributions of circumstellar matter exhibiting temperature
gradients.  These broadband slopes were
modeled for many of our targets in previous work
\citep[e.g.,][]{EISNER+07a,EISNER+09}.  
However the observed $V^2$ for V1685 Cyg show unusually large changes
with wavelength, including a non-monotonic dependence.  
We illustrate this behavior
in Figure \ref{fig:v1685cyg}, comparing V1685 Cyg to AS 442, the
latter representative of the spectral $V^2$ behavior across the sample.
We do not have a clear explanation for the $V^2$ observed from V1685
Cyg, but suggest below that binarity may help to explain the peculiar
continuum shape.

Separating the continuum
and line emission components, we determine approximate spatial
locations of the first two CO overtone bandheads (Table
\ref{tab:cosizes}).   For AS 353 A and V1331
Cyg, the inferred sizes of the emission arising in the first two overtone
bandheads are comparable to the continuum sizes.   
For V1685 Cyg the CO emission appears more
widely distributed than the continuum emission, although the estimated
sizes of CO-emitting regions have large uncertainties.  In contrast,
the CO emission in DG Tau and DO Tau appears more compactly
distributed than the continuum emission.

The centroid offsets determined for the CO spectral region are noisy
(Figure \ref{fig:codp}), and it is difficult to discern clear
signatures at the wavelengths of the CO overtone bandheads.  However
one target, V1685 Cyg, does appear to have emission from the CO
bandheads that is spatially offset from the continuum emission.

\section{Kinematic Modeling \label{sec:modeling}}
Following \citet{EISNER+10} we model our KI data with both disk and
infall/outflow models.  The properties of these models are similar to
\citet{EISNER+10}, although we explore a substantially larger
range of parameter values here.  We describe the models below, and
refer to \citet{EISNER+10} for details of how each model parameter
affects synthetic data.  We will use the same models developed for the
KI data to interpret the NIRSPEC spectra.  

The disk and outflow models are both fitted to the KI Br$\gamma$ data.  
The KI data in the CO spectral region are typically noisier than in
the Br$\gamma$ region, making it difficult to constrain models well.
We will therefore pursue a less
rigorous modeling effort in this spectral region, restricting our
attention to disk models, and comparing to
data rather than performing rigorous fits.

As in \citet{EISNER+10}  we make the simple assumption that continuum
emission is confined to a ring whose annular width is 20\% of its
inner radius.  This assumption is consistent with models where
continuum emission traces dust near the sublimation radius
\citep[e.g.,][]{DDN01,EISNER+04,IN05}.
The size of the continuum ring is determined from a fit
of a uniform ring model to $V^2$ data in spectral regions adjacent to
those where Br$\gamma$ or CO emission is observed (Tables
\ref{tab:brgsizes} and \ref{tab:cosizes}).  We determine the ring
radius, $R_{\rm ring}$,
for all position angles and inclinations considered for
our gaseous disk models. The temperature of the ring is set so that 
the continuum flux level matches the observed continuum fluxes.
Neither $R_{\rm ring}$ nor $T_{\rm ring}$ are free parameters in the
models discussed below.

\subsection{Keplerian Disk Models \label{sec:disk}}
We assume that while the continuum emission is confined to a ring, the
gaseous emission resides in a disk extending from $R_{\rm in}$ to
$R_{\rm out}$.  Both the disk and the ring of continuum emission have a
common  inclination,
$i$, and position angle, $PA$, that are  free parameters.

The brightness profile of the gaseous disk is parameterized with 
a power-law, 
\begin{equation}
B_{\rm disk} (R) = B_{\rm in} \left(\frac{R}{R_{\rm in}}\right)^{-\alpha}. 
\end{equation}
The value of $\alpha$ depends on the temperature profile and surface
density profile
of the disk.  For optically thin gas following a surface density
profile similar to the minimum mass solar nebula \citep{WEID+77}, and
heated by thermal radiation from the central star, $\alpha \approx 2$.
However the temperature and surface density profiles are
not well-constrained, and so we leave 
$\alpha$ as a free parameter in our modeling.   Instead of using
$B_{\rm in}$, we normalize the brightness profile so that the
resulting spectrum has a specified line-to-continuum ratio, $L/C$.  
$L/C$ is defined as the ratio of the total flux of the gaseous
emission, integrated over space and velocity, to the continuum flux
level.

We assume the gas to be in Keplerian rotation, with a radial velocity profile,
\begin{equation}
v_{\rm obs}(R) = \sqrt{\frac{G M_{\ast}}{R}} \cos (\theta) \sin (i).
\end{equation}
Here, $M_{\ast}$ is the stellar mass, $\theta$ is the azimuthal angle
in the disk for a given $(x,y)$ in cartesian coordinates,
and $i$ is the disk inclination.  For simplicity,
we do not enter exact values of $M_{\ast}$ for
each source into the model (these are not determined
to high accuracy for most objects).  Rather, we assume
a stellar mass of 1 M$_{\odot}$ for the T Tauri stars in our sample, 
3 M$_{\odot}$ for the Herbig Ae stars, and 10 M$_{\odot}$ for
the Herbig Be stars V1685 Cyg and MWC 1080.

To minimize the number of free parameters in our model grid we set
$R_{\rm out}=R_{\rm ring}$.  Thus we assume that any gaseous emission
that may exist at stellocentric radii larger than $R_{\rm ring}$ is
hidden by the optically thick dust disk.  This may not be an accurate
assumption, especially for CO emission where models suggest excitation
in disk surface layers is possible \citep[e.g.,][]{GNI04,GH08}.  We
therefore relax this assumption when modeling the CO emission (Section
\ref{sec:co}).

We explore the following values of free parameters in our grid of
Keplerian disk models: $R_{\rm in} = 0.01,0.02,0.03,0.04,$ and 0.05
AU; position angle = $0,25,50,75,$ and 90$^{\circ}$; inclination =
$5,25,50,$ and $75^{\circ}$; $L/C = 0.1,0.25,0.5,0.75,$ and 1.0;
and $\alpha = 2,3,$ and 4.  Our grid contains 1500 models, compared to
the 243 disk models computed in \citet{EISNER+10}.

\subsection{Infall/Outflow Models}
We construct models consisting of a face-on ring of continuum emission
(described above) and a bipolar conical gas infall/outflow.
We assume an infall/outflow cone with an
opening angle of $5^{\circ}$, a position angle, $PA$, and an
inclination with respect to the plane of the sky, $\phi$.  We allow the infall/outflow
to extend from an outer radius, $R_{\rm out}$ to an inner radius,
$R_{\rm in}$.  In contrast to the disk model considered above, $R_{\rm
  out} \ne R_{\rm ring}$ here.

The velocity of material in this cone is described as
a radial power-law:
\begin{equation}
v_{\rm obs}(R) = v_{\rm in} \left(\frac{R}{R_{\rm in}}\right)^{-\beta} \sin \phi.
\label{eq:outflow_vr}
\end{equation}
Here the velocity of material at the inner edge of the infall/outflow
structure, $v_{\rm in}$, is chosen to produce a specified linewidth of
the emission,
\begin{equation}
\Delta v = v_{\rm in} \sin \phi.
\label{eq:outflow_vin}
\end{equation} 
Examination of Equations \ref{eq:outflow_vr} and \ref{eq:outflow_vin} shows that the
velocity profile depends on $\Delta v$, $R_{\rm in}$, and $\beta$, but not directly
on $\phi$.   The geometry of the outflow of the sky depends on $PA$, as well
as on $R_{\rm in}$ and $R_{\rm out}$, but again not explicitly on $\phi$.  We thus
fix $\phi=45^{\circ}$ in our models.

We include another cone, reflected through the
origin, with the same velocity profile multiplied by $-1$.  
The brightness distribution of the infall/outflow cones is
\begin{equation}
B_{\rm infall/outflow}(R) =  B_{\rm in} \left(\frac{R}{R_{\rm in}}\right)^{-\alpha},
\end{equation}
where $B_{\rm in}$ is chosen to reproduce a specified line-to-continuum ratio.
As above, $L/C$ is defined as the total, integrated flux of the gaseous emission
over the continuum flux.
$\alpha$ combines the temperature and surface
density profile of the infall/outflow structure into a single parameter.
Finally, we include as a free parameter a factor by which the flux in 
one of the two cones or ``poles'' or the infall/outflow may be scaled.
We denote this factor as $f_{\rm a}$, since it represents an asymmetry in the model.

We explore the following values of free parameters in our grid of
infall/outflow models: $R_{\rm in} = 0.01,0.02,0.03,$ and 0.04,
AU; $R_{\rm out} = 0.05,0.1,0.5,$ and 1 AU; position angle =
$0,20,40,60,$ and 80$^{\circ}$; $\beta = 1,2,$ and 3; $\Delta v = 250,
375,$ and 500 km s$^{-1}$; $L/C = 0.1,0.5,$ and 1.0; $\alpha = 2,3,$
and 4; and $f_{\rm a} = 0.1,0.5,$ and 1.0.  
The computed grid includes 19440 infall/outflow models, compared to
the 2916 models included in \citet{EISNER+10}.

\subsection{Results for the Br$\gamma$ Spectral Region \label{sec:brg}}

After computing synthetic fluxes, $V^2$, and $\Delta \phi$ values for
grids of both disk and infall/outflow models, we compute 
the $\chi^2$ residuals between these and the observed quantities from KI.
The total $\chi^2$ is given by
\begin{equation}
\chi^2_{\rm tot} = \sqrt{(\chi^2_{\rm flux})^2 + (\chi^2_{\rm V^2})^2 +
 (\chi^2_{\rm \Delta \phi})^2}.
\end{equation}
Finally, we minimize $\chi^2_{\rm tot}$ to determine the ``best-fit'' model.
Even though this grid of models samples significantly more parameter
values than previous work, it remains sparsely sampled, and hence we
can not give rigorous error intervals on the fitted parameters.

The best-fit models are illustrated in Figure \ref{fig:brgmods}.
Reduced $\chi^2$ values, and parameters of the best-fitting models,
are listed in Table \ref{tab:results}.  Disk models and infall/outflow
models generally produce fits of comparable quality.  This differs
from the modeling in \citet{EISNER+10}, and reflects the better fits
of disk models that are achieved with the larger grid of models used
here.  However, we do confirm that for objects with the highest
signal-to-noise data, infall/outflow models are generally preferred.
This preference is based largely on the differential phase data,
which can constrain asymmetric structures compatible with
infall/outflow models but inconsistent with disk models.
For the brighter sources in our sample the differential phase data
have high enough signal-to-noise to discern such asymmetries.  

For a handful of targets neither model provides a particularly good
fit to the combined KI dataset.  Formally the worst fits are obtained
for RW Aur, V2508 Oph, AS 353 A, and V1685 Cyg (Table
\ref{tab:results}), although the $\chi_{\rm r}^2$ values are all
smaller than unity, suggesting that none of the fits are terrible.
Examination of Figure \ref{fig:brgmods} suggests that the fits for RW
Aur, V2508 Oph, and AS 353 A may have higher $\chi_{\rm r}^2$
values simply because of somewhat higher scatter in the data.  However
V1685 Cyg is a bright source with high signal-to-noise, and the fits
are clearly inconsistent with the observed $V^2$ data.  The model
places Br$\gamma$ emission on compact scales, while the lack of a
clear $V^2$ signature suggests a similar spatial distribution of line
and continuum emission.  The best-fit model selected the largest
values of $R_{\rm in}$ and $R_{\rm out}$ available, 0.04 AU and 1 AU,
respectively.  For comparison, the continuum size is 1.17 AU (Table
\ref{tab:brgsizes}).  A better fit to the data would likely be
achieved with a Br$\gamma$ distribution centered closer to the
continuum emission region.

We noted above that the Br$\gamma$ spectra measured with NIRSPEC have
higher line-to-continuum ratios than the KI spectra.  For this reason
we did not attempt to model the NIRSPEC data at the same time as the
KI dataset.  We will now attempt to reconcile the best-fit
models, determined for the KI data, with the NIRSPEC spectra.

\begin{figure}
\plotone{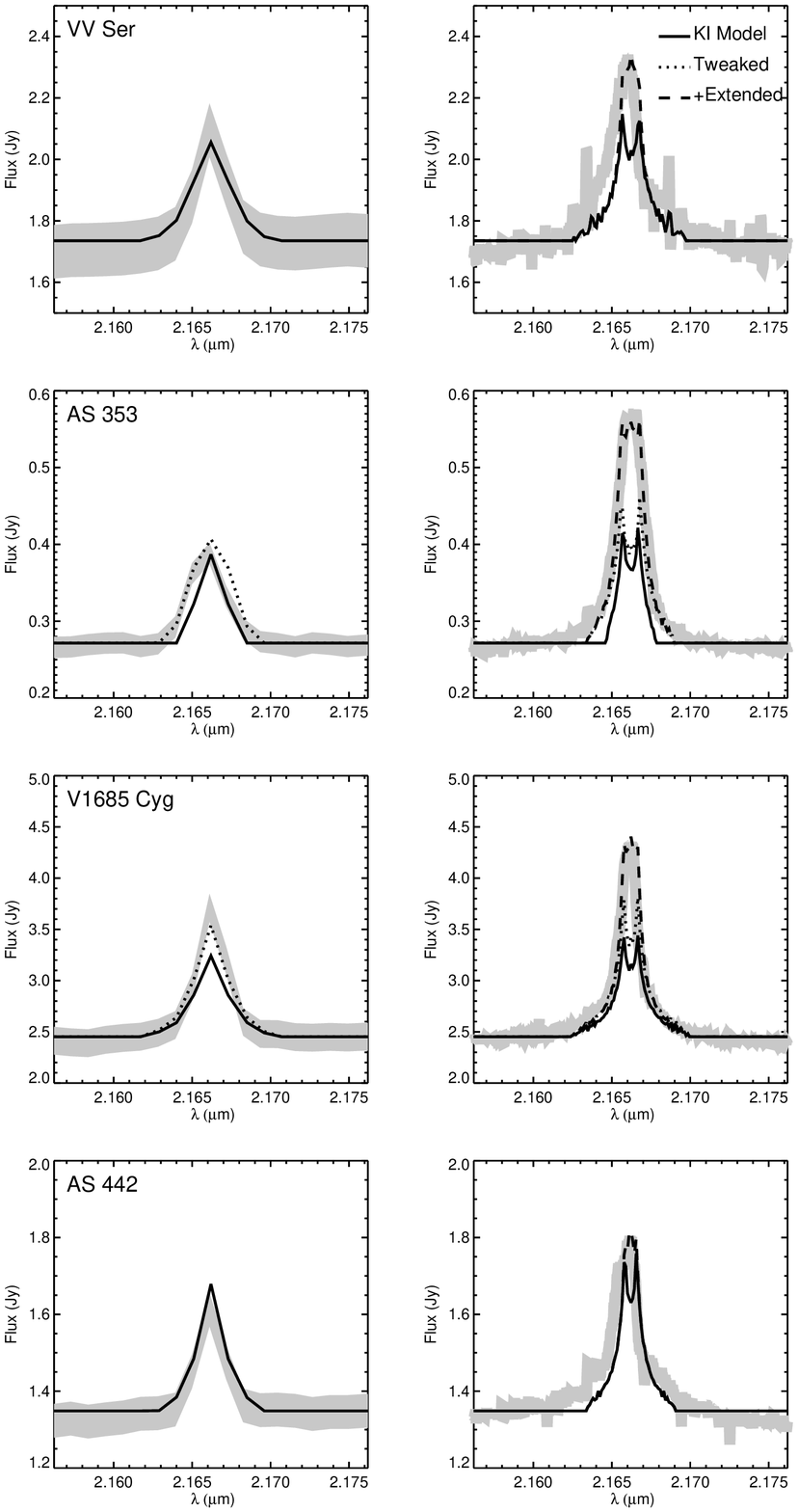}
\caption{Observed and synthetic Br$\gamma$ spectra for targets
  observed by NIRSPEC.
  Left panels show KI spectra and right panels show
  NIRSPEC spectra (gray regions).  Solid curves are
  the predicted spectra for the best-fit model determined from the KI
  dataset.   In some cases, we made minor tweaks to this model to
  better fit the NIRSPEC data, without materially affecting the
  quality of the fits to the KI data (dotted curves).  In all cases,
  the best-fit models do not fit the NIRSPEC data well.  To improve
  the fits, for each source we add a Gaussian 
  with FWHM between 25 and 75 km s$^{-1}$.  This range of FWHM
  corresponds to Keplerian velocities at
  stellocentric radii of a few AU for our targets. The Gaussian component
  thus represents extended emission that might be outside of the KI
  FOV, but well within the NIRSPEC FOV.
\label{fig:nspecmod_brg}}
\end{figure}

First, we compute the expected Br$\gamma$ spectra for our best-fit
models at the spectral resolution of the NIRSPEC data.  Given the
similar quality of the disk and outflow model fits for these objects
(Table \ref{tab:results}), we adopt the disk models for computational
expediency.   We adjusted $L/C$ to maintain a constant line flux
between the KI and NIRSPEC data, but left all other model parameters
unchanged.   Synthetic NIRSPEC spectra for the best-fit disk models
are shown as solid curves in Figure \ref{fig:nspecmod_brg}.  

For AS 353 A and V1685 Cyg, the best-fit model did not fit
the KI spectra perfectly (owing to limited model grid sampling), and so we
tweaked the model to yield superior fits to the NIRSPEC data without
altering the fit quality to the KI data.  For AS 353 A we changed
$R_{\rm in}$ from 0.01 to 0.003 AU, and for V1685 Cyg we increased
$L/C$ by $\sim 30\%$.   These ``tweaked'' models are shown with dotted
lines in Figure \ref{fig:nspecmod_brg}.  

As expected, synthetic spectra from our best-fit models, after minor
tweaking where appropriate, under-predict the Br$\gamma$ lines seen in
NIRSPEC data.  We speculated above that NIRSPEC may be sensitive to
spatially extended Br$\gamma$ emission on scales beyond 
the sensitivity of KI.  The relevant spatial scales would be beyond
a few AU, which would correspond to Keplerian velocities of $\sim
50$ km s$^{-1}$ for these objects.  We added Gaussians with FWHM
between 25 and 75 km s$^{-1}$ to the synthetic spectra, to simulate
the addition of an extended emission component.  The dashed curves in
Figure \ref{fig:nspecmod_brg} show that adding these low-velocity,
possibly spatially extended emission components, can indeed produce
resultant spectra similar to the observed NIRSPEC spectra.

It is interesting to note that the NIRSPEC data are sensitive not only
to larger spatial scales, but also potentially to smaller spatial
scales than the KI data.  The higher dispersion of NIRSPEC means that
more extreme velocities--which likely trace matter at the smallest
stellocentric radii--can be constrained.  The tweaked model for AS 353 A
is a case-in-point, where the NIRSPEC data compel us to consider a
model extending to higher velocities and smaller spatial scales.

Figure \ref{fig:nspecmod_brg} shows that the best-fit disk
models produce double-peaked line profiles at the spectral resolution
of the NIRSPEC data.  With the addition of spatially extended
emission, the central dip in the line profiles is filled in, and the
observed NIRSPEC data can be reproduced.  If our explanation of the
discrepancy between NIRSPEC and KI spectra as the result of extended
emission is incorrect, then disk models would be ruled out. 
As discussed in
\citet{EISNER+10}, infall/outflow models can produce narrower line
profiles while still reproducing the KI $V^2$ and differential phase
data.  However infall/outflow models would still
struggle to simultaneously reproduce the NIRSPEC and KI data (which
are not consistent with each other),
precluding a simple alternative explanation.

\subsection{Modeling Results for the CO Spectral Region \label{sec:co}}

Given the relatively poor quality of the KI data in the CO spectral
region, we do not attempt a rigorous $\chi^2$ minimization over our
model grid.  Instead we assume that the CO emission traces Keplerian
disks in our targets \citep[as suggested by previous line profile
modeling; e.g.,][]{NAJITA+96}, and perform a by-eye minimization over varied
model parameters.  We do not consider infall/outflow models for the CO
data.  We perform this analysis for the sub-sample of objects with
detected CO overtone emission, listed in Table \ref{tab:cosizes}.

For each source we use as a starting point the
best-fit disk model determined for the Br$\gamma$ spectral region
(Table \ref{tab:results}). For the Br$\gamma$ modeling, we
described the radial brightness distribution with a singe parameter,
$\alpha$.  This is appropriate, since the Br$\gamma$ emission arises
from a single transition, and so there is no explicit dependence of
the synthetic spectrum on temperature (although the transition
requires population of the $n=7$ state of hydrogen, which implies a
gas temperature $\ga 10,000$ K).  

In contrast, the 
CO bandheads are made up of multiple transitions, and  
the synthetic spectrum depends explicitly on
temperature.  We therefore replace $\alpha$ with exponents on
power-law profiles of gas temperature and surface density.  We assume
that the temperature power law has an exponent of $-0.5$, similar to
values used in previous modeling of CO overtone emission \citep[e.g.,
$\sim -0.4 - -0.75$ used by][]{NAJITA+96,CTN04}.
We take the surface density power law exponent to be
$-1.5$, as calculated for the protosolar nebula \citep{WEID+77}; this
value is also consistent with previous modeling of spatially resolved
observations of inner disk emission  across a broad wavelength range
\citep[e.g.,][]{KPO08}.
These profiles are used with CO line opacities from HITEMP
\citep{ROTHMAN+05} to calculate the emergent spectrum as a function of
disk radius.

\begin{figure}
\plotone{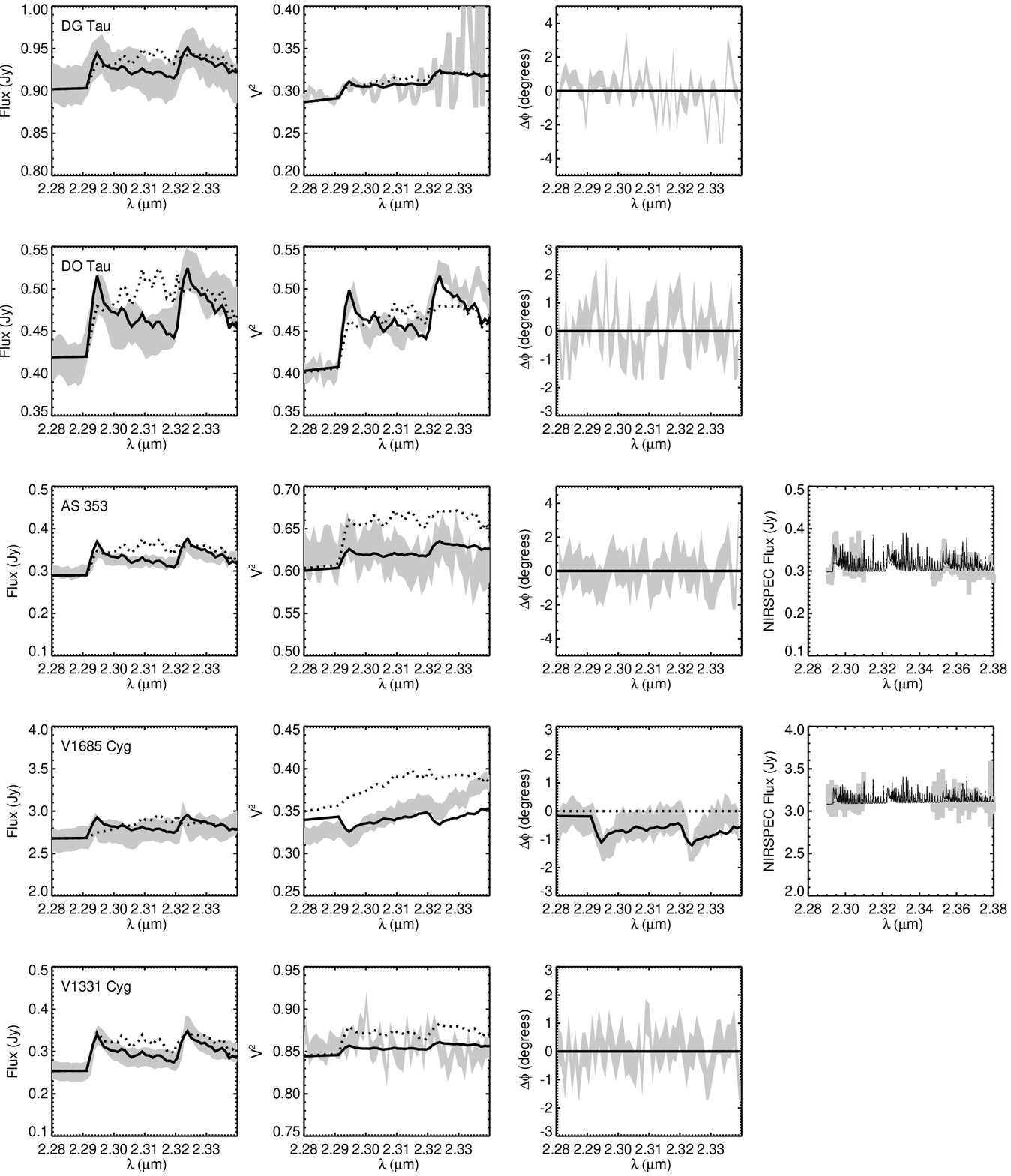}
\caption{Synthetic spectra, $V^2$, and $\Delta \phi$, and
  computed for disk models, compared to KI data and NIRSPEC spectra
  (where available).  Dotted curves show synthetic
  data for the best-fit models determined for the Br$\gamma$ emission,
  listed in Table \ref{tab:results}.  To better fit the data, we
  adjusted these models by varying $R_{\rm in}$ and $R_{\rm out}$ for
  all sources, and
  introducing a spatial offset between CO and continuum for V1685 Cyg
  (solid curves).  As described in Section \ref{sec:co}, these models
  confine CO emission to disk regions with fractional widths of 20\%
  and temperatures $\ga 3000$ K.
\label{fig:comodels}}  
\end{figure}

Synthetic spectra, $V^2$, and $\Delta \phi$ for these models 
are shown as dotted curves in Figure \ref{fig:comodels}.   These
models do not match the data well. In all cases the models produce too
much emission in between the first two overtone bandheads.  The lines
in this spectral region are relatively stronger in cooler CO gas, at
temperatures $\la 2,000$ K.  Because our models assume a radial
temperature gradient, they all include contributions from such cool
gas at larger radii.

The synthetic data for the models generated from Table
\ref{tab:results} are also discrepant from the $V^2$ data in most
cases.  For DO Tau, the model $V^2$ fall below the data, suggesting
that the model CO emission is more extended than implied by the
observations.  For AS 353 A, V1685 Cyg, and V1331 Cyg, the opposite is
true; models imply CO distributions more compact than the data.  One
target, V1685 Cyg, also shows centroid offsets that are not reproduced
with this simple disk model.

To find models that reproduce the observations better, we allow
$R_{\rm in}$ and $R_{\rm out}$ to vary.  Given that cool CO emission
is inconsistent with the data, we make the further assumption that
$R_{\rm out} = 1.2 R_{\rm in}$;  i.e., we restrict the CO emission to
a narrow spatial and temperature distribution.  Synthetic data from
models with CO confined to such rings are shown as solid curves in
Figure \ref{fig:comodels}.  These models provide superior fits
to the data than the ones based entirely on the Br$\gamma$ emission
geometry.

For all sources, our models produce CO emission at temperatures $\ga
3000$ K; since the HITEMP opacities do not cover higher temperatures,
we can not constrain $T$ more precisely.   The inner radius of the CO
emission in models for AS 353 A and V1331 Cyg are comparable to the
continuum radii.  For V1685 Cyg, the model places the CO emission at
radii $\sim 30\%$ larger than the continuum emission.  In contrast,
the CO emission for DO Tau appears to be located at much smaller radii
than the continuum: Figure \ref{fig:comodels} shows a model with CO
located at $R_{\rm in}= 0.01$ AU.  Finally, for DG Tau, the
CO emission is located interior to the continuum emission, although
this is not well-constrained by the data.  The ring model shown in
Figure \ref{fig:comodels} represents CO at a radius of $\sim 0.1$ AU.

V1685 Cyg also
shows a clear differential phase signature in the CO bandhead
wavelengths.  To model this, we introduce a spatial offset of $\sim
0.5$ mas ($\approx 0.5$ AU)
between the centroid of the CO emission ring and the centroid of the
continuum emission.  Physically, such an offset seems difficult to
explain with a model where gas and dust trace the same physical
component.  However if the CO traces a different source than the
continuum, for example a binary companion at $\sim 0.5$ AU separation,
then such a large offset could arise (see Section \ref{sec:v1685cyg}).

\section{Discussion}

\subsection{Continuum Emission \label{sec:cont}}
The inferred sizes of continuum emission regions for our sample are
between $\sim 0.1$ and 1.3 AU (Table \ref{tab:brgsizes}), 
compatible with previous measurements
\citep[e.g.,][]{EISNER+07b,EISNER+09}.  In Figure \ref{fig:gasdist},
we plot the continuum sizes for each source, ordered by  source
luminosity.  The luminosity is the sum of the stellar luminosity (from the
literature) and the accretion luminosity listed in Table \ref{tab:brg}.
The continuum size generally increases with source luminosity, and the
relationship is consistent with an
origin of the continuum emission in dust sublimation fronts
\citep[e.g.,][]{MM02,EISNER+04,MONNIER+05}.

However the continuum may not trace dust only.  Previous
work suggested that the continuum emission from young stars---including
many in our sample---contained contributions from matter hotter than
dust sublimation temperatures, perhaps tracing free-free emission from H or
H$^-$ \citep[e.g.,][]{EISNER+09}.  If some of the continuum emission is
due to opacity from hot gas, measured continuum radii may
lie between the compactly distributed  hot gas and cooler dust at
larger radii.

\subsection{Br$\gamma$ Emission}
Most objects in our sample show Br$\gamma$ emission with a compact
distribution.  Inferred sizes of the Br$\gamma$ emission are typically
$\la 0.05$ AU, considerably smaller than continuum sizes (Table
\ref{tab:brgsizes}).  Kinematic modeling confirms that Br$\gamma$
emission extends in to such small stellocentric radii for nearly all
targets (Table \ref{tab:results}). 

Magnetospheric accretion models predict that
matter falling in along stellar magnetic 
field lines will glow brightly near to the stellar surface, where it converts most of its 
gravitational potential energy to radiation, and so we would expect to see Br$\gamma$ 
emission from very close to the stellar surface in this case
\citep[e.g.,][]{MHC98}.  For accretion via a Keplerian disk extending
to the stellar surface, Br$\gamma$ emission would likely trace the
shock at the disk/stellar surface boundary layer \citep[e.g.,][]{LP74}.
The compact Br$\gamma$ emission, 
and the asymmetric differential phases for some sources (Figure
\ref{fig:dp}), suggest asymmetric Br$\gamma$ emission
morphologies compatible with bipolar infall models.  Asymmetric
emission could arise from a tilted dipole accreting preferentially from
one side of the disk, or from symmetric magnetospheric infall where
one lobe is partially obscured by the disk.

Further support for magnetospheric accretion columns as the origin of
the Br$\gamma$ emission comes from
analysis of line profiles observed at higher spectral (but lower
spatial) resolution.  While most of our targets exhibit H$\alpha$
emission showing blueshifted absorption components or P Cygni-like
line profiles indicative of winds \citep[e.g.,][and
references therein]{AVD05,NCT96}, the Br$\gamma$ line profiles are
more symmetric and/or blueshifted, and often show a ``blue shoulder''
\citep{NCT96,FE01,EISNER+07c}.  These properties are all consistent
with infalling material rather than winds \citep{NCT96}.  The high
velocity components of Br$\gamma$ emission seen in our NIRSPEC data,
most prominently in AS 353 A, also suggest an origin deep in the
potential well of the star, pointing towards an accretion column as
the likely origin.

The outer radii of best-fit kinematic models (which may extend to 1 AU;
Table \ref{tab:results}), and the average emission size scales
listed in Table \ref{tab:brgsizes}, suggest that many objects have some
Br$\gamma$ emission distributed beyond the magnetospheric scale.
Indeed, we argued that Br$\gamma$ emission observed by NIRSPEC may
include a component on $\ga$ AU scales.  
For the sample of objects
with more extended average Br$\gamma$ emission distributions (Table
\ref{tab:brgsizes})---which are also those targets with best-fit
models exhibiting shallower brightness profiles (Table
\ref{tab:results})---Br$\gamma$ emission probably traces: 
infalling material over a range of radii; 
outflowing material that is magnetospherically launched from small
radii;  or a combination of both infalling and outflowing material.
In the latter case, the infalling material could produce very compact
emission while more extended winds could fill in the emission
profiles at larger radii.  

\begin{figure}
\plottwo {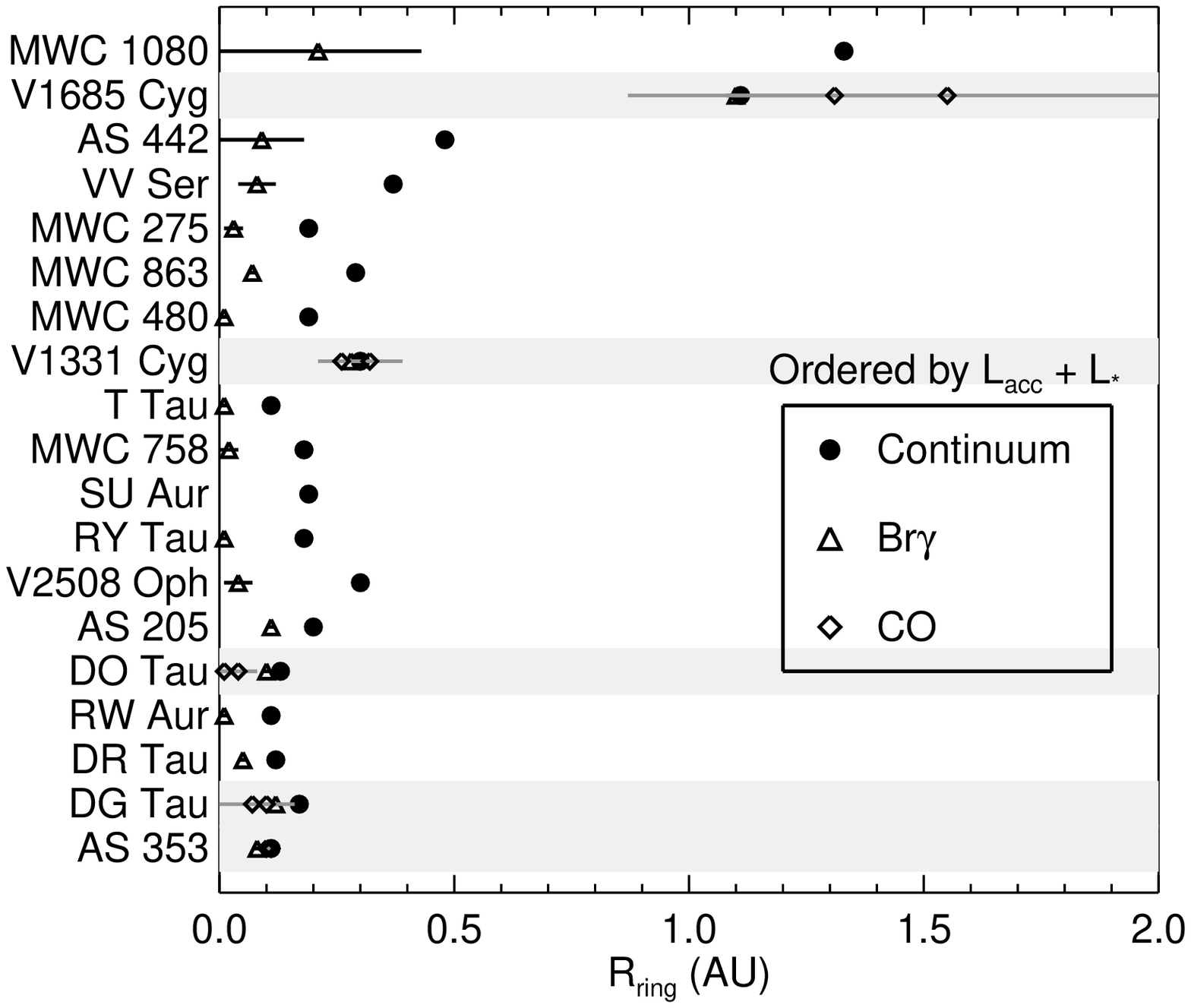}{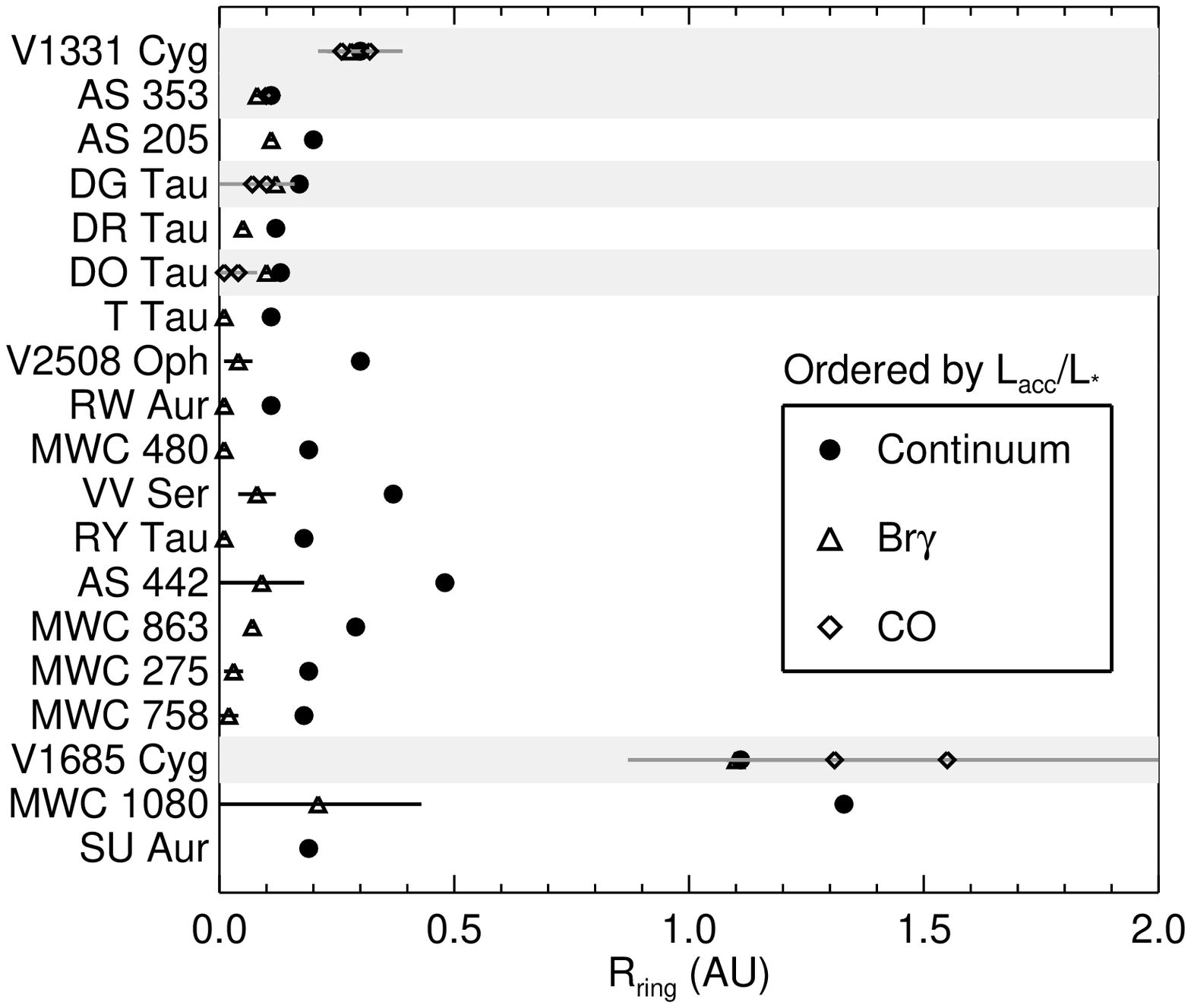}
\caption{Spatial distributions of continuum, Br$\gamma$, and CO
  emission for our sample, based on fitted uniform ring radii (Tables
  \ref{tab:brgsizes} and \ref{tab:cosizes}).  The error bars for the
  sizes of emission in the $v=2\rightarrow 0$ and $v=3\rightarrow 1$
  CO bandheads are plotted as gray lines, while those of the continuum and
  Br$\gamma$ sizes are plotted as black lines.  The error
  bars are often smaller than the symbol sizes.  In the left panel, we
  plot objects in order of total luminosity (the sum of stellar and
  accretion luminosity).  In the right panel, we order objects by
  $L_{\rm acc}/L_{\ast}$.  We highlight the sub-sample of CO overtone
  emitters with light gray shading.
\label{fig:gasdist}}
\end{figure}

We see no obvious correlation between the inferred spatial
distribution of Br$\gamma$ emission and stellar 
properties.  Figure \ref{fig:gasdist} shows that most objects in our
sample, including low-luminosity T Tauri stars and the most luminous
Herbig Be star in the sample, have Br$\gamma$ emission distributed on
more compact scales than the continuum emission.  Similarly, 
the spatial distribution of  Br$\gamma$ emission
shows no clear dependence on accretion luminosity
alone, or on stellar mass.  There is some dependence of the
relative sizes of Br$\gamma$ and continuum emission on the ratio of
accretion to stellar luminosity:  accretion-dominated sources tend to
have relatively more extended Br$\gamma$ emission
(Figure \ref{fig:gasdist}).  However the clearest correlation is that
objects where the Br$\gamma$ emission is distributed on scales
comparable to the continuum all exhibit CO overtone emission.

\subsection{CO Overtone Emission}
CO overtone emission, when detected in our sample, generally has a
spatial distribution similar to the continuum emission
(Table \ref{tab:cosizes}).  For DG Tau and DO Tau, we inferred a more
compact distribution of CO emission relative to the continuum,
although the difference is not highly significant for DG Tau.
Furthermore, uncertainties arising from subtraction of late-type
stellar photospheric CO
absorption in DO Tau may lead to an inflated line-to-continuum ratio,
and perhaps an underestimate of the stellocentric radius of the
circumstellar CO emission.

Figure \ref{fig:gasdist} shows that the location of the  CO overtone
emission is generally similar to the continuum distribution.
 Moreover, the
Br$\gamma$ emission is distributed on spatial scales comparable to
both the CO and continuum emission for these objects.  With the
relatively large uncertainties in the CO emission sizes (recall that
the data in this spectral region are noisier), the line and continuum
emission distributions appear roughly comparable across the sub-sample
of CO overtone emitters.

The coincidence of continuum and line emission is consistent with an
origin of the line emission in disk surface layers.  While the
inferred CO distributions for DG Tau and DO Tau allow the possibility
of gas in an optically thin disk midplane within the dust sublimation
radius, the extended Br$\gamma$ emission in these sources argues that
hot gas may also be co-located with the dust disk.  Furthermore, the
inferred narrow distribution of CO emission (Section \ref{sec:co})
is compatible with the expected widths of dust inner rims
\citep[e.g.,][]{IN05}.  Thus, one might interpret the data for these
objects as a dust sublimation front that emits continuum emission, and
a  hot surface layer atop the dust where excited gas produces line
emission.

Dust emission may not be confined to inner rims, but could arise in
centrifugally-launched winds \citep{BK12}. Indeed, dusty wind models
can produce continuum emission distributed over
a range of radii similar to that predicted for inner rims.  Gaseous
emission may also arise in jets or winds close to the star, and such
outflows are traced by forbidden line emission for DG Tau, DO Tau, AS
353 A, and V1331 Cyg \citep[e.g.,][]{HAMANN94,HEG95}.  For DG Tau, [Fe II]
emission has been inferred to originate in a disk wind 
within $0\rlap{.}''2$ of the central star \citep{PYO+03}.  A dusty
wind beneath a hot,  dust-free wind region
may provide a viable alternative explanation for co-spatial line and
continuum emission.

Spatially extended emission from Br$\gamma$ or CO requires an
excitation source.  Sources showing CO
emission, requiring excitation to $\ga 3000$ K, might also be able to
excite nearby gas to $\ga 10,000$ K, thereby producing Br$\gamma$
emission. Such high temperature gas could also produce
free-free emission, perhaps emitting the continuum that is
roughly spatially coincident with both the CO overtone and Br$\gamma$
emission in these objects.  However the fact
that inferred continuum sizes for these sources are compatible with
expected dust sublimation radii (Section \ref{sec:cont}) suggests that
the continuum may trace dust.

While thermal emission from the central star is insufficient 
to heat gas at these
stellocentric radii to such high temperatures, non-thermal
X-ray or UV radiation can excite CO and other gaseous
emission \citep[e.g.,][]{GNI04,GH08}.  Perhaps DG Tau, DO Tau, 
AS 353 A, V1685 Cyg, and
V1331 Cyg have particularly high accretion rates or active accretion
shocks, producing larger X-ray or UV fluxes than other sources in our
sample.  

\subsection{The peculiar case of V1685 Cyg \label{sec:v1685cyg}}
The differential phase data for V1685 Cyg suggest a
large spatial offset, $\sim 0.5$ AU, between the CO emission and the
continuum emission (Figure \ref{fig:comodels}).   If the CO and
continuum arise at different radii, a disk warp might help explain the
observed centroid offset \citep[see e.g.,][]{EH11}.  Alternatively, if
the continuum and CO emission arise from the same stellocentric
radius, but at different heights, geometric effects may explain the
centroid offset.  For example, if the continuum traces an optically
thick dusty rim then some parts of the rim may be obscured
\citep[e.g.,][]{DDN01,IN05}; in contrast, an optically thin surface
layer of CO might be completely visible.  Finally, if the CO 
traces outflowing matter while the continuum traces a
disk that could obscure part of the outflow, the two components could
be spatially offset.

The offset between the CO and continuum emission centroids may be
linked to the peculiar $V^2$ versus wavelength exhibited by V1685 Cyg 
over the $K$ band (Figure \ref{fig:v1685cyg}).  Explaining the $V^2$
is difficult because of the lack of a corresponding signature in the
flux spectrum; the spectrum for V1685 Cyg increases monotonically
across the $K$-band, similar to other sample
objects.  Whatever causes the large, non-monotonic changes in
$V^2$ with wavelength does not produce a corresponding spectral
signature in the flux data.

The shape of the spectral size distribution for V1685 Cyg 
is similar to the expected spectrum for hot H$_2$O vapor and CO
\citep[as noted previously by][based on spatially resolved data in 5
bins across the $K$-band]{EISNER+07a}.  These opacity sources are
known to produce
absorption in the extended molecular envelopes of AGB stars, which
leads to $V^2$ signatures that are the inverse of that seen for V1685
Cyg \citep[e.g.,][]{EISNER+07b}.  We speculate that a binary model,
where the secondary resembles an AGB star, could explain the data for
V1685 Cyg.  

A faint secondary might not
strongly affect the total flux of the system.  However if placed far 
from the primary, such a companion could produce resolved
visibilities.  At a separation of $\sim 5$ mas, a secondary with 5\% of the
primary flux would lead to a $\sim 20\%$ reduction in $V^2$. 
Absorption features in the secondary spectrum would
change the flux ratio as a function of wavelength, leading to higher
(i.e., less resolved) $V^2$ where the molecular opacity is strongest.  If CO
overtone absorption were present in the companion, this scenario might
also help to explain the extended spatial distribution and large
centroid offset of the CO observed towards V1685 Cyg.

\section{Summary}
We presented $K$-band observations of a sample of 21 young stars at an
angular resolution of a few milliarcseconds and a spectral resolution of
2000 from the Keck Interferometer.  
We also presented seeing-limited NIRSPEC spectra of four of these
targets with a dispersion of 24,000.  From these data we derived flux
spectra, emission size scale versus wavelength, and emission centroids
versus wavelength.  We obtained these quantities across the entire
$K$-band, but restricted our analysis to the spectral regions around
the Br$\gamma$ line and the CO overtone bandheads.

We analyzed the Br$\gamma$ emission in the context of both disk and
infall/outflow models.   We generated a larger grid of models than in
previous work, which allowed us to more fully explore the parameter
space.  We found that disk and infall/outflow models typically
produced fits of comparable quality, but that sources with the highest
signal-to-noise data preferred infall/outflow models.   Based on this
preference, and other lines of evidence in the literature, we argued
that infall/outflow models are more likely for our sample in general.

All best-fit models to the Br$\gamma$ observations indicated gas
extending to small stellocentric radii, $\la 0.05$ AU, consistent with
accreting matter on the magnetospheric scale.  However some
Br$\gamma$-emitting gas also seems to located at radii $\ga 0.1$ AU,
perhaps tracing the inner regions of magnetically launched outflows.
The fitted radial brightness profiles of the Br$\gamma$ emission
varied across our sample, reproducing the range of average inferred
sizes of the Br$\gamma$ emitting regions. 

The average size scales of the
Br$\gamma$ emission for our sample are inferred to be smaller than
the continuum emission distribution in most cases, consistent with
previous work.  However for a handful of objects the Br$\gamma$ and
continuum emission distributions are nearly coincident.   These
objects join the small number of other targets with previously
detected extended Br$\gamma$ emission
\citep{MALBET+07,TATULLI+07,WEIGELT+11}.
The sources with relatively extended Br$\gamma$ emission are also the
objects in our sample that show CO overtone emission.


CO overtone emission is detected from five objects.  Strong emission
is detected from AS 353 A and V1331 Cyg.  Weaker emission is detected
from V1685 Cyg, but strong signatures are seen in the spatially
resolved KI data.  Emission features are seen in DG Tau and DO Tau,
although these detections depend on the subtraction of photospheric
absorption features from the late-type central stars.  The CO emission is
distributed on scales comparable to continuum and Br$\gamma$ emission
for AS 353 A, V1331 Cyg, and V1685 Cyg.  For DG Tau and DO Tau, the CO
appears to be located at smaller radii, although there is 
additional uncertainty on this result related to the subtraction of
photospheric CO features.  

For each of these sources, we computed
synthetic spectra, $V^2$ and differential phases for a disk
model, and compared with KI and NIRSPEC observations across the CO
bandheads.  To fit the data
we required a narrow spatial distribution of CO with a temperature $\ga
3000$ K.  Models with the CO confined to disk regions with fractional
widths of 20\% matched the data well.

The near-coincidence of CO, Br$\gamma$, and continuum emission for the
sub-sample of objects identified as CO emitters suggests these
systems share peculiar properties.  We speculated that these
objects may have unusually active accretion processes that can
generate non-thermal excitation of gas at large stellocentric radii.
We also argued that while the continuum emission may trace dust
sublimation fronts, the line emission may arise in hot,
non-thermally-excited, atmospheres above these dusty rims.  
Alternatively, dusty winds with overlying, non-thermally-excited, 
dust-free regions may explain the line and continuum data.

Finally, we discussed the unusual behavior of V1685 Cyg.  This target
shows a large spatial offset between the CO and continuum emission,
as well as a peculiar broadband behavior of the $V^2$.  We 
suggested that a faint binary companion with strong molecular
absorption features (perhaps resembling an AGB star) might explain
these observations.

\noindent{\bf Acknowledgments:}
This work was supported by NASA Origins grant NNXX11AK57G.  JAE is
also grateful for support from an Alfred P. Sloan Research Fellowship.
The ASTRA program, which enabled the
observations presented here, was made possible by the ASTRA team and
by funding from NSF MRI grant AST-0619965. 
The authors also wish to recognize and acknowledge the cultural role
and reverence that the summit of Mauna Kea has always had within the
indigenous Hawaiian community. We are most fortunate to have the
opportunity to conduct observations from this mountain.

\clearpage
\begin{deluxetable}{lccccccc}
\tabletypesize{\tiny}
\tablewidth{0pt}
\tablecaption{Target and Calibrator Properties
\label{tab:sample}}
\tablehead{\colhead{Source} & \colhead{$\alpha$} 
& \colhead{$\delta$} & \colhead{$d$} & \colhead{Spectral Type} & 
\colhead{$m_{V}$} & \colhead{$m_{K}$} & \colhead{References} \\
 & (J2000) & (J2000) & (pc) & & & &}
\startdata
\multicolumn{8}{c}{Target Stars} \\
\hline
RY Tau & 04 21 57.409 & +28 26 35.56 & 140 & G1 & 10.2 & 5.4 & 1 \\
T Tau A & 04 21 59.434 & +19 32 06.42 & 140 & K0 & 9.9 & 5.3 & 2 \\
DG Tau & 04 27 04.700 & +26 06 16.20 & 140 & K3 & 12.4 & 7.0 & 2 \\
DK Tau A & 04 30 44.28 & +26 01 24.6 & 140 & K9 & 12.6 & 7.1 &  3 \\
GK Tau & 04 33 34.560 & +21 21 05.85 & 140 & K7 & 12.0 & 7.5 & 2 \\
DO Tau & 04 38 28.582 & +26 10 49.44 & 140 & M0 & 13.5 & 7.3 & 4,5,6,7\\
DR Tau & 04 47 06.21 & +16 58 42.8 & 140 & K4 & 13.6 & 6.9 & 2 \\
SU Aur & 04 55 59.385 & +30 34 01.52 & 140 & G2 & 9.4 & 6.0 & 4 \\
MWC 480 & 04 58 46.266 & +29 50 37.00 & 140 & A2 & 7.7 & 5.5 & 8 \\ 
RW Aur A & 05 07 49.568 & +30 24 05.161 & 140 & K2 & 10.5 & 7.0 & 2 \\  
MWC 758 & 05 30 27.530 & +25 19 57.08 & 140 & A3 & 8.3 & 5.8 & 8 \\
AS 205 A & 16 11 31.402 & -18 38 24.54 & 160 & K5 & 12.1 & 6.0 & 10 \\
MWC 863 A & 16 40 17.922 & -23 53 45.18 & 150 & A2 & 8.9 & 5.5 & 9 \\
V2508 Oph & 16 48 45.62 & -14 16 35.9 & 160 & K6 & 13.5 & 7.0 & 10 \\
MWC 275 & 17 56 21.288 & -21 57 21.88 & 122 & A1 & 6.9 & 4.8 & 9 \\
VV Ser & 18 28 47.865 & +00 08 39.76 & 310 & A0 & 11.9 & 6.3 & 8 \\
AS 353 A & 19 20 30.992 & +11 01 54.550 & 150 & F8 & 12.5 & 8.4 & 11\\
V1685 Cyg & 20 20 28.245 & +41 21 51.56 & 1000 & B2 & 10.7 & 5.9 & 8 \\
AS 442 & 20 47 37.469 & +43 47 24.89 & 600 & B8 & 11.0 & 6.5 & 8 \\
V1331 Cyg & 21 01 09.21 & +50 21 44.8 & 700 & G5 & 11.8 & 8.6 & 13 \\
MWC 1080 & 23 17 25.574 & +60 50 43.34 & 1000 & B0 & 11.6 & 4.7 & 8 \\
\hline
\multicolumn{7}{c}{Calibrator Stars} & Applied to: \\
\hline
HD 23258 &  03 44 28.204 & +20 55 43.452 & 78 & A0V & 6.1 & 6.1 & T
Tau \\ 
HD23642 & 03 47 29.453 & +24 17 18.04 &  110 & A0V & 6.8 & 6.8 & DG
Tau,DK Tau A,GK Tau,DO Tau,DR Tau,RW Aur A\\
HD23632 & 03 47 20.969 & +23 48 12.05 & 120 & A1V & 7.0 & 7.0 & DG
Tau,DK Tau A,GK Tau,DO Tau,DR Tau,RW Aur A\\
HD 23753 & 03 48 20.816 & +23 25 16.499  & 104 & B8V & 5.4 & 5.7 & RY Tau,MWC 480,MWC 758 \\
HD 27149 & 04 18 01.839 & +18 15 24.498 & 46 & G5V & 7.5 & 5.9 & T Tau A \\ 
HD 27777 & 04 24 29.155 & +34 07 50.73 & 187 & B8V & 5.7 & 6.0 & RY
Tau,SU Aur,MWC 480,MWC 758 \\
HD 283668 & 04 27 52.933 & +24 26 41.238 & 42 & K3V & 9.4 & 7.0 & GK
Tau, DO Tau \\
HD31464 & 04 57 06.426 & +24 45 07.90 & 45 & G5V & 8.6 & 7.0 & DG
Tau,DK Tau A,GK Tau,DO Tau,DR Tau,RW Aur A\\
HD139364 & 15 38 25.358 & -19 54 47.45 & 53 & F3V & 6.7 & 5.7 & AS 205 A \\
HD141465 & 15 49 52.297 & -17 54 07.007 & 43  & F3V & 6.8 & 5.9 & AS 205 A \\ 
HD144821 & 16 08 16.582 & -13 46 08.582 & 76 &  G2V  & 7.5 & 6.0 & AS 205 A,V2508 Oph \\ 
HD148968 & 16 32 08.085 & -12 25 53.910 & 146 & A0V & 7.0 & 7.0 & AS 205 A, V2508 Oph \\
HD 149013 & 16 32 38.133 & -15 59 15.12 & 41 & F8V & 7.0 & 5.7 & MWC 863A \\
HD 163955 & 17 59 47.553 & -23 48 58.08 & 134 & B9V & 4.7 & 4.9 & MWC 275 \\
HD 170657 & 18 31 18.960 & -18 54 31.72 & 13 & K1V & 6.8 & 4.7 & MWC 275 \\
HD 174240 & 18 49 37.193 & +00 50 10.310 & 180 & A1V &  6.2 & 6.2 & VV
Ser \\ 
HD 174719 & 18 51 48.423 & +03 01 51.680 & 28 & G6V & 7.5 & 5.9 & VV
Ser \\ 
HD183442 & 19 29 30.077 & +03 05 23.607  & & B7V & 8.1 & 8.4 & AS 353 A\\
HD 184152 & 19 32 54.242 & +07 24 16.018 & 78 & G5V & 9.4 & 7.8 & AS 353 A \\
HD 185195 & 19 37 17.840 & +15 15 02.407 & 248  & B8V & 8.1 & 8.3 & AS 353 A \\
HD 189178 & 19 57 13.868 & +40 22 04.166 & 340 & B5V & 5.5 & 5.9 &
V1685 Cyg \\
HD192985 & 20 16 00.615 & +45 34 46.291 & 35 &  F5V& 5.9 & 4.8 & V1057 Cyg \\ 
HD195050 & 20 27 34.258 & +38 26 25.194 & 83 & A3V & 5.6 & 5.5  & V1057 Cyg \\
HD198182 & 20 46 53.060 & +47 06 41.502 & 185 & A1V & 7.8 & 7.8 & V1331 Cyg \\ 
HIP 102667 & 20 48 16.305 & +44 04 35.857 & 394 & K2III & 8.8 & 5.6 &
V1685 Cyg, AS 442 \\
HD 199099 & 20 53 26.390 & +42 24 36.722 & 140 & A1V & 6.7 & 6.7 & AS 442 \\ 
HD219623 & 23 16 42.303 & +53 12 48.512  & 20 & F7V & 5.6 & 4.3 & MWC 1080 \\
\enddata
\tablerefs{(1) \citet{CALVET+04}; (2) \citet{WG01}; (3)
  \citet{MMD98}; (4)  \citet{MUZEROLLE+03}; (5) \citet{BSC07}; (6)
  \citet{BSC07}; (7) \citet{HH14}; (8) \citet{EISNER+04};
(9) \citet{MONNIER+06}; (10) \citet{EISNER+05}; (11) \citet{PGS03}; (12) \citet{HPD03}; (13) \citet{EISNER+07c}.  
Calibrator star distances are
based on Hipparcos parallax measurements \citep{PERRYMAN+97}.}
\end{deluxetable}

\begin{deluxetable}{lccc}
\tabletypesize{\scriptsize}
\tablewidth{0pt}
\tablecaption{Log of KI Observations
\label{tab:obs}}
\tablehead{\colhead{Source}
& \colhead{Date (UT)} & \colhead{$u$ (m)} & \colhead{$v$ (m)}}
\startdata
RY Tau & 2008 November 18 & 56,56,56,56,47,46,45 & 57,58,58,59,71,71,72 \\
T Tau A & 2011 September 13 & 52,54 & 52,54 \\
DG Tau & 2008 November 17 & 55,56,47,30 & 52,55,70,78 \\
 & 2010 November 25 & 46,26 & 71,79 \\
& 2011 September 13 &  56,56 & 54,55 \\
 & 2011 November 7 &  14 &  81 \\
DK Tau A & 2008 November 17 & 56,19 & 56,80 \\
 & 2011 September 13 & 56 & 59 \\
 & 2011 November 7 & 56 & 55 \\
GK Tau & 2011 September 13 & 55 & 63 \\
 & 2011 November 7 & 57 & 58 \\
DO Tau & 2011 September 13 & 55 & 63 \\
 & 2011 November 7 & 56 & 59 \\
DR Tau & 2008 November 17 & 56,56,41,40 & 60,61,72,72 \\
SU Aur & 2011 November 7 & 54 & 62 \\
MWC 480 & 2008 November 18 & 56,46,46 & 56,70,71 \\
RW Aur A & 2008 November 17 & 56,56,33,22 & 55,58,77,80 \\
& 2008 November 18 & 35 & 77 \\
& 2010 November 25 & 21 & 81 \\
 & 2011 September 13 & 55 & 61 \\
MWC 758 & 2008 November 18 & 56,56 & 59,59 \\
AS 205 A & 2008 April 25 & 33,32,32 & 45,45,45 \\
 & 2009 July 15 & 54,53,53,51,51 & 54,54,54,52,52 \\
 & 2010 July 21 & 51,50,49 & 52,52,51 \\
MWC 863 A & 2009 July 15 & 44,44 & 44,43 \\
V2508 Oph & 2008 April 25 & 35,35,31 & 51,51,50 \\
MWC 275 & 2008 April 25 & 53,53,51,51 & 51,51,50,49 \\
 & 2009 July 15 & 54,54,50,50 & 53,53,49,49 \\
VV Ser & 2011 September 13 & 51,47 & 64,64 \\
AS 353 A & 2009 July 15 & 52 & 66 \\
 & 2011 September 13 & 44,38 & 69,70 \\
V1685 Cyg & 2011 September 13 & 25,19 & 81,83 \\
 & 2011 September 14 & 56,56,38 & 51,54,76 \\
AS 442 & 2011 September 13 & 41,36,20 & 72,76,82 \\
V1331 Cyg & 2009 July 15 & 44,43,40,39 & 67,67,71,71 \\
MWC 1080 & 2009 July 15 & 48,48,45,44 & 55,56,60,61 \\
\enddata
\end{deluxetable}

\begin{deluxetable}{lcccccccc}
\tabletypesize{\scriptsize}
\tablewidth{0pt}
\tablecaption{Properties Derived from KI Br$\gamma$ Spectra
\label{tab:brg}}
\tablehead{\colhead{Source} & \colhead{EW (\AA)} & \colhead{$L_{\rm Br \gamma}$/($10^{-4}$ L$_{\odot}$)} & 
\colhead{$L_{\rm acc}$/L$_{\odot}$} & \colhead{$L_{\rm acc}$/L$_{\ast}$} }
\startdata
RY Tau &    -2.4 &     4.5 &     1.6 &     0.2 \\
T Tau A &    -7.0 &    13.9 &     6.8 &     0.9 \\
DG Tau &    -6.9 &     3.2 &     1.1 &     1.2 \\
DK Tau A &    -1.3 &     0.4 &     0.1 &     0.1 \\
GK Tau &    -3.5 &     0.9 &     0.2 &     0.2 \\
DO Tau &   -11.9 &     4.1 &     1.5 &     1.1 \\
DR Tau &    -7.6 &     3.4 &     1.1 &     1.2 \\
SU Aur &    -0.0 &     0.0 &     0.0 &     0.0 \\
MWC 480 &    -8.1 &    10.2 &     6.8 &     0.4 \\
RW Aur A &    -5.7 &     2.2 &     0.7 &     0.4 \\
MWC 758 &    -1.8 &     2.5 &     1.9 &     0.2 \\
AS 205 A &    -3.7 &     5.3 &     2.0 &     1.5 \\
MWC 863 A &    -6.2 &     8.1 &     5.5 &     0.3 \\
V2508 Oph &    -6.5 &     4.7 &     1.7 &     0.5 \\
MWC 275 &    -6.3 &    15.3 &     9.7 &     0.2 \\
VV Ser &    -5.1 &    18.2 &     11.4 &     0.2 \\
AS 353 A &   -13.1 &     1.6 &     1.2 &   ... \\
V1685 Cyg  &   -12.1 &   674.1 &   304.9 &     0.1 \\
AS 442 &    -4.5 &    53.0 &    30.1 &     0.2 \\
V1331 Cyg &   -14.0 &    31.8 &    19.0 &  ... \\
MWC 1080 &    -5.2 &  1196.3 &  513.8 &     0.1 &  
\enddata
\tablecomments{The stellar parameters are not known for AS 353 A and
  V1331 Cyg, and hence we cannot estimate $L_{\ast}$ for these
  objects.  In our analysis, we assume the total luminosity is equal
  to the accretion luminosity for these two targets.}
\end{deluxetable}

\begin{deluxetable}{lcc|cc}
\tabletypesize{\scriptsize}
\tablewidth{0pt}
\tablecaption{Inferred sizes of Br$\gamma$ emission regions
\label{tab:brgsizes}}
\tablehead{\colhead{Source}
& \colhead{$\theta_{\rm Br \gamma}$ (mas)} &  \colhead{$R_{\rm
    Br\gamma}$ (AU)} & \colhead{$\theta_{\rm continuum}$ (mas)} &
\colhead{$R_{\rm cont}$ (AU) }}
\startdata
RY Tau & $<$0.10 & $<$0.01 & 2.57 $\pm$ 0.02 & 0.18 $\pm$ 0.01 \\ 
T Tau A & $<$0.19 & $<$0.01 & 1.54 $\pm$ 0.07 & 0.11 $\pm$ 0.01 \\ 
DG Tau & 1.67 $\pm$ 0.06 & 0.12 $\pm$ 0.01 & 2.40 $\pm$ 0.03 & 0.17 $\pm$ 0.01 \\ 
DK Tau A$^\ast$ & 1.49 & 0.10 & 1.86 $\pm$ 0.05 & 0.13 $\pm$ 0.01 \\ 
GK Tau$^\ast$ & $<$0.10 & $<$0.01 & 2.08 $\pm$ 0.03 & 0.15 $\pm$ 0.01 \\ 
DO Tau & 1.36 $\pm$ 0.07 & 0.10 $\pm$ 0.01 & 2.54 $\pm$ 0.01 & 0.18 $\pm$ 0.01 \\ 
DR Tau & 0.68 $\pm$ 0.16 & 0.05 $\pm$ 0.01 & 1.72 $\pm$ 0.04 & 0.12 $\pm$ 0.01 \\ 
SU Aur$^\ast$ & $<$5.00 & $<$0.35 & 2.67 $\pm$ 0.02 & 0.19 $\pm$ 0.01 \\ 
MWC 480 & $<$0.10 & $<$0.01 & 2.75 $\pm$ 0.01 & 0.19 $\pm$ 0.01 \\ 
RW Aur A & $<$0.16 & $<$0.01 & 1.51 $\pm$ 0.06 & 0.11 $\pm$ 0.01 \\ 
MWC 758 & $<$0.21 & $<$0.02 & 2.36 $\pm$ 0.02 & 0.18 $\pm$ 0.01 \\ 
AS 205 A & 1.42 $\pm$ 0.10 & 0.11 $\pm$ 0.01 & 2.52 $\pm$ 0.04 & 0.20 $\pm$ 0.01 \\ 
MWC 863 A & 0.97 $\pm$ 0.19 & 0.07 $\pm$ 0.01 & 3.84 $\pm$ 0.01 & 0.29 $\pm$ 0.01 \\ 
V2508 Oph & 0.56 $\pm$ 0.42 & 0.04 $\pm$ 0.03 & 3.81 $\pm$ 0.02 & 0.30 $\pm$ 0.01 \\ 
MWC 275 & 0.45 $\pm$ 0.40 & 0.03 $\pm$ 0.02 & 3.13 $\pm$ 0.02 & 0.19 $\pm$ 0.01 \\ 
VV Ser & 0.52 $\pm$ 0.23 & 0.08 $\pm$ 0.04 & 2.39 $\pm$ 0.03 & 0.37 $\pm$ 0.01 \\ 
AS 353 A & 1.06 $\pm$ 0.09 & 0.08 $\pm$ 0.01 & 1.47 $\pm$ 0.06 & 0.11 $\pm$ 0.01 \\ 
V1685 Cyg & 2.19 $\pm$ 0.03 & 1.10 $\pm$ 0.02 & 2.34 $\pm$ 0.03 & 1.17 $\pm$ 0.02 \\ 
AS 442 & $<$0.31 & $<$0.09 & 1.60 $\pm$ 0.05 & 0.48 $\pm$ 0.02 \\ 
V1331 Cyg & 0.80 $\pm$ 0.13 & 0.28 $\pm$ 0.05 & 0.86 $\pm$ 0.12 & 0.30 $\pm$ 0.04 \\ 
MWC 1080 & 0.42 $\pm$ 0.44 & 0.21 $\pm$ 0.22 & 2.66 $\pm$ 0.03 & 1.33 $\pm$ 0.02 \\ 
\enddata
\tablecomments{Angular ring diameters ($\theta$) are
  converted into linear ring radii using the distances listed in Table
\ref{tab:sample}.  $^\ast$ Note that DK Tau A, GK Tau, and SU Aur do not show
strong Br$\gamma$ emission features (Figure
\ref{fig:spectra}), and so the Br$\gamma$ sizes listed for those
objects are not particularly meaningful.}
\end{deluxetable}

\begin{deluxetable}{lcc|cc|cc}
\tabletypesize{\scriptsize}
\tablewidth{0pt}
\tablecaption{Inferred sizes of CO emission regions
\label{tab:cosizes}}
\tablehead{\colhead{Source}
& \colhead{$\theta_{\rm v=2\rightarrow 0}$ (mas)} &  \colhead{$R_{\rm
    v=2\rightarrow 0}$ (AU)}  & \colhead{$\theta_{\rm v=3\rightarrow
    1}$ (mas)} &  \colhead{$R_{\rm
    v=3\rightarrow 1}$ (AU)} &\colhead{$\theta_{\rm continuum}$ (mas)} &
\colhead{$R_{\rm cont}$ (AU) }}
\startdata
DG Tau & 1.44 $\pm$ 0.18 & 0.10 $\pm$ 0.01 & 1.00 $\pm$ 1.25 & 0.07 $\pm$ 0.09 & 2.37 $\pm$ 0.03 & 0.17 $\pm$ 0.01 \\ 
DO Tau & $<$0.52 & $<$0.04 & $<$0.12 & $<$0.01 & 1.87 $\pm$ 0.04 & 0.13 $\pm$ 0.01 \\ 
AS 353 A & 1.38 $\pm$ 0.17 & 0.10 $\pm$ 0.01 & 1.46 $\pm$ 0.13 & 0.11 $\pm$ 0.01 & 1.51 $\pm$ 0.06 & 0.11 $\pm$ 0.01 \\ 
V1685 Cyg & 2.61 $\pm$ 0.44 & 1.31 $\pm$ 0.22 & 3.10 $\pm$ 1.36 & 1.55 $\pm$ 0.68 & 2.21 $\pm$ 0.03 & 1.11 $\pm$ 0.02 \\ 
V1331 Cyg & 0.74 $\pm$ 0.15 & 0.26 $\pm$ 0.05 & 0.90 $\pm$ 0.19 & 0.32 $\pm$ 0.07 & 0.87 $\pm$ 0.12 & 0.30 $\pm$ 0.04 \\ 
\enddata
\tablecomments{Angular ring diameters ($\theta$) are
  converted into linear ring radii using the distances listed in Table
\ref{tab:sample}. } 
\end{deluxetable}

\begin{deluxetable}{l|cccccc|ccccccccc}
\tabletypesize{\scriptsize}
\tablewidth{0pt}
\tablecaption{Results of Kinematic Modeling of KI Br$\gamma$ Data
\label{tab:results}}
\tablehead{\colhead{Source}
& \colhead{$\chi_r^2$} & \colhead{$R_{\rm in}$}  & \colhead{$PA$} & 
\colhead{$i$} & \colhead{$L/C$} & \colhead{$\alpha$} &
 \colhead{$\chi_r^2$} & \colhead{$R_{\rm in}$} & \colhead{$R_{\rm out}$} &
\colhead{$PA$} & \colhead{$\beta$} & \colhead{$\Delta v$} & \colhead{$L/C$} & 
\colhead{$f_{\rm a}$} & \colhead{$\alpha$} \\
 & & (AU) & ($^{\circ}$) & ($^{\circ}$) & & & & (AU) & (AU) & ($^{\circ}$) & & (km s$^{-1}$)}
\startdata
 & \multicolumn{6}{c}{Disk Models} & \multicolumn{8}{c}{Infall/Outflow Models} \\
\hline
RY Tau & 0.11 & 0.01 & 25 & 75 & 0.2 &  2 & 0.16 & 0.04 & 0.10 & 40 &  3 & 250 & 0.1 & 1.0 &  3 \\
T Tau A & 0.48 & 0.01 & 90 & 25 & 0.8 &  4 & 0.55 & 0.01 & 0.50 &  0 &  3 & 250 & 0.5 & 1.0 &  3 \\
DG Tau & 0.56 & 0.01 & 75 & 50 & 0.8 &  2 & 0.66 & 0.02 & 0.50 &  0 &  1 & 375 & 1.0 & 1.0 &  2 \\
DK Tau A & 0.21 & 0.01 & 90 & 25 & 0.1 &  4 & 0.22 & 0.01 & 0.50 & 40 &  3 & 250 & 0.1 & 1.0 &  3 \\
GK Tau & 0.49 & 0.01 & 75 & 75 & 0.2 &  2 & 0.52 & 0.04 & 0.50 & 60 &  1 & 250 & 0.1 & 1.0 &  2 \\
DO Tau & 0.49 & 0.01 & 90 & 25 & 1.0 &  3 & 0.48 & 0.01 & 0.10 & 20 &  2 & 250 & 1.0 & 0.5 &  2 \\
DR Tau & 0.40 & 0.01 & 75 & 25 & 0.5 &  3 & 0.40 & 0.03 & 0.50 &  0 &  3 & 250 & 0.5 & 1.0 &  3 \\
SU Aur & 0.04 & 0.01 & 50 & 50 & 0.2 &  3 & 0.05 & 0.01 & 0.05 & 20 &  3 & 250 & 0.1 & 1.0 &  2 \\
MWC 480 & 0.67 & 0.01 & 90 & 25 & 0.8 &  3 & 0.47 & 0.02 & 0.05 & 20 &  3 & 250 & 0.5 & 0.1 &  2 \\
RW Aur A & 0.89 & 0.02 & 90 & 75 & 0.2 &  2 & 0.93 & 0.01 & 0.50 & 60 &  2 & 500 & 0.5 & 1.0 &  3 \\
MWC 758 & 0.20 & 0.01 &  0 & 25 & 0.1 &  2 & 0.20 & 0.02 & 0.50 &  0 &  2 & 250 & 0.1 & 0.5 &  3 \\
AS 205 A & 0.58 & 0.01 &  0 & 75 & 0.2 &  3 & 0.63 & 0.01 & 1.00 &  0 &  1 & 500 & 0.5 & 1.0 &  2 \\
MWC 863 A & 0.39 & 0.01 &  0 & 25 & 0.5 &  2 & 0.33 & 0.01 & 0.05 &  0 &  2 & 250 & 0.5 & 0.1 &  3 \\
V2508 Oph & 0.73 & 0.04 & 50 & 25 & 0.5 &  4 & 0.76 & 0.01 & 0.50 &  0 &  3 & 250 & 0.5 & 1.0 &  1 \\
MWC 275 & 0.37 & 0.01 &  0 & 25 & 0.5 &  3 & 0.35 & 0.01 & 0.05 &  0 &  2 & 250 & 0.5 & 0.5 &  3 \\
VV Ser & 0.21 & 0.01 & 90 & 75 & 0.5 &  2 & 0.21 & 0.02 & 0.50 &  0 &  1 & 500 & 0.5 & 1.0 &  2 \\
AS 353 A & 0.92 & 0.01 & 90 & 25 & 0.8 &  2 & 0.87 & 0.02 & 0.10 & 60 &  2 & 375 & 1.0 & 1.0 &  2 \\
V1685 Cyg & 0.74 & 0.02 & 90 & 50 & 0.8 &  2 & 0.68 & 0.04 & 1.00 & 80 &  1 & 500 & 0.5 & 1.0 &  1 \\
AS 442 & 0.34 & 0.01 & 90 & 50 & 0.5 &  2 & 0.35 & 0.01 & 0.05 &  0 &  3 & 250 & 0.5 & 1.0 &  3 \\
V1331 Cyg & 0.55 & 0.05 & 25 &  5 & 0.8 &  2 & 0.48 & 0.01 & 0.50 & 60 &  2 & 375 & 1.0 & 1.0 &  1 \\
MWC 1080 & 0.27 & 0.01 & 50 & 25 & 0.5 &  2 & 0.26 & 0.04 & 0.50 & 40 &  2 & 500 & 0.5 & 1.0 &  2 \\
\enddata
\end{deluxetable}


\clearpage

\epsscale{1.0}

\epsscale{0.7}
\begin{figure}
\plotone{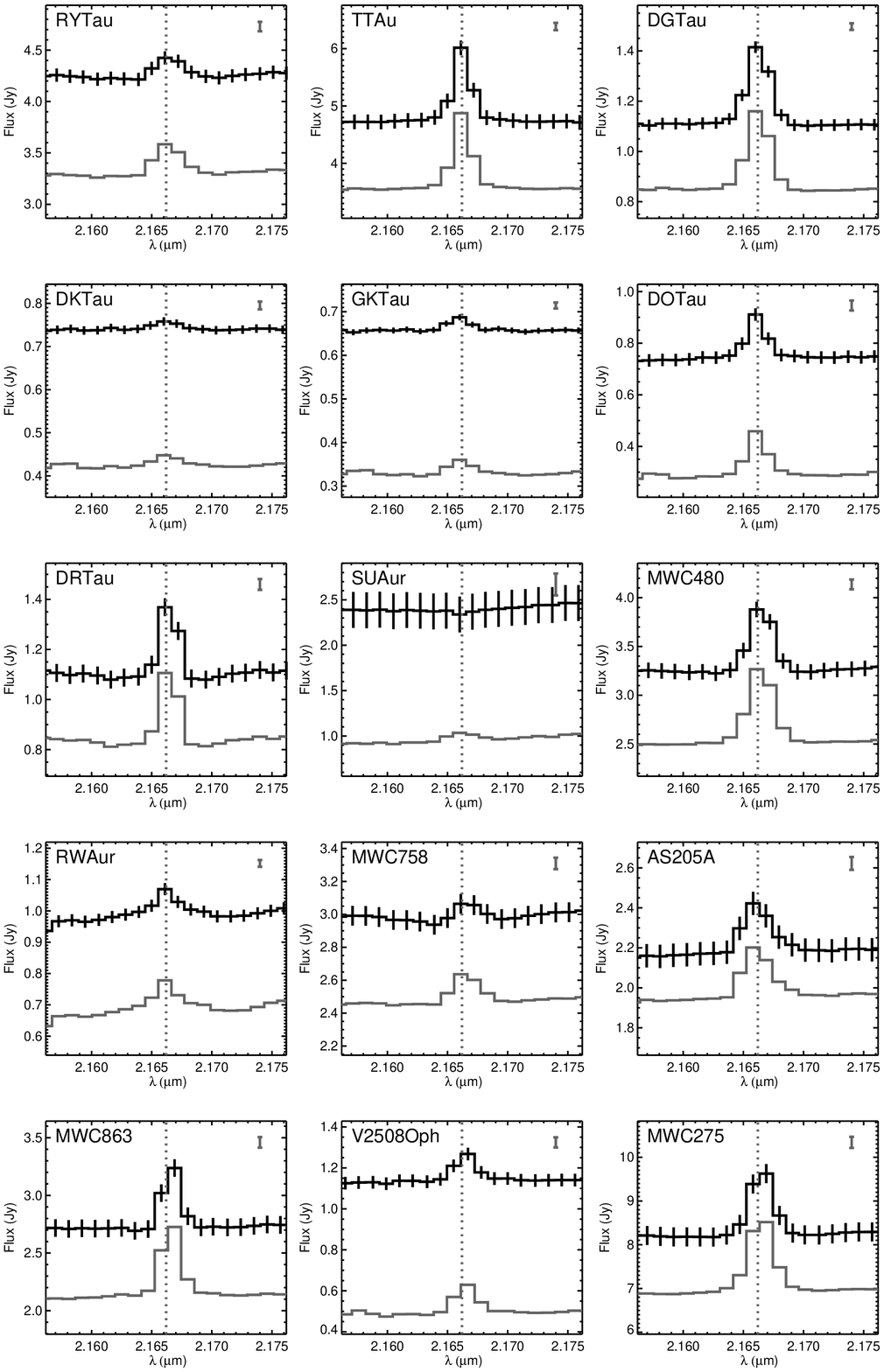}
\caption{Flux versus wavelength for our sample objects measured with
  KI.  This figure focuses on the spectral region around the
  Br$\gamma$ transition.  The solid histogram shows the observed
  spectrum for the target, and the gray histogram shows the flux of
  the circumstellar component only.  For
clarity of presentation, we have not plotted the error bars associated
with the circumstellar fluxes; the magnitudes of the uncertainties are
indicated in the upper right corners.  Dotted
gray lines indicate the
central (rest) wavelength of the Br$\gamma$ transition, 2.1662 $\mu$m.
\label{fig:spectra}}
\end{figure}

\begin{figure}
\ContinuedFloat
\plotone{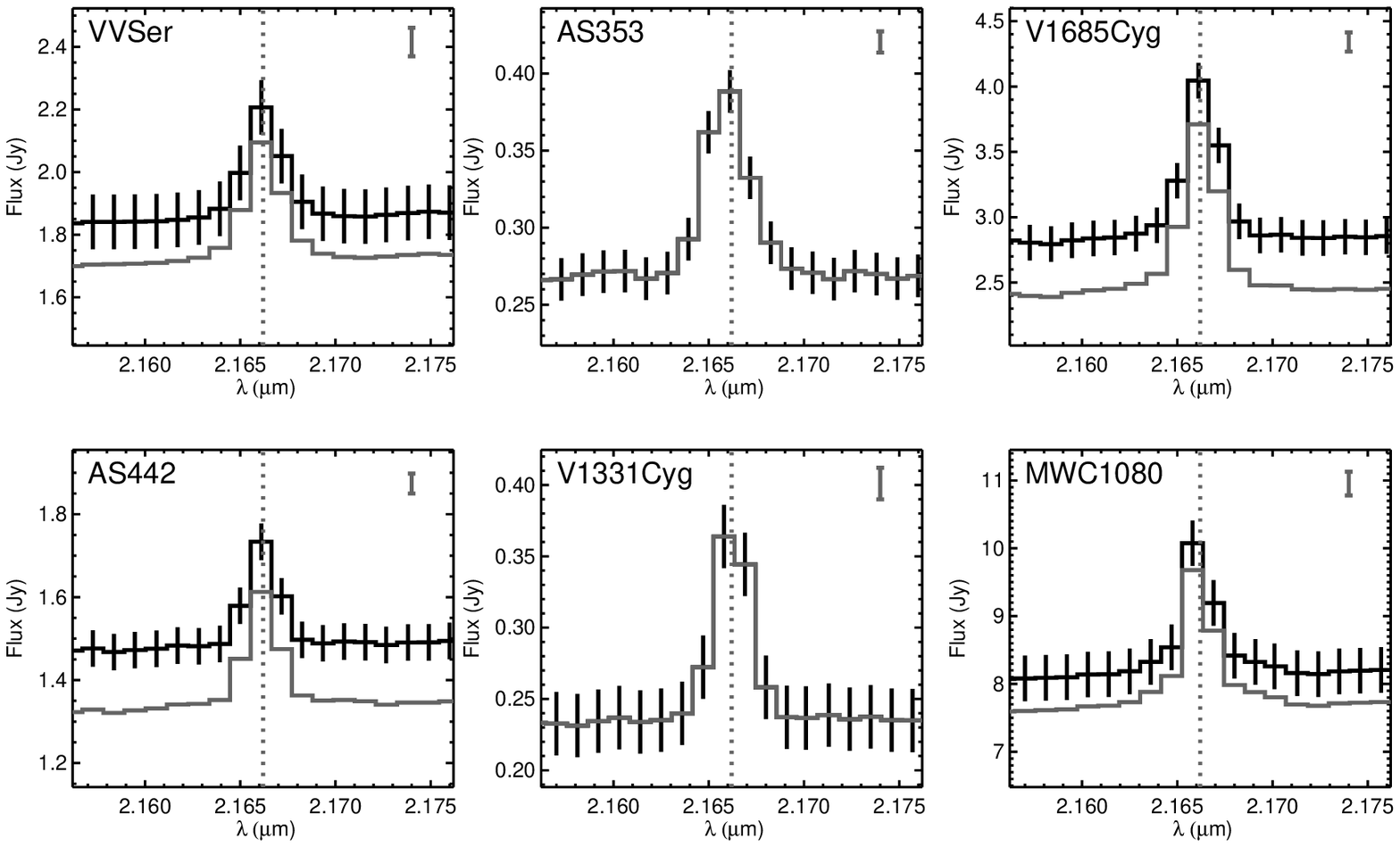}
\caption{continued.
\label{fig:spectra2}}
\end{figure}

\epsscale{1.0}

\clearpage

\epsscale{0.7}
\begin{figure}
\plotone{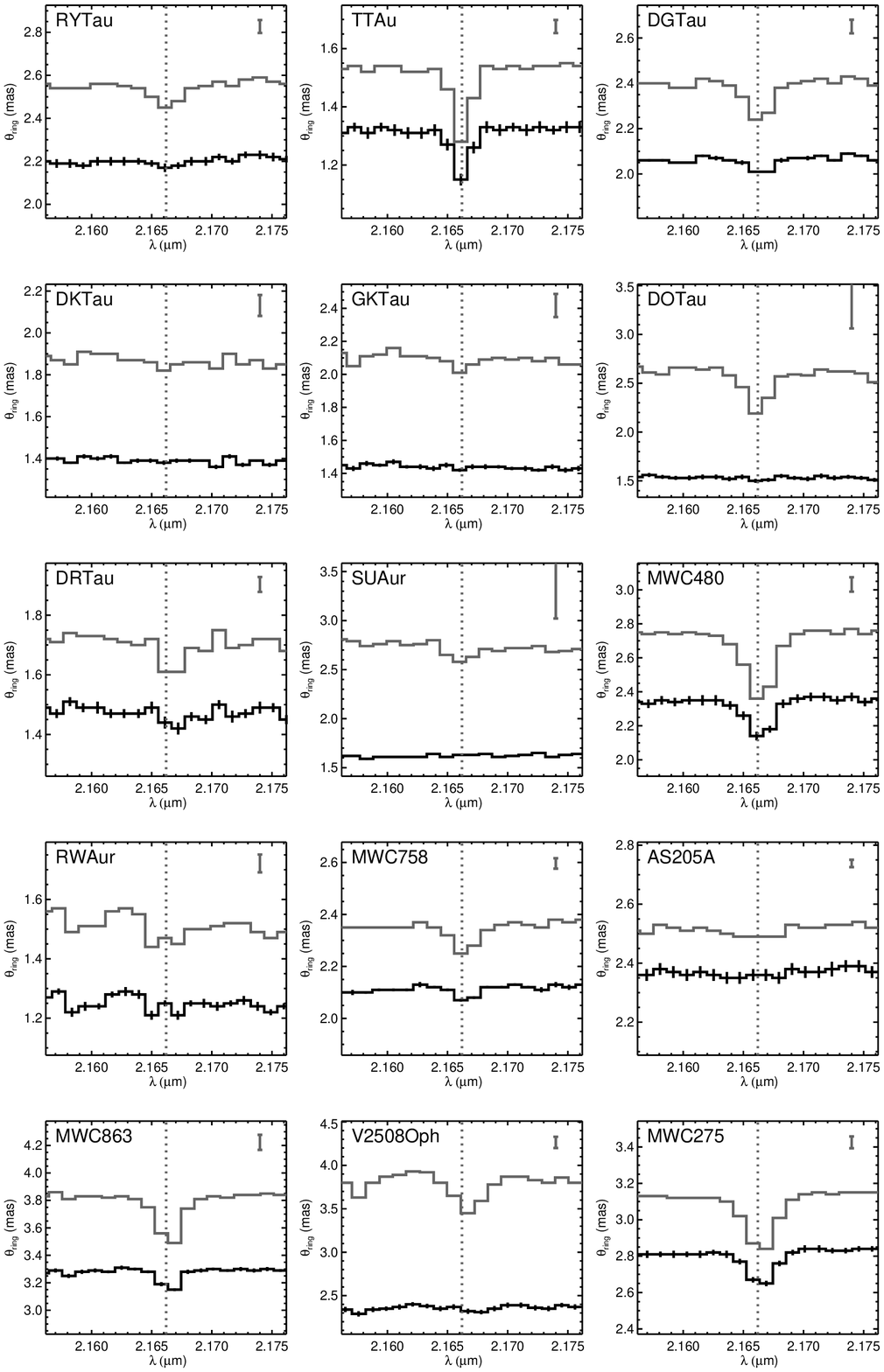}
\caption{Uniform ring angular diameters of our sample, plotted in the spectral region
around Br$\gamma$.  Angular sizes computed directly from the observed $V^2$ are plotted
with black histograms.  Angular sizes of only
the circumstellar emission, determined using Equation \ref{eq:vdisk}, are 
shown with gray histograms. 
\label{fig:v2}}
\end{figure}

\begin{figure}
\ContinuedFloat
\plotone{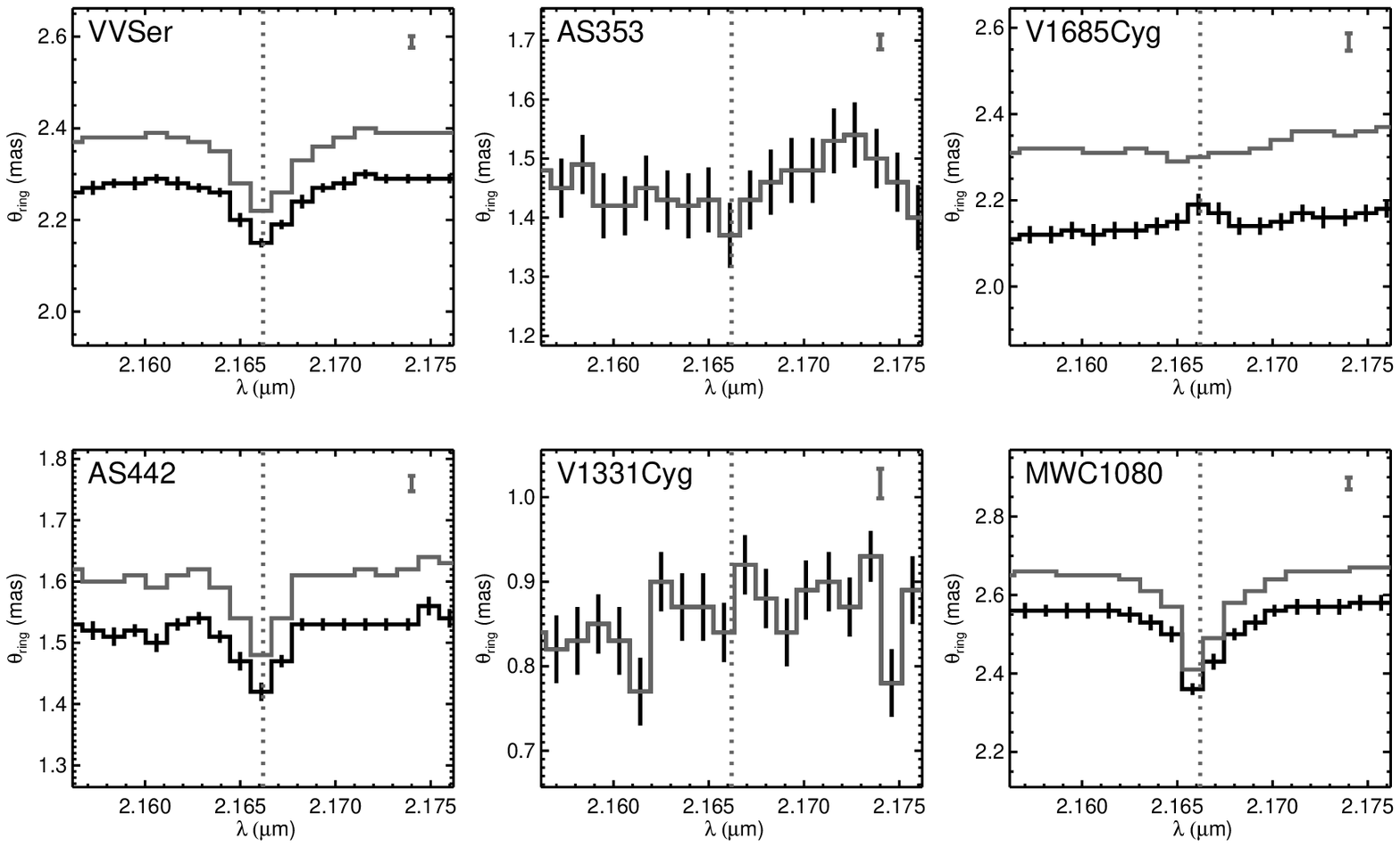}
\caption{ continued. 
\label{fig:v22}}
\end{figure}

\begin{figure}
\plotone{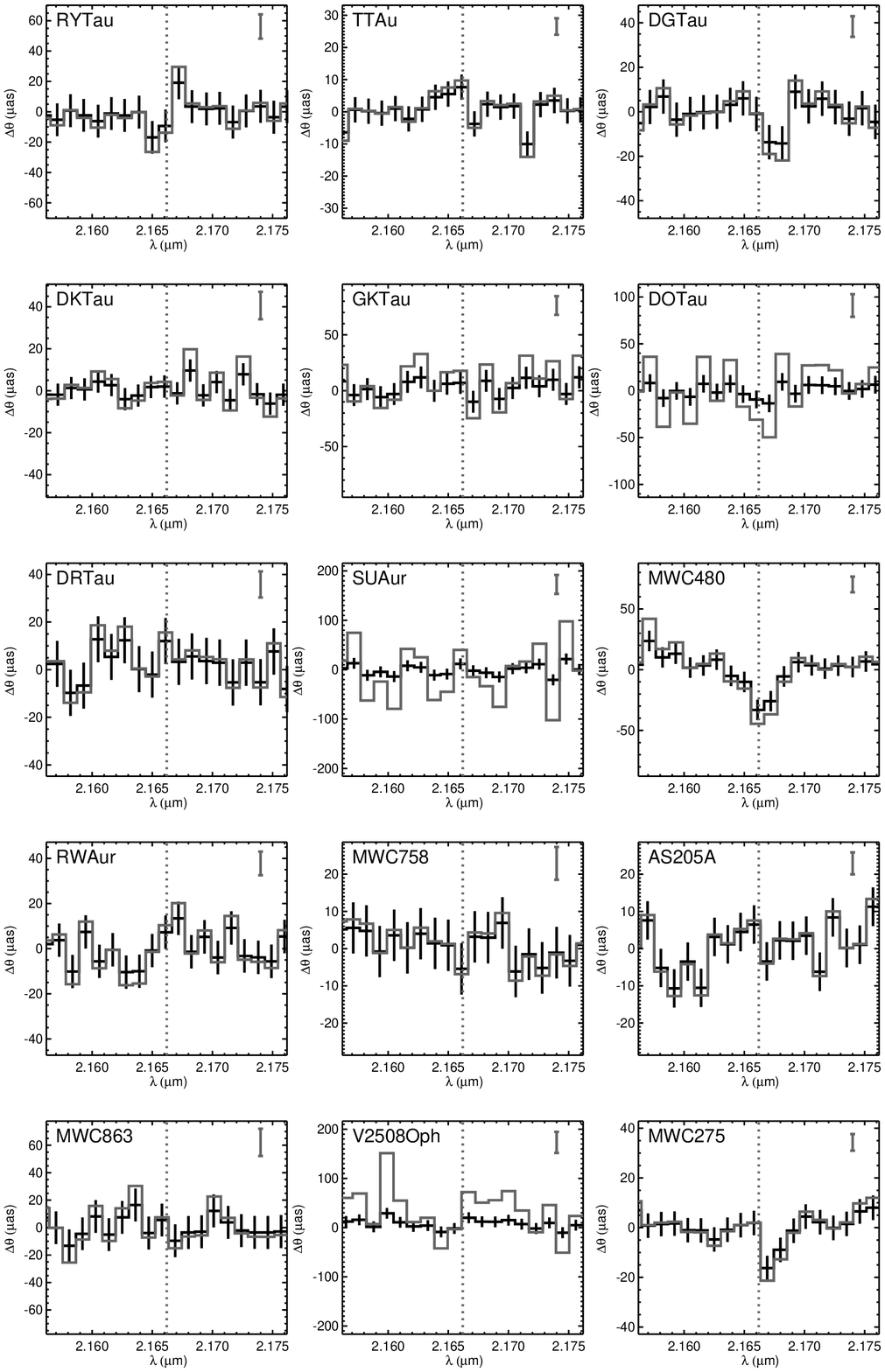}
\caption{Centroid offsets of our sample, plotted in the spectral region
around Br$\gamma$.  Black histograms show the offsets derived for the observed data, and gray histograms
show the centroid offsets for the circumstellar component of the
emission, determined using Equation \ref{eq:dpcirc}.
\label{fig:dp}}
\end{figure}

\begin{figure}
\ContinuedFloat
\plotone{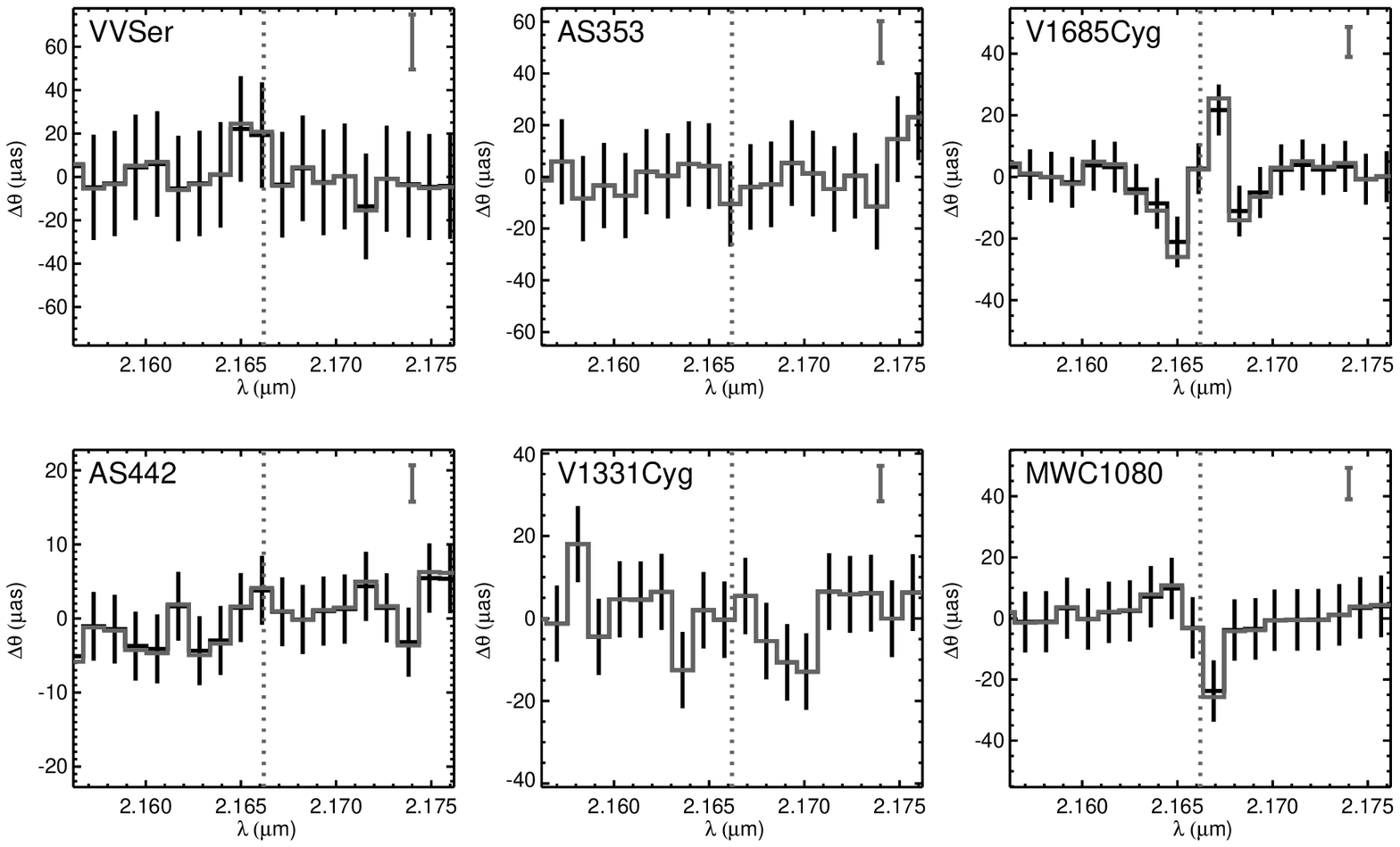}
\caption{continued. 
\label{fig:dp2}}
\end{figure}

\begin{figure}
\plotone{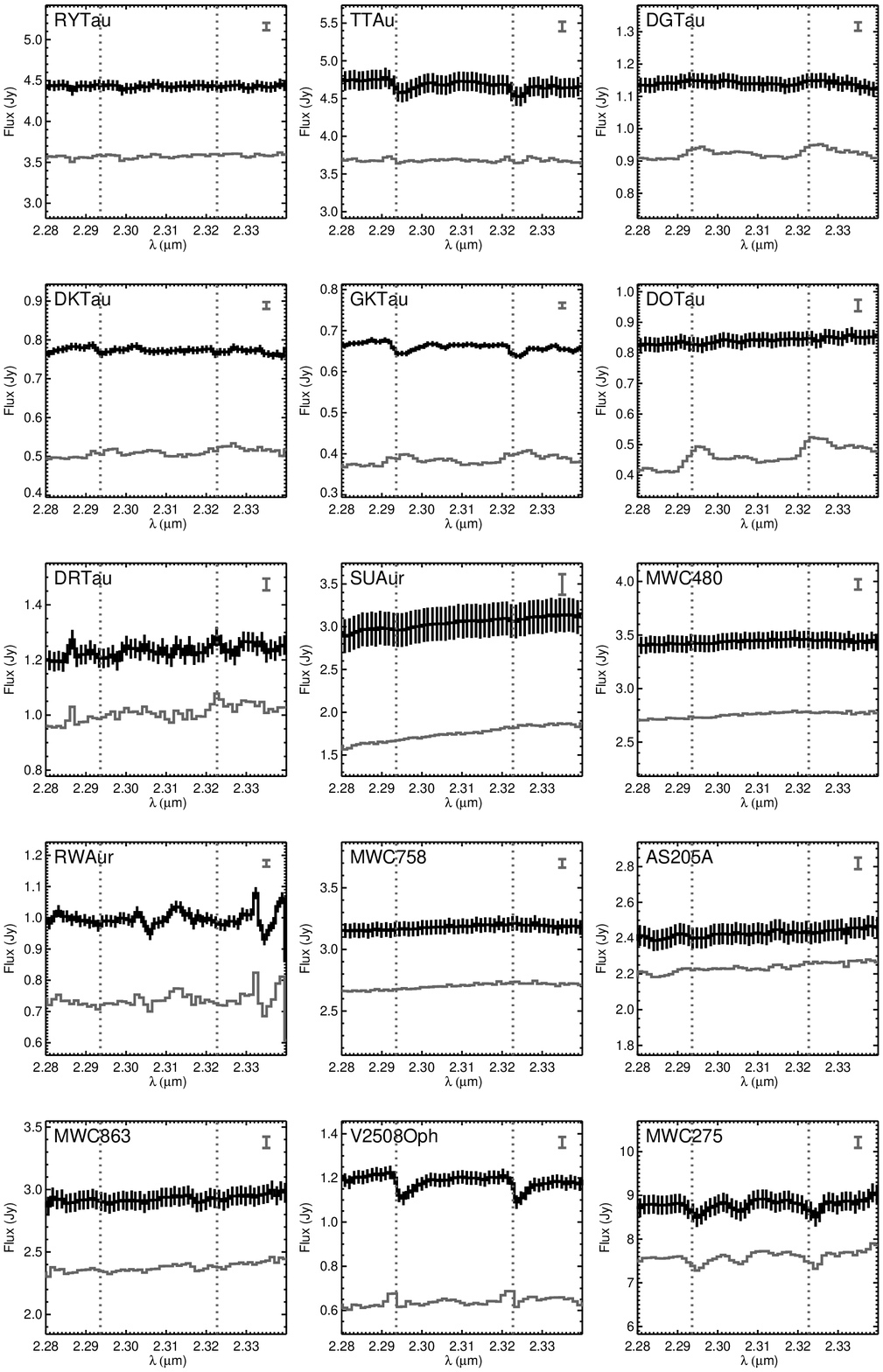}
\caption{Flux versus wavelength for our sample objects measured with
  KI.  This figure focuses on the spectral region around the
  CO rovibrational overtone transitions.  The solid histogram shows the observed
  spectrum for the target, and the gray histogram shows the flux of
  the circumstellar component only.  For
  clarity of presentation, we have not plotted the error bars associated
  with the circumstellar fluxes; the magnitudes of the uncertainties are
  indicated in the upper right corners.  Dotted
  gray lines indicate the
  central (rest) wavelengths of the $v=2\rightarrow 0$ bandhead at
  2.2936 $\mu$m and the $v=3 \rightarrow 1$ bandhead at 2.3227
  $\mu$m. 
\label{fig:cospectra}}
\end{figure}

\begin{figure}
\ContinuedFloat
\plotone{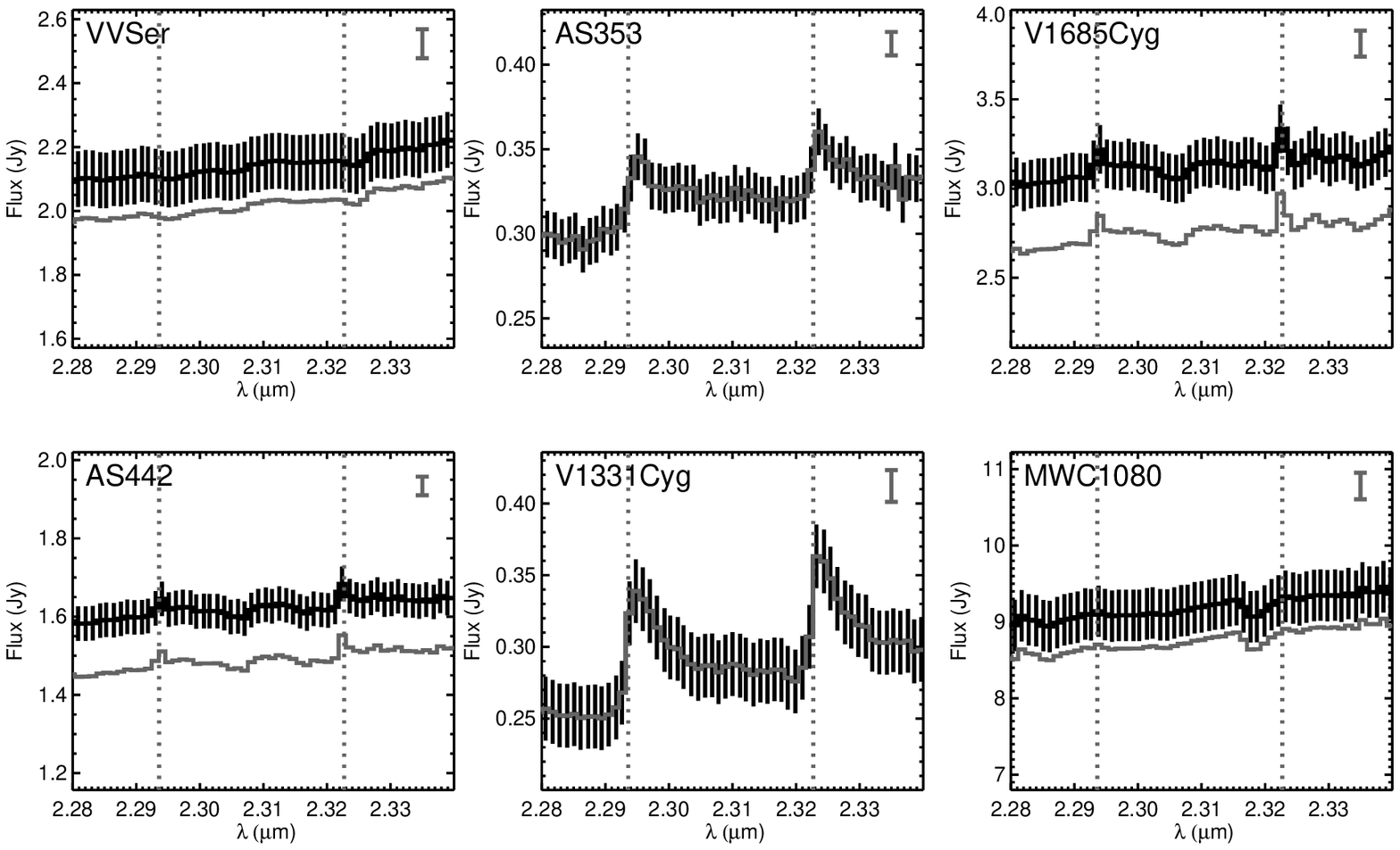}
\caption{continued.
\label{fig:cospectra2}}
\end{figure}

\begin{figure}
\plotone{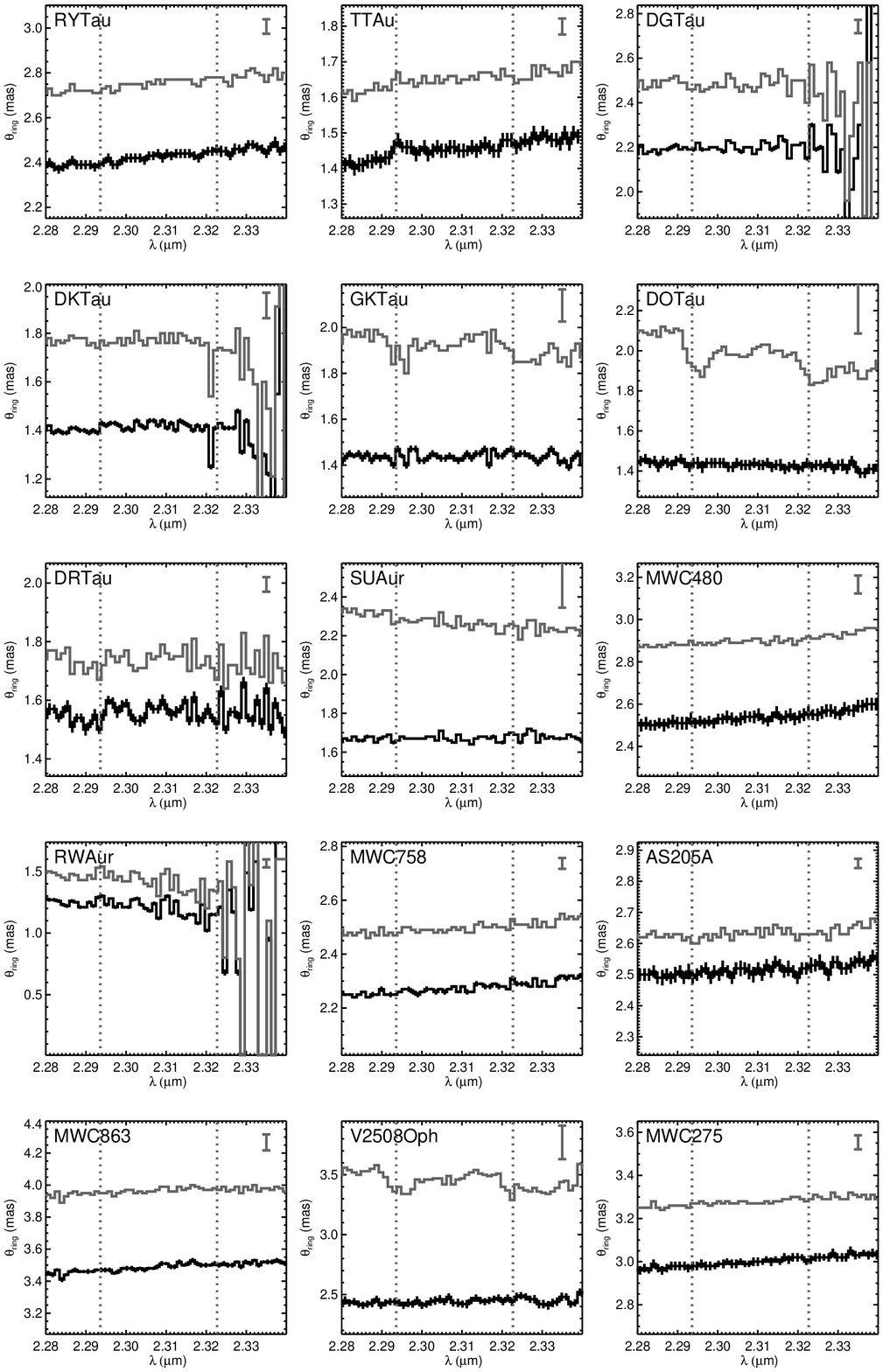}
\caption{Uniform ring angular diameters of our sample, plotted in the spectral region
around the CO overtone bandheads.  
Angular sizes computed directly from the observed $V^2$ are plotted
with black histograms.  Angular sizes of only
the circumstellar emission, determined using Equation \ref{eq:vdisk}, are 
shown with gray histograms. 
\label{fig:cov2}}
\end{figure}

\begin{figure}
\ContinuedFloat
\plotone{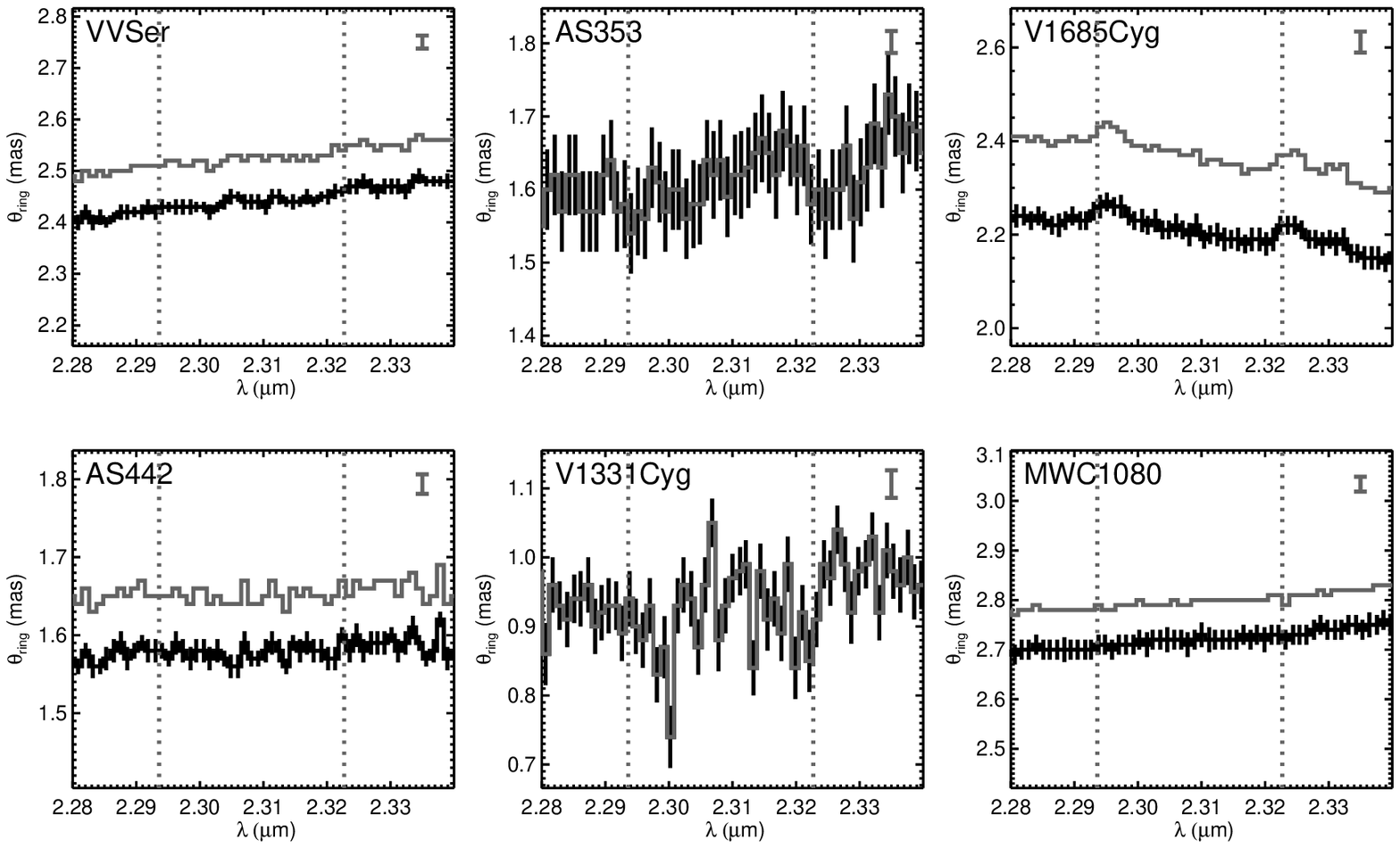}
\caption{continued. 
\label{fig:vco22}}
\end{figure}

\begin{figure}
\plotone{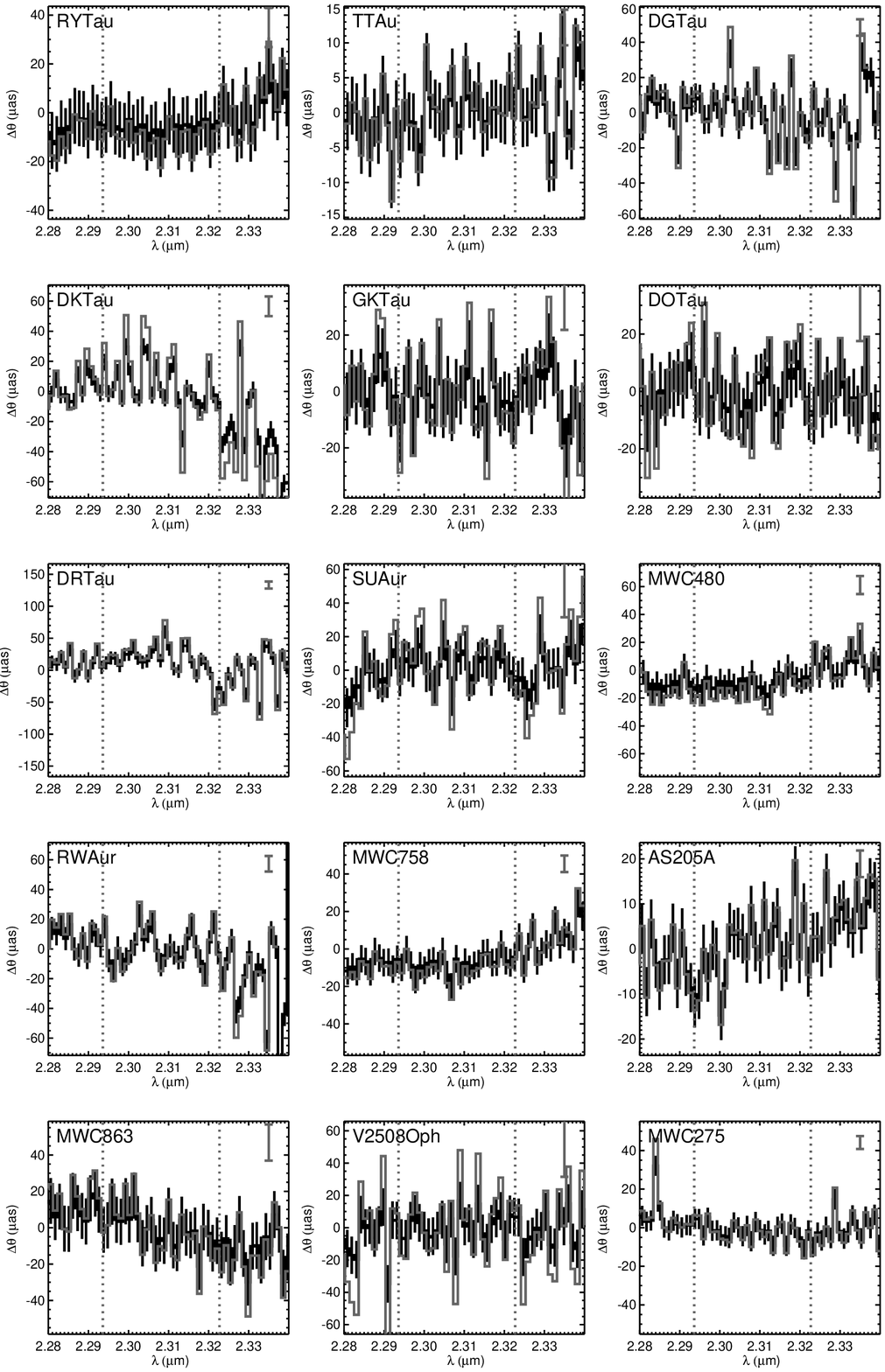}
\caption{Centroid offsets of our sample, plotted in the spectral region
around the CO overtone bandheads.  
Black histograms show the offsets derived for the observed data, and gray histograms
show the centroid offsets for the circumstellar component of the
emission, determined using Equation \ref{eq:dpcirc}. 
\label{fig:codp}}
\end{figure}

\begin{figure}
\ContinuedFloat
\plotone{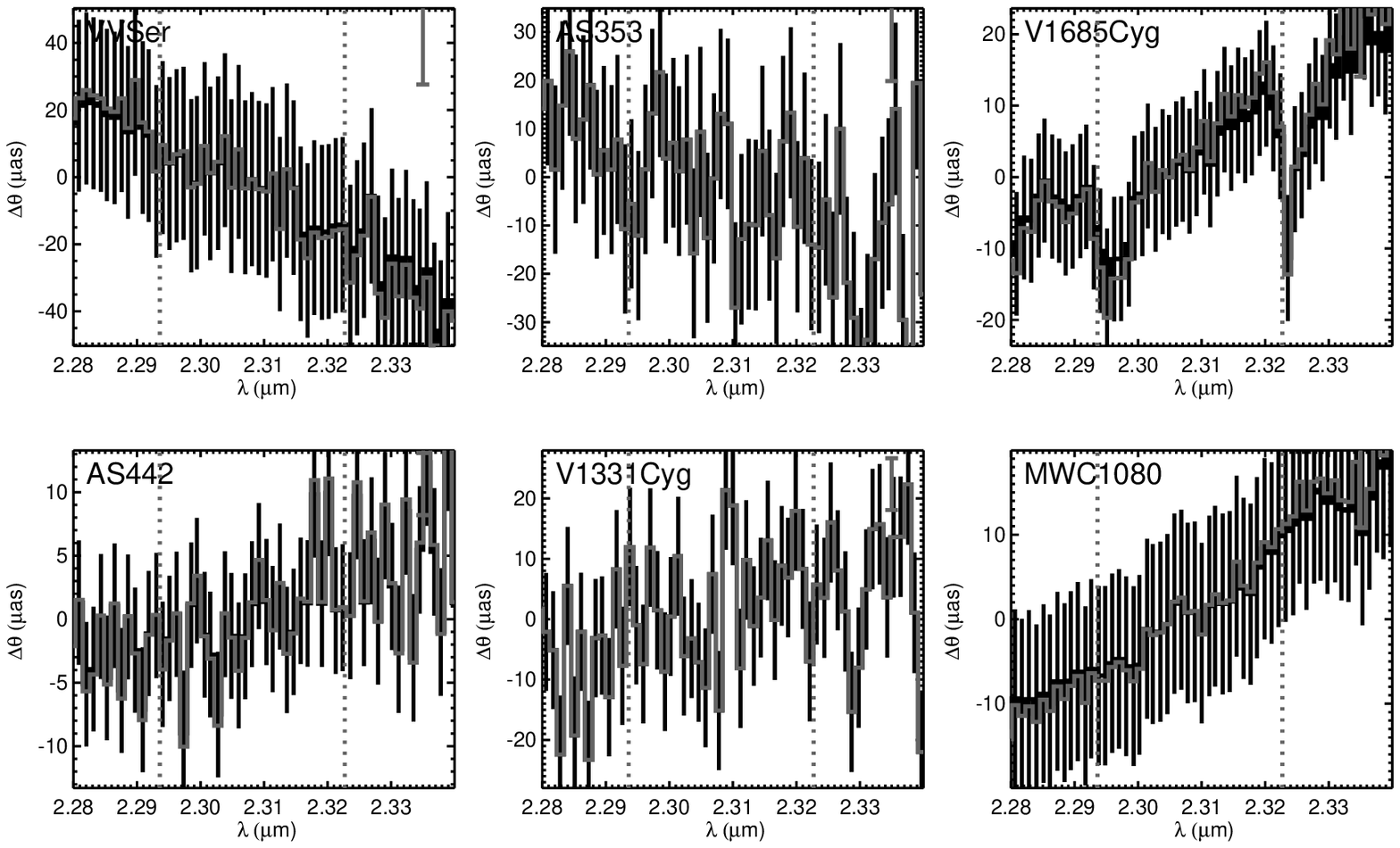}
\caption{continued. 
\label{fig:codp2}}
\end{figure}

\clearpage

\epsscale{0.8}
\begin{figure}
\plotone{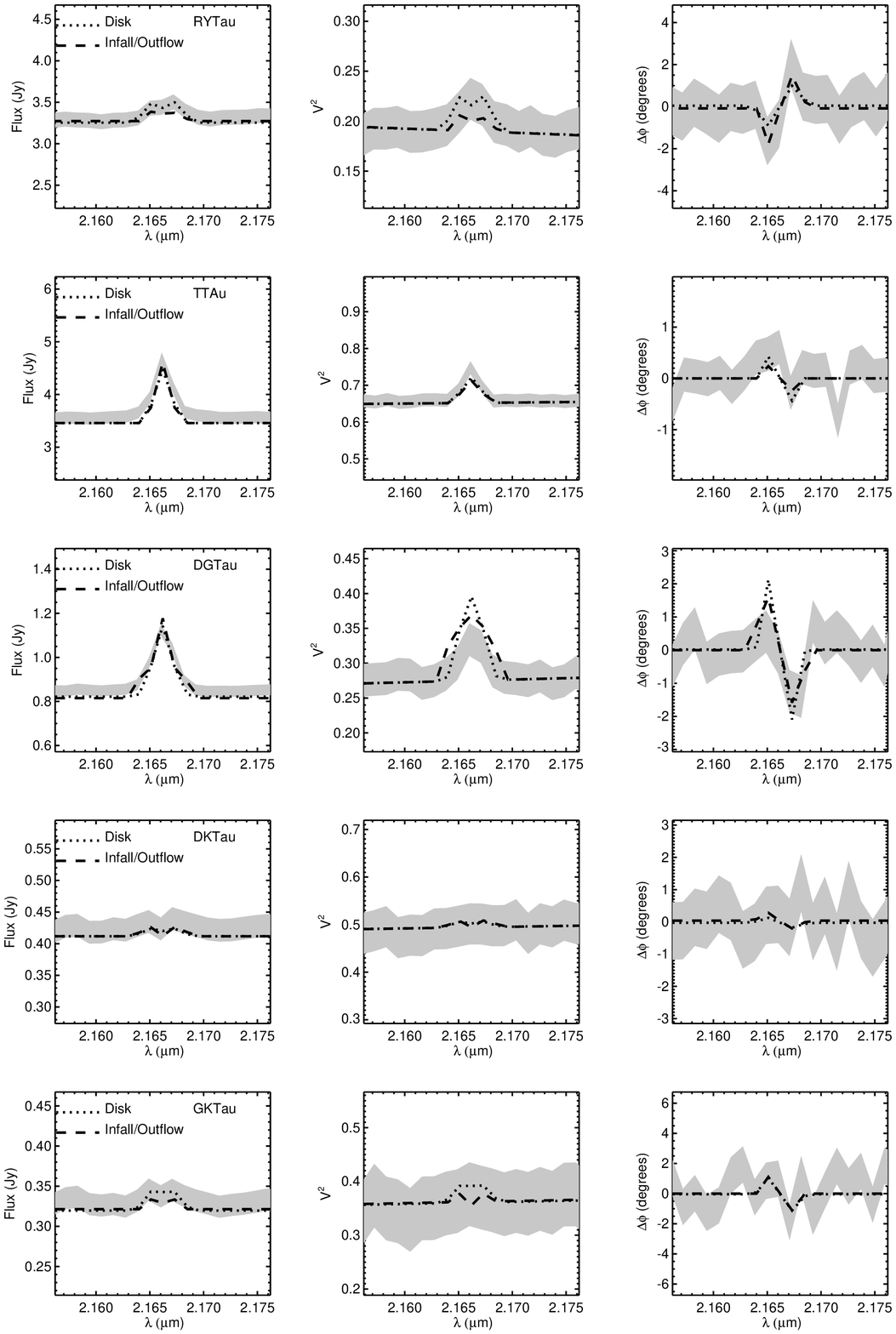}
\caption{Synthetic data for best-fit disk and infall/outflow models (dotted
  and dashed curves) plotted
  atop the 1$\sigma$ region allowed by the observed data (gray shaded
  regions).
\label{fig:brgmods}}
\end{figure}

\begin{figure}
\ContinuedFloat
\plotone{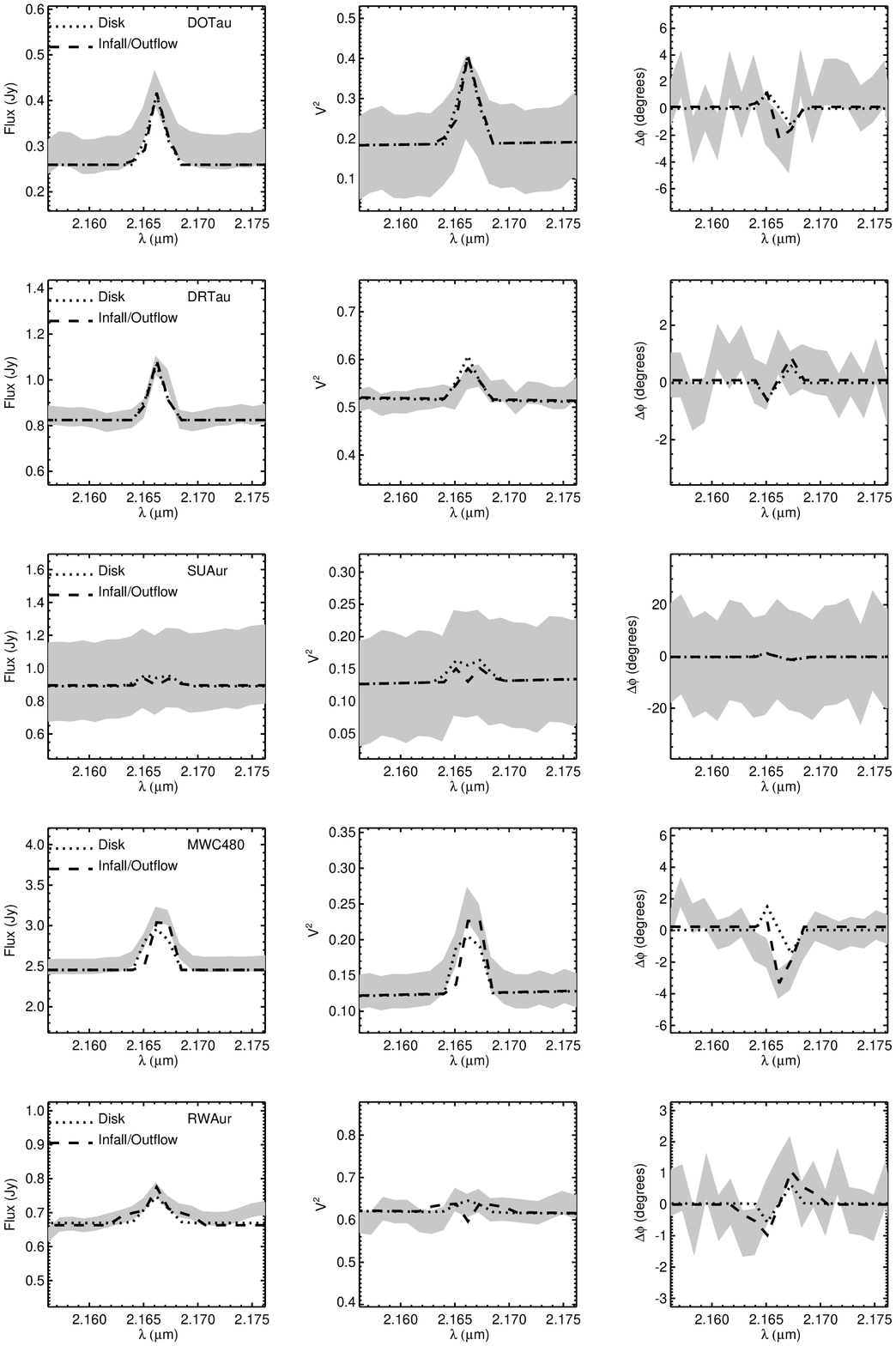}
\caption{continued.}
\end{figure}
\begin{figure}
\ContinuedFloat
\plotone{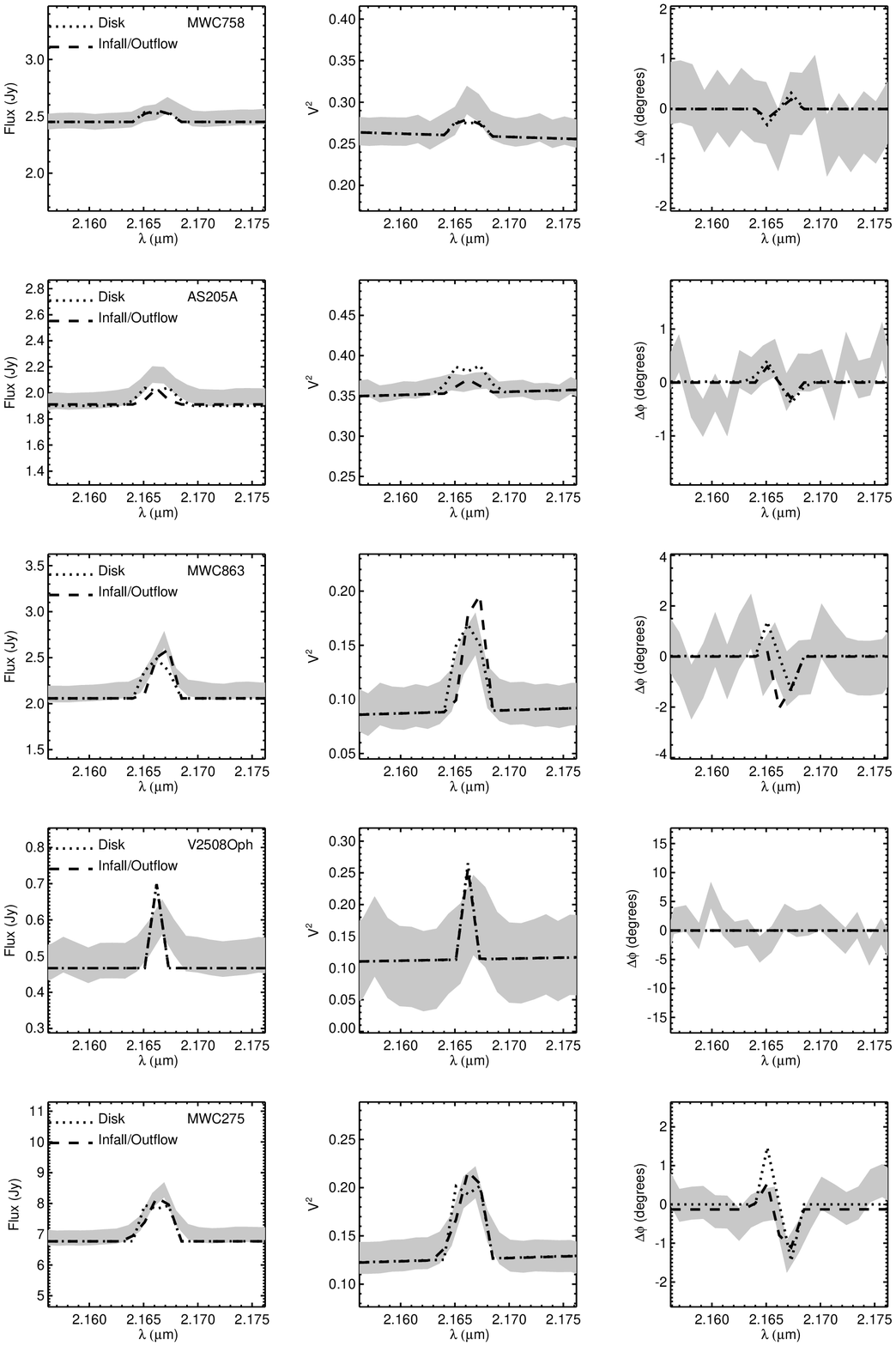}
\caption{ continued.}
\end{figure}
\begin{figure}
\ContinuedFloat
\plotone{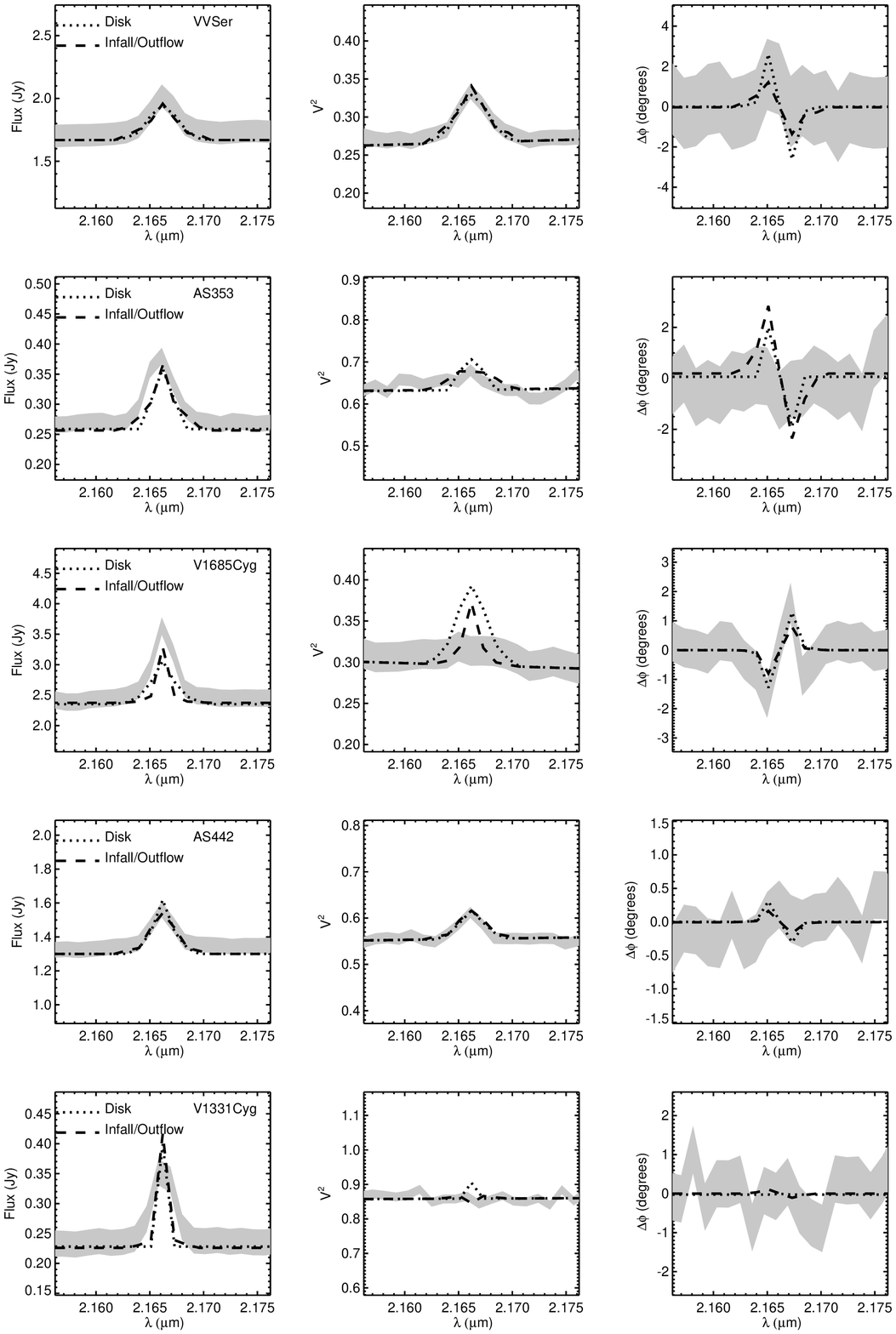}
\caption{continued.}
\end{figure}
\begin{figure}
\ContinuedFloat
\plotone{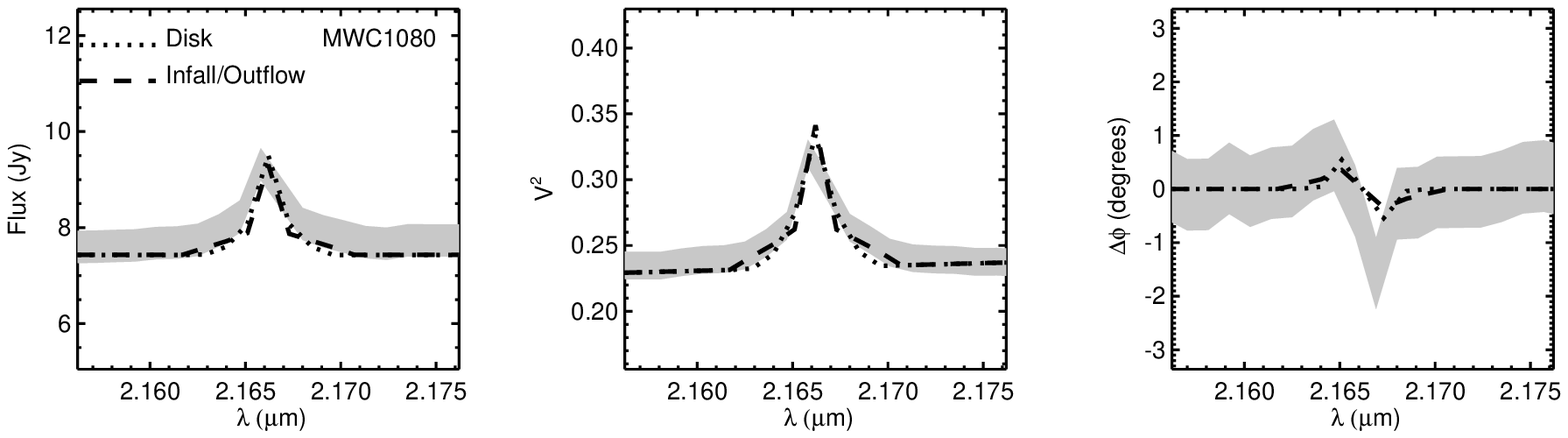}
\caption{continued.}
\end{figure}

\epsscale{0.5}

\epsscale{0.9}

\epsscale{1.0}


\begin{thebibliography}{81}
\expandafter\ifx\csname natexlab\endcsname\relax\def\natexlab#1{#1}\fi

\bibitem[{{Acke} {et~al.}(2005){Acke}, {van den Ancker}, \&
  {Dullemond}}]{AVD05}
{Acke}, B., {van den Ancker}, M.~E., \& {Dullemond}, C.~P. 2005, \aap, 436, 209

\bibitem[{{Bans} \& {K{\"o}nigl}(2012)}]{BK12}
{Bans}, A. \& {K{\"o}nigl}, A. 2012, \apj, 758, 100

\bibitem[{{Beck} {et~al.}(2010){Beck}, {Bary}, \& {McGregor}}]{BBM10}
{Beck}, T.~L., {Bary}, J.~S., \& {McGregor}, P.~J. 2010, \apj, 722, 1360

\bibitem[{{Benisty} {et~al.}(2010){Benisty}, {Natta}, {Isella}, {Berger},
  {Massi}, {Le Bouquin}, {M{\'e}rand}, {Duvert}, {Kraus}, {Malbet}, {Olofsson},
  {Robbe-Dubois}, {Testi}, {Vannier}, \& {Weigelt}}]{BENISTY+10}
{Benisty}, M., {Natta}, A., {Isella}, A., {Berger}, J., {Massi}, F., {Le
  Bouquin}, J., {M{\'e}rand}, A., {Duvert}, G., {Kraus}, S., {Malbet}, F.,
  {Olofsson}, J., {Robbe-Dubois}, S., {Testi}, L., {Vannier}, M., \& {Weigelt},
  G. 2010, \aap, 511, 74

\bibitem[{{Berthoud}(2008)}]{BERTHOUD08}
{Berthoud}, M.~G. 2008, PhD thesis, Cornell University

\bibitem[{{Berthoud} {et~al.}(2007){Berthoud}, {Keller}, {Herter}, {Richter},
  \& {Whelan}}]{BERTHOUD+07}
{Berthoud}, M.~G., {Keller}, L.~D., {Herter}, T.~L., {Richter}, M.~J., \&
  {Whelan}, D.~G. 2007, \apj, 660, 461

\bibitem[{{Bertout} {et~al.}(2007){Bertout}, {Siess}, \& {Cabrit}}]{BSC07}
{Bertout}, C., {Siess}, L., \& {Cabrit}, S. 2007, \aap, 473, L21

\bibitem[{{Biscaya} {et~al.}(1997){Biscaya}, {Rieke}, {Narayanan}, {Luhman}, \&
  {Young}}]{BISCAYA+97}
{Biscaya}, A.~M., {Rieke}, G.~H., {Narayanan}, G., {Luhman}, K.~L., \& {Young},
  E.~T. 1997, \apj, 491, 359

\bibitem[{{Boden} {et~al.}(1998){Boden}, {Colavita}, {van Belle}, \&
  {Shao}}]{BODEN+98}
{Boden}, A.~F., {Colavita}, M.~M., {van Belle}, G.~T., \& {Shao}, M. 1998, in
  Proc. SPIE Vol. 3350, p. 872-880, Astronomical Interferometry, Robert D.
  Reasenberg; Ed., 872--880

\bibitem[{{Brittain} {et~al.}(2007){Brittain}, {Simon}, {Najita}, \&
  {Rettig}}]{BRITTAIN+07}
{Brittain}, S.~D., {Simon}, T., {Najita}, J.~R., \& {Rettig}, T.~W. 2007, \apj,
  659, 685

\bibitem[{{Calvet} {et~al.}(2004){Calvet}, {Muzerolle}, {Brice{\~n}o},
  {Hern{\'a}ndez}, {Hartmann}, {Saucedo}, \& {Gordon}}]{CALVET+04}
{Calvet}, N., {Muzerolle}, J., {Brice{\~n}o}, C., {Hern{\'a}ndez}, J.,
  {Hartmann}, L., {Saucedo}, J.~L., \& {Gordon}, K.~D. 2004, \aj, 128, 1294

\bibitem[{{Carr}(1989)}]{CARR89}
{Carr}, J.~S. 1989, \apj, 345, 522

\bibitem[{{Carr} {et~al.}(2004){Carr}, {Tokunaga}, \& {Najita}}]{CTN04}
{Carr}, J.~S., {Tokunaga}, A.~T., \& {Najita}, J. 2004, \apj, 603, 213

\bibitem[{{Chini}(1981)}]{CHINI81}
{Chini}, R. 1981, \aap, 99, 346

\bibitem[{{Colavita} {et~al.}(2003){Colavita}, {Akeson}, {Wizinowich}, {Shao},
  {Acton}, {Beletic}, {Bell}, {Berlin}, {Boden}, {Booth}, {Boutell}, {Chaffee},
  {Chan}, {Chock}, {Cohen}, {Crawford}, {Creech-Eakman}, {Eychaner},
  {Felizardo}, {Gathright}, {Hardy}, {Henderson}, {Herstein}, {Hess},
  {Hovland}, {Hrynevych}, {Johnson}, {Kelley}, {Kendrick}, {Koresko}, {Kurpis},
  {Le Mignant}, {Lewis}, {Ligon}, {Lupton}, {McBride}, {Mennesson},
  {Millan-Gabet}, {Monnier}, {Moore}, {Nance}, {Neyman}, {Niessner}, {Palmer},
  {Reder}, {Rudeen}, {Saloga}, {Sargent}, {Serabyn}, {Smythe}, {Stomski},
  {Summers}, {Swain}, {Swanson}, {Thompson}, {Tsubota}, {Tumminello}, {van
  Belle}, {Vasisht}, {Vause}, {Walker}, {Wallace}, \& {Wehmeier}}]{COLAVITA+03}
{Colavita}, M., {Akeson}, R., {Wizinowich}, P., {Shao}, M., {Acton}, S.,
  {Beletic}, J., {Bell}, J., {Berlin}, J., {Boden}, A., {Booth}, A., {Boutell},
  R., {Chaffee}, F., {Chan}, D., {Chock}, J., {Cohen}, R., {Crawford}, S.,
  {Creech-Eakman}, M., {Eychaner}, G., {Felizardo}, C., {Gathright}, J.,
  {Hardy}, G., {Henderson}, H., {Herstein}, J., {Hess}, M., {Hovland}, E.,
  {Hrynevych}, M., {Johnson}, R., {Kelley}, J., {Kendrick}, R., {Koresko}, C.,
  {Kurpis}, P., {Le Mignant}, D., {Lewis}, H., {Ligon}, E., {Lupton}, W.,
  {McBride}, D., {Mennesson}, B., {Millan-Gabet}, R., {Monnier}, J., {Moore},
  J., {Nance}, C., {Neyman}, C., {Niessner}, A., {Palmer}, D., {Reder}, L.,
  {Rudeen}, A., {Saloga}, T., {Sargent}, A., {Serabyn}, E., {Smythe}, R.,
  {Stomski}, P., {Summers}, K., {Swain}, M., {Swanson}, P., {Thompson}, R.,
  {Tsubota}, K., {Tumminello}, A., {van Belle}, G., {Vasisht}, G., {Vause}, J.,
  {Walker}, J., {Wallace}, K., \& {Wehmeier}, U. 2003, \apjl, 592, L83

\bibitem[{{Colavita}(1999)}]{COLAVITA99}
{Colavita}, M.~M. 1999, \pasp, 111, 111

\bibitem[{{Colavita} \& {Wizinowich}(2003)}]{CW03}
{Colavita}, M.~M. \& {Wizinowich}, P.~L. 2003, in Interferometry for Optical
  Astronomy II. Edited by Wesley A. Traub. Proceedings of the SPIE, Volume
  4838, pp. 79-88 (2003)., 79--88

\bibitem[{{Colavita} {et~al.}(2013){Colavita}, {Wizinowich}, {Akeson},
  {Ragland}, {Woillez}, {Millan-Gabet}, {Serabyn}, {Abajian}, {Acton},
  {Appleby}, {Beletic}, {Beichman}, {Bell}, {Berkey}, {Berlin}, {Boden},
  {Booth}, {Boutell}, {Chaffee}, {Chan}, {Chin}, {Chock}, {Cohen}, {Cooper},
  {Crawford}, {Creech-Eakman}, {Dahl}, {Eychaner}, {Fanson}, {Felizardo},
  {Garcia-Gathright}, {Gathright}, {Hardy}, {Henderson}, {Herstein}, {Hess},
  {Hovland}, {Hrynevych}, {Johansson}, {Johnson}, {Kelley}, {Kendrick},
  {Koresko}, {Kurpis}, {Le Mignant}, {Lewis}, {Ligon}, {Lupton}, {McBride},
  {Medeiros}, {Mennesson}, {Moore}, {Morrison}, {Nance}, {Neyman}, {Niessner},
  {Paine}, {Palmer}, {Panteleeva}, {Papin}, {Parvin}, {Reder}, {Rudeen},
  {Saloga}, {Sargent}, {Shao}, {Smith}, {Smythe}, {Stomski}, {Summers},
  {Swain}, {Swanson}, {Thompson}, {Tsubota}, {Tumminello}, {Tyau}, {van Belle},
  {Vasisht}, {Vause}, {Vescelus}, {Walker}, {Wallace}, {Wehmeier}, \&
  {Wetherell}}]{COLAVITA+13}
{Colavita}, M.~M., {Wizinowich}, P.~L., {Akeson}, R.~L., {Ragland}, S.,
  {Woillez}, J.~M., {Millan-Gabet}, R., {Serabyn}, E., {Abajian}, M., {Acton},
  D.~S., {Appleby}, E., {Beletic}, J.~W., {Beichman}, C.~A., {Bell}, J.,
  {Berkey}, B.~C., {Berlin}, J., {Boden}, A.~F., {Booth}, A.~J., {Boutell}, R.,
  {Chaffee}, F.~H., {Chan}, D., {Chin}, J., {Chock}, J., {Cohen}, R., {Cooper},
  A., {Crawford}, S.~L., {Creech-Eakman}, M.~J., {Dahl}, W., {Eychaner}, G.,
  {Fanson}, J.~L., {Felizardo}, C., {Garcia-Gathright}, J.~I., {Gathright},
  J.~T., {Hardy}, G., {Henderson}, H., {Herstein}, J.~S., {Hess}, M.,
  {Hovland}, E.~E., {Hrynevych}, M.~A., {Johansson}, E., {Johnson}, R.~L.,
  {Kelley}, J., {Kendrick}, R., {Koresko}, C.~D., {Kurpis}, P., {Le Mignant},
  D., {Lewis}, H.~A., {Ligon}, E.~R., {Lupton}, W., {McBride}, D., {Medeiros},
  D.~W., {Mennesson}, B.~P., {Moore}, J.~D., {Morrison}, D., {Nance}, C.,
  {Neyman}, C., {Niessner}, A., {Paine}, C.~G., {Palmer}, D.~L., {Panteleeva},
  T., {Papin}, M., {Parvin}, B., {Reder}, L., {Rudeen}, A., {Saloga}, T.,
  {Sargent}, A., {Shao}, M., {Smith}, B., {Smythe}, R.~F., {Stomski}, P.,
  {Summers}, K.~R., {Swain}, M.~R., {Swanson}, P., {Thompson}, R., {Tsubota},
  K., {Tumminello}, A., {Tyau}, C., {van Belle}, G.~T., {Vasisht}, G., {Vause},
  J., {Vescelus}, F., {Walker}, J., {Wallace}, J.~K., {Wehmeier}, U., \&
  {Wetherell}, E. 2013, \pasp, 125, 1226

\bibitem[{{Corcoran} \& {Ray}(1997)}]{CR97}
{Corcoran}, M. \& {Ray}, T.~P. 1997, \aap, 321, 189

\bibitem[{{Dullemond} {et~al.}(2001){Dullemond}, {Dominik}, \& {Natta}}]{DDN01}
{Dullemond}, C.~P., {Dominik}, C., \& {Natta}, A. 2001, \apj, 560, 957

\bibitem[{{Dullemond} \& {Monnier}(2010)}]{DM10}
{Dullemond}, C.~P. \& {Monnier}, J.~D. 2010, \araa, 48, 205

\bibitem[{{Eisner}(2007)}]{EISNER07}
{Eisner}, J.~A. 2007, \nat, 447, 562

\bibitem[{{Eisner} {et~al.}(2007{\natexlab{a}}){Eisner}, {Chiang}, {Lane}, \&
  {Akeson}}]{EISNER+07a}
{Eisner}, J.~A., {Chiang}, E.~I., {Lane}, B.~F., \& {Akeson}, R.~L.
  2007{\natexlab{a}}, \apj, 657, 347

\bibitem[{{Eisner} {et~al.}(2007{\natexlab{b}}){Eisner}, {Graham}, {Akeson},
  {Ligon}, {Colavita}, {Basri}, {Summers}, {Ragland}, \& {Booth}}]{EISNER+07b}
{Eisner}, J.~A., {Graham}, J.~R., {Akeson}, R.~L., {Ligon}, E.~R., {Colavita},
  M.~M., {Basri}, G., {Summers}, K., {Ragland}, S., \& {Booth}, A.
  2007{\natexlab{b}}, \apjl, 654, L77

\bibitem[{{Eisner} {et~al.}(2009){Eisner}, {Graham}, {Akeson}, \&
  {Najita}}]{EISNER+09}
{Eisner}, J.~A., {Graham}, J.~R., {Akeson}, R.~L., \& {Najita}, J. 2009, \apj,
  692, 309

\bibitem[{{Eisner} \& {Hillenbrand}(2011)}]{EH11}
{Eisner}, J.~A. \& {Hillenbrand}, L.~A. 2011, \apj, 738, 9

\bibitem[{{Eisner} {et~al.}(2005){Eisner}, {Hillenbrand}, {White}, {Akeson}, \&
  {Sargent}}]{EISNER+05}
{Eisner}, J.~A., {Hillenbrand}, L.~A., {White}, R.~J., {Akeson}, R.~L., \&
  {Sargent}, A.~I. 2005, \apj, 623, 952

\bibitem[{{Eisner} {et~al.}(2007{\natexlab{c}}){Eisner}, {Hillenbrand},
  {White}, {Bloom}, {Akeson}, \& {Blake}}]{EISNER+07c}
{Eisner}, J.~A., {Hillenbrand}, L.~A., {White}, R.~J., {Bloom}, J.~S.,
  {Akeson}, R.~L., \& {Blake}, C.~H. 2007{\natexlab{c}}, \apj, 669, 1072

\bibitem[{{Eisner} {et~al.}(2004){Eisner}, {Lane}, {Hillenbrand}, {Akeson}, \&
  {Sargent}}]{EISNER+04}
{Eisner}, J.~A., {Lane}, B.~F., {Hillenbrand}, L., {Akeson}, R., \& {Sargent},
  A. 2004, \apj, 613, 1049

\bibitem[{{Eisner} {et~al.}(2010){Eisner}, {Monnier}, {Woillez}, {Akeson},
  {Millan-Gabet}, {Graham}, {Hillenbrand}, {Pott}, {Ragland}, \&
  {Wizinowich}}]{EISNER+10}
{Eisner}, J.~A., {Monnier}, J.~D., {Woillez}, J., {Akeson}, R.~L.,
  {Millan-Gabet}, R., {Graham}, J.~R., {Hillenbrand}, L.~A., {Pott}, J.,
  {Ragland}, S., \& {Wizinowich}, P. 2010, \apj, 718, 774

\bibitem[{{Eisner} {et~al.}(2013){Eisner}, {Rieke}, {Rieke}, {Flaherty},
  {Arnold}, {Stone}, {Cortes}, {Cox}, {Hawkins}, {Cole}, {Zajac}, \&
  {Rudolph}}]{EISNER+13}
{Eisner}, J.~A., {Rieke}, G.~H., {Rieke}, M.~J., {Flaherty}, K.~M., {Arnold},
  T.~J., {Stone}, J.~M., {Cortes}, S.~R., {Cox}, E., {Hawkins}, C., {Cole}, A.,
  {Zajac}, S., \& {Rudolph}, A.~L. 2013, \mnras, 434, 407

\bibitem[{{Folha} \& {Emerson}(2001)}]{FE01}
{Folha}, D.~F.~M. \& {Emerson}, J.~P. 2001, \aap, 365, 90

\bibitem[{{Glassgold} {et~al.}(2004){Glassgold}, {Najita}, \& {Igea}}]{GNI04}
{Glassgold}, A.~E., {Najita}, J., \& {Igea}, J. 2004, \apj, 615, 972

\bibitem[{{Gorti} \& {Hollenbach}(2008)}]{GH08}
{Gorti}, U. \& {Hollenbach}, D. 2008, \apj, 683, 287

\bibitem[{{Hamann}(1994)}]{HAMANN94}
{Hamann}, F. 1994, \apjs, 93, 485

\bibitem[{{Hartigan} {et~al.}(1995){Hartigan}, {Edwards}, \&
  {Ghandour}}]{HEG95}
{Hartigan}, P., {Edwards}, S., \& {Ghandour}, L. 1995, \apj, 452, 736

\bibitem[{{Hartigan} {et~al.}(1986){Hartigan}, {Mundt}, \& {Stocke}}]{HMS86}
{Hartigan}, P., {Mundt}, R., \& {Stocke}, J. 1986, \aj, 91, 1357

\bibitem[{{Hauschildt} {et~al.}(1999){Hauschildt}, {Allard}, \&
  {Baron}}]{HAB99}
{Hauschildt}, P.~H., {Allard}, F., \& {Baron}, E. 1999, \apj, 512, 377

\bibitem[{{Herbig} \& {Jones}(1983)}]{HJ83}
{Herbig}, G.~H. \& {Jones}, B.~F. 1983, \aj, 88, 1040

\bibitem[{{Herbig} {et~al.}(2003){Herbig}, {Petrov}, \& {Duemmler}}]{HPD03}
{Herbig}, G.~H., {Petrov}, P.~P., \& {Duemmler}, R. 2003, \apj, 595, 384

\bibitem[{{Herczeg} \& {Hillenbrand}(2014)}]{HH14}
{Herczeg}, G.~J. \& {Hillenbrand}, L.~A. 2014, ArXiv e-prints

\bibitem[{{Isella} \& {Natta}(2005)}]{IN05}
{Isella}, A. \& {Natta}, A. 2005, \aap, 438, 899

\bibitem[{{Isella} {et~al.}(2008){Isella}, {Tatulli}, {Natta}, \&
  {Testi}}]{ISELLA+08}
{Isella}, A., {Tatulli}, E., {Natta}, A., \& {Testi}, L. 2008, \aap, 483, L13

\bibitem[{{Kenyon} {et~al.}(1994){Kenyon}, {Dobrzycka}, \& {Hartmann}}]{KDH94}
{Kenyon}, S.~J., {Dobrzycka}, D., \& {Hartmann}, L. 1994, \aj, 108, 1872

\bibitem[{{K\"{o}nigl}(1991)}]{KONIGL91}
{K\"{o}nigl}, A. 1991, \apjl, 370, L39

\bibitem[{{Konigl} \& {Pudritz}(2000)}]{KP00}
{Konigl}, A. \& {Pudritz}, R.~E. 2000, Protostars and Planets IV, 759

\bibitem[{{Kraus} {et~al.}(2008{\natexlab{a}}){Kraus}, {Hofmann}, {Benisty},
  {Berger}, {Chesneau}, {Isella}, {Malbet}, {Meilland}, {Nardetto}, {Natta},
  {Preibisch}, {Schertl}, {Smith}, {Stee}, {Tatulli}, {Testi}, \&
  {Weigelt}}]{KRAUS+08}
{Kraus}, S., {Hofmann}, K.-H., {Benisty}, M., {Berger}, J.-P., {Chesneau}, O.,
  {Isella}, A., {Malbet}, F., {Meilland}, A., {Nardetto}, N., {Natta}, A.,
  {Preibisch}, T., {Schertl}, D., {Smith}, M., {Stee}, P., {Tatulli}, E.,
  {Testi}, L., \& {Weigelt}, G. 2008{\natexlab{a}}, \aap, 489, 1157

\bibitem[{{Kraus} {et~al.}(2008{\natexlab{b}}){Kraus}, {Preibisch}, \&
  {Ohnaka}}]{KPO08}
{Kraus}, S., {Preibisch}, T., \& {Ohnaka}, K. 2008{\natexlab{b}}, \apj, 676,
  490

\bibitem[{{Kuhi}(1964)}]{KUHI64}
{Kuhi}, L.~V. 1964, \apj, 140, 1409

\bibitem[{{Kurosawa} {et~al.}(2006){Kurosawa}, {Harries}, \&
  {Symington}}]{KHS06}
{Kurosawa}, R., {Harries}, T.~J., \& {Symington}, N.~H. 2006, \mnras, 370, 580

\bibitem[{{Lynden-Bell} \& {Pringle}(1974)}]{LP74}
{Lynden-Bell}, D. \& {Pringle}, J.~E. 1974, \mnras, 168, 603

\bibitem[{{Malbet} {et~al.}(2007){Malbet}, {Benisty}, {de Wit}, {Kraus},
  {Meilland}, {Millour}, {Tatulli}, {Berger}, {Chesneau}, {Hofmann}, {Isella},
  {Natta}, {Petrov}, {Preibisch}, {Stee}, {Testi}, {Weigelt}, {Antonelli},
  {Beckmann}, {Bresson}, {Chelli}, {Dugu{\'e}}, {Duvert}, {Gennari},
  {Gl{\"u}ck}, {Kern}, {Lagarde}, {Le Coarer}, {Lisi}, {Perraut}, {Puget},
  {Rantakyr{\"o}}, {Robbe-Dubois}, {Roussel}, {Zins}, {Accardo}, {Acke},
  {Agabi}, {Altariba}, {Arezki}, {Aristidi}, {Baffa}, {Behrend}, {Bl{\"o}cker},
  {Bonhomme}, {Busoni}, {Cassaing}, {Clausse}, {Colin}, {Connot},
  {Delboulb{\'e}}, {Domiciano de Souza}, {Driebe}, {Feautrier}, {Ferruzzi},
  {Forveille}, {Fossat}, {Foy}, {Fraix-Burnet}, {Gallardo}, {Giani}, {Gil},
  {Glentzlin}, {Heiden}, {Heininger}, {Hernandez Utrera}, {Kamm}, {Kiekebusch},
  {Le Contel}, {Le Contel}, {Lesourd}, {Lopez}, {Lopez}, {Magnard}, {Marconi},
  {Mars}, {Martinot-Lagarde}, {Mathias}, {M{\`e}ge}, {Monin}, {Mouillet},
  {Mourard}, {Nussbaum}, {Ohnaka}, {Pacheco}, {Perrier}, {Rabbia}, {Rebattu},
  {Reynaud}, {Richichi}, {Robini}, {Sacchettini}, {Schertl}, {Sch{\"o}ller},
  {Solscheid}, {Spang}, {Stefanini}, {Tallon}, {Tallon-Bosc}, {Tasso},
  {Vakili}, {von der L{\"u}he}, {Valtier}, {Vannier}, \& {Ventura}}]{MALBET+07}
{Malbet}, F., {Benisty}, M., {de Wit}, W.-J., {Kraus}, S., {Meilland}, A.,
  {Millour}, F., {Tatulli}, E., {Berger}, J.-P., {Chesneau}, O., {Hofmann},
  K.-H., {Isella}, A., {Natta}, A., {Petrov}, R.~G., {Preibisch}, T., {Stee},
  P., {Testi}, L., {Weigelt}, G., {Antonelli}, P., {Beckmann}, U., {Bresson},
  Y., {Chelli}, A., {Dugu{\'e}}, M., {Duvert}, G., {Gennari}, S., {Gl{\"u}ck},
  L., {Kern}, P., {Lagarde}, S., {Le Coarer}, E., {Lisi}, F., {Perraut}, K.,
  {Puget}, P., {Rantakyr{\"o}}, F., {Robbe-Dubois}, S., {Roussel}, A., {Zins},
  G., {Accardo}, M., {Acke}, B., {Agabi}, K., {Altariba}, E., {Arezki}, B.,
  {Aristidi}, E., {Baffa}, C., {Behrend}, J., {Bl{\"o}cker}, T., {Bonhomme},
  S., {Busoni}, S., {Cassaing}, F., {Clausse}, J.-M., {Colin}, J., {Connot},
  C., {Delboulb{\'e}}, A., {Domiciano de Souza}, A., {Driebe}, T., {Feautrier},
  P., {Ferruzzi}, D., {Forveille}, T., {Fossat}, E., {Foy}, R., {Fraix-Burnet},
  D., {Gallardo}, A., {Giani}, E., {Gil}, C., {Glentzlin}, A., {Heiden}, M.,
  {Heininger}, M., {Hernandez Utrera}, O., {Kamm}, D., {Kiekebusch}, M., {Le
  Contel}, D., {Le Contel}, J.-M., {Lesourd}, T., {Lopez}, B., {Lopez}, M.,
  {Magnard}, Y., {Marconi}, A., {Mars}, G., {Martinot-Lagarde}, G., {Mathias},
  P., {M{\`e}ge}, P., {Monin}, J.-L., {Mouillet}, D., {Mourard}, D.,
  {Nussbaum}, E., {Ohnaka}, K., {Pacheco}, J., {Perrier}, C., {Rabbia}, Y.,
  {Rebattu}, S., {Reynaud}, F., {Richichi}, A., {Robini}, A., {Sacchettini},
  M., {Schertl}, D., {Sch{\"o}ller}, M., {Solscheid}, W., {Spang}, A.,
  {Stefanini}, P., {Tallon}, M., {Tallon-Bosc}, I., {Tasso}, D., {Vakili}, F.,
  {von der L{\"u}he}, O., {Valtier}, J.-C., {Vannier}, M., \& {Ventura}, N.
  2007, \aap, 464, 43

\bibitem[{{McLean} {et~al.}(2003){McLean}, {McGovern}, {Burgasser},
  {Kirkpatrick}, {Prato}, \& {Kim}}]{MCLEAN+03}
{McLean}, I.~S., {McGovern}, M.~R., {Burgasser}, A.~J., {Kirkpatrick}, J.~D.,
  {Prato}, L., \& {Kim}, S.~S. 2003, \apj, 596, 561

\bibitem[{{Mendigut{\'{\i}}a} {et~al.}(2011){Mendigut{\'{\i}}a}, {Calvet},
  {Montesinos}, {Mora}, {Muzerolle}, {Eiroa}, {Oudmaijer}, \&
  {Mer{\'{\i}}n}}]{MENDIGUTIA+11b}
{Mendigut{\'{\i}}a}, I., {Calvet}, N., {Montesinos}, B., {Mora}, A.,
  {Muzerolle}, J., {Eiroa}, C., {Oudmaijer}, R.~D., \& {Mer{\'{\i}}n}, B. 2011,
  \aap, 535, A99

\bibitem[{{Monin} {et~al.}(1998){Monin}, {Menard}, \& {Duchene}}]{MMD98}
{Monin}, J.-L., {Menard}, F., \& {Duchene}, G. 1998, \aap, 339, 113

\bibitem[{{Monnier} {et~al.}(2006){Monnier}, {Berger}, {Millan-Gabet}, {Traub},
  {Schloerb}, {Pedretti}, {Benisty}, {Carleton}, {Haguenauer}, {Kern},
  {Labeye}, {Lacasse}, {Malbet}, {Perraut}, {Pearlman}, \& {Zhao}}]{MONNIER+06}
{Monnier}, J.~D., {Berger}, J.-P., {Millan-Gabet}, R., {Traub}, W.~A.,
  {Schloerb}, F.~P., {Pedretti}, E., {Benisty}, M., {Carleton}, N.~P.,
  {Haguenauer}, P., {Kern}, P., {Labeye}, P., {Lacasse}, M.~G., {Malbet}, F.,
  {Perraut}, K., {Pearlman}, M., \& {Zhao}, M. 2006, \apj, 647, 444

\bibitem[{{Monnier} \& {Millan-Gabet}(2002)}]{MM02}
{Monnier}, J.~D. \& {Millan-Gabet}, R. 2002, \apj, 579, 694

\bibitem[{{Monnier} {et~al.}(2005){Monnier}, {Millan-Gabet}, {Billmeier},
  {Akeson}, {Wallace}, {Berger}, {Calvet}, {D'Alessio}, {Danchi}, {Hartmann},
  {Hillenbrand}, {Kuchner}, {Rajagopal}, {Traub}, {Tuthill}, {Boden}, {Booth},
  {Colavita}, {Gathright}, {Hrynevych}, {Le Mignant}, {Ligon}, {Neyman},
  {Swain}, {Thompson}, {Vasisht}, {Wizinowich}, {Beichman}, {Beletic},
  {Creech-Eakman}, {Koresko}, {Sargent}, {Shao}, \& {van Belle}}]{MONNIER+05}
{Monnier}, J.~D., {Millan-Gabet}, R., {Billmeier}, R., {Akeson}, R.~L.,
  {Wallace}, D., {Berger}, J.-P., {Calvet}, N., {D'Alessio}, P., {Danchi},
  W.~C., {Hartmann}, L., {Hillenbrand}, L.~A., {Kuchner}, M., {Rajagopal}, J.,
  {Traub}, W.~A., {Tuthill}, P.~G., {Boden}, A., {Booth}, A., {Colavita}, M.,
  {Gathright}, J., {Hrynevych}, M., {Le Mignant}, D., {Ligon}, R., {Neyman},
  C., {Swain}, M., {Thompson}, R., {Vasisht}, G., {Wizinowich}, P., {Beichman},
  C., {Beletic}, J., {Creech-Eakman}, M., {Koresko}, C., {Sargent}, A., {Shao},
  M., \& {van Belle}, G. 2005, \apj, 624, 832

\bibitem[{{Mundt} \& {Eisl{\"o}ffel}(1998)}]{ME98}
{Mundt}, R. \& {Eisl{\"o}ffel}, J. 1998, \aj, 116, 860

\bibitem[{{Muzerolle} {et~al.}(2003){Muzerolle}, {Calvet}, {Hartmann}, \&
  {D'Alessio }}]{MUZEROLLE+03}
{Muzerolle}, J., {Calvet}, N., {Hartmann}, L., \& {D'Alessio }, P. 2003, \apj,
  597, L865

\bibitem[{{Muzerolle} {et~al.}(1998){Muzerolle}, {Hartmann}, \&
  {Calvet}}]{MHC98}
{Muzerolle}, J., {Hartmann}, L., \& {Calvet}, N. 1998, \aj, 116, 2965

\bibitem[{{Najita} {et~al.}(1996{\natexlab{a}}){Najita}, {Carr}, {Glassgold},
  {Shu}, \& {Tokunaga}}]{NAJITA+96}
{Najita}, J., {Carr}, J.~S., {Glassgold}, A.~E., {Shu}, F.~H., \& {Tokunaga},
  A.~T. 1996{\natexlab{a}}, \apj, 462, 919

\bibitem[{{Najita} {et~al.}(1996{\natexlab{b}}){Najita}, {Carr}, \&
  {Tokunaga}}]{NCT96}
{Najita}, J., {Carr}, J.~S., \& {Tokunaga}, A.~T. 1996{\natexlab{b}}, \apj,
  456, 292

\bibitem[{{Najita} {et~al.}(2007){Najita}, {Carr}, {Glassgold}, \&
  {Valenti}}]{NAJITA+06}
{Najita}, J.~R., {Carr}, J.~S., {Glassgold}, A.~E., \& {Valenti}, J.~A. 2007,
  in Protostars and Planets V, B. Reipurth, D. Jewitt, and K. Keil (eds.),
  University of Arizona Press, Tucson, 951 pp., 2007., p.507-522, ed.
  B.~{Reipurth}, D.~{Jewitt}, \& K.~{Keil}, 507--522

\bibitem[{{Najita} {et~al.}(2009){Najita}, {Doppmann}, {Carr}, {Graham}, \&
  {Eisner}}]{NAJITA+09}
{Najita}, J.~R., {Doppmann}, G.~W., {Carr}, J.~S., {Graham}, J.~R., \&
  {Eisner}, J.~A. 2009, \apj, 691, 738

\bibitem[{{Najita} {et~al.}(2000){Najita}, {Edwards}, {Basri}, \&
  {Carr}}]{NAJITA+00}
{Najita}, J.~R., {Edwards}, S., {Basri}, G., \& {Carr}, J. 2000, Protostars and
  Planets IV, 457

\bibitem[{{Palla} \& {Stahler}(1993)}]{PS93}
{Palla}, F. \& {Stahler}, S.~W. 1993, \apj, 418, 414

\bibitem[{{Perryman} {et~al.}(1997){Perryman}, {Lindegren}, {Kovalevsky},
  {Hoeg}, {Bastian}, {Bernacca}, {Cr{\'e}z{\'e}}, {Donati}, {Grenon}, {van
  Leeuwen}, {van der Marel}, {Mignard}, {Murray}, {Le Poole}, {Schrijver},
  {Turon}, {Arenou}, {Froeschl{\'e}}, \& {Petersen}}]{PERRYMAN+97}
{Perryman}, M.~A.~C., {Lindegren}, L., {Kovalevsky}, J., {Hoeg}, E., {Bastian},
  U., {Bernacca}, P.~L., {Cr{\'e}z{\'e}}, M., {Donati}, F., {Grenon}, M., {van
  Leeuwen}, F., {van der Marel}, H., {Mignard}, F., {Murray}, C.~A., {Le
  Poole}, R.~S., {Schrijver}, H., {Turon}, C., {Arenou}, F., {Froeschl{\'e}},
  M., \& {Petersen}, C.~S. 1997, \aap, 323, L49

\bibitem[{{Prato} {et~al.}(2003){Prato}, {Greene}, \& {Simon}}]{PGS03}
{Prato}, L., {Greene}, T.~P., \& {Simon}, M. 2003, \apj, 584, 853

\bibitem[{{Pyo} {et~al.}(2003){Pyo}, {Kobayashi}, {Hayashi}, {Terada}, {Goto},
  {Takami}, {Takato}, {Gaessler}, {Usuda}, {Yamashita}, {Tokunaga}, {Hayano},
  {Kamata}, {Iye}, \& {Minowa}}]{PYO+03}
{Pyo}, T.-S., {Kobayashi}, N., {Hayashi}, M., {Terada}, H., {Goto}, M.,
  {Takami}, H., {Takato}, N., {Gaessler}, W., {Usuda}, T., {Yamashita}, T.,
  {Tokunaga}, A.~T., {Hayano}, Y., {Kamata}, Y., {Iye}, M., \& {Minowa}, Y.
  2003, \apj, 590, 340

\bibitem[{{Rothman} {et~al.}(2005){Rothman}, {Jacquemart}, {Barbe}, {Benner},
  {Birk}, {Brown}, {Carleer}, {Chackerian}, {Chance}, {Coudert}, {Dana},
  {Devi}, {Flaud}, {Gamache}, {Goldman}, {Hartmann}, {Jucks}, {Maki}, {Mandin},
  {Massie}, {Orphal}, {Perrin}, {Rinsland}, {Smith}, {Tennyson}, {Tolchenov},
  {Toth}, {Vander Auwera}, {Varanasi}, \& {Wagner}}]{ROTHMAN+05}
{Rothman}, L.~S., {Jacquemart}, D., {Barbe}, A., {Benner}, D.~C., {Birk}, M.,
  {Brown}, L.~R., {Carleer}, M.~R., {Chackerian}, C., {Chance}, K., {Coudert},
  L.~H., {Dana}, V., {Devi}, V.~M., {Flaud}, J.~M., {Gamache}, R.~R.,
  {Goldman}, A., {Hartmann}, J.~M., {Jucks}, K.~W., {Maki}, A.~G., {Mandin},
  J.~Y., {Massie}, S.~T., {Orphal}, J., {Perrin}, A., {Rinsland}, C.~P.,
  {Smith}, M.~A.~H., {Tennyson}, J., {Tolchenov}, R.~N., {Toth}, R.~A., {Vander
  Auwera}, J., {Varanasi}, P., \& {Wagner}, G. 2005, Journal of Quantitative
  Spectroscopy and Radiative Transfer, 96, 139

\bibitem[{{Shu} {et~al.}(1994){Shu}, {Najita}, {Ostriker}, {Wilkin}, {Ruden},
  \& {Lizano}}]{SHU+94}
{Shu}, F., {Najita}, J., {Ostriker}, E., {Wilkin}, F., {Ruden}, S., \&
  {Lizano}, S. 1994, \apj, 429, 781

\bibitem[{{Tannirkulam} {et~al.}(2008){Tannirkulam}, {Monnier}, {Millan-Gabet},
  {Harries}, {Pedretti}, {ten Brummelaar}, {McAlister}, {Turner}, {Sturmann},
  \& {Sturmann}}]{TANNIRKULAM+08}
{Tannirkulam}, A., {Monnier}, J.~D., {Millan-Gabet}, R., {Harries}, T.~J.,
  {Pedretti}, E., {ten Brummelaar}, T.~A., {McAlister}, H., {Turner}, N.,
  {Sturmann}, J., \& {Sturmann}, L. 2008, \apjl, 677, L51

\bibitem[{{Tatulli} {et~al.}(2007){Tatulli}, {Isella}, {Natta}, {Testi},
  {Marconi}, {Malbet}, {Stee}, {Petrov}, {Millour}, {Chelli}, {Duvert},
  {Antonelli}, {Beckmann}, {Bresson}, {Dugu{\'e}}, {Gennari}, {Gl{\"u}ck},
  {Kern}, {Lagarde}, {Le Coarer}, {Lisi}, {Perraut}, {Puget}, {Rantakyr{\"o}},
  {Robbe-Dubois}, {Roussel}, {Weigelt}, {Zins}, {Accardo}, {Acke}, {Agabi},
  {Altariba}, {Arezki}, {Aristidi}, {Baffa}, {Behrend}, {Bl{\"o}cker},
  {Bonhomme}, {Busoni}, {Cassaing}, {Clausse}, {Colin}, {Connot},
  {Delboulb{\'e}}, {Domiciano de Souza}, {Driebe}, {Feautrier}, {Ferruzzi},
  {Forveille}, {Fossat}, {Foy}, {Fraix-Burnet}, {Gallardo}, {Giani}, {Gil},
  {Glentzlin}, {Heiden}, {Heininger}, {Hernandez Utrera}, {Hofmann}, {Kamm},
  {Kiekebusch}, {Kraus}, {Le Contel}, {Le Contel}, {Lesourd}, {Lopez}, {Lopez},
  {Magnard}, {Mars}, {Martinot-Lagarde}, {Mathias}, {M{\`e}ge}, {Monin},
  {Mouillet}, {Mourard}, {Nussbaum}, {Ohnaka}, {Pacheco}, {Perrier}, {Rabbia},
  {Rebattu}, {Reynaud}, {Richichi}, {Robini}, {Sacchettini}, {Schertl},
  {Sch{\"o}ller}, {Solscheid}, {Spang}, {Stefanini}, {Tallon}, {Tallon-Bosc},
  {Tasso}, {Vakili}, {von der L{\"u}he}, {Valtier}, {Vannier}, \&
  {Ventura}}]{TATULLI+07}
{Tatulli}, E., {Isella}, A., {Natta}, A., {Testi}, L., {Marconi}, A., {Malbet},
  F., {Stee}, P., {Petrov}, R.~G., {Millour}, F., {Chelli}, A., {Duvert}, G.,
  {Antonelli}, P., {Beckmann}, U., {Bresson}, Y., {Dugu{\'e}}, M., {Gennari},
  S., {Gl{\"u}ck}, L., {Kern}, P., {Lagarde}, S., {Le Coarer}, E., {Lisi}, F.,
  {Perraut}, K., {Puget}, P., {Rantakyr{\"o}}, F., {Robbe-Dubois}, S.,
  {Roussel}, A., {Weigelt}, G., {Zins}, G., {Accardo}, M., {Acke}, B., {Agabi},
  K., {Altariba}, E., {Arezki}, B., {Aristidi}, E., {Baffa}, C., {Behrend}, J.,
  {Bl{\"o}cker}, T., {Bonhomme}, S., {Busoni}, S., {Cassaing}, F., {Clausse},
  J.-M., {Colin}, J., {Connot}, C., {Delboulb{\'e}}, A., {Domiciano de Souza},
  A., {Driebe}, T., {Feautrier}, P., {Ferruzzi}, D., {Forveille}, T., {Fossat},
  E., {Foy}, R., {Fraix-Burnet}, D., {Gallardo}, A., {Giani}, E., {Gil}, C.,
  {Glentzlin}, A., {Heiden}, M., {Heininger}, M., {Hernandez Utrera}, O.,
  {Hofmann}, K.-H., {Kamm}, D., {Kiekebusch}, M., {Kraus}, S., {Le Contel}, D.,
  {Le Contel}, J.-M., {Lesourd}, T., {Lopez}, B., {Lopez}, M., {Magnard}, Y.,
  {Mars}, G., {Martinot-Lagarde}, G., {Mathias}, P., {M{\`e}ge}, P., {Monin},
  J.-L., {Mouillet}, D., {Mourard}, D., {Nussbaum}, E., {Ohnaka}, K.,
  {Pacheco}, J., {Perrier}, C., {Rabbia}, Y., {Rebattu}, S., {Reynaud}, F.,
  {Richichi}, A., {Robini}, A., {Sacchettini}, M., {Schertl}, D.,
  {Sch{\"o}ller}, M., {Solscheid}, W., {Spang}, A., {Stefanini}, P., {Tallon},
  M., {Tallon-Bosc}, I., {Tasso}, D., {Vakili}, F., {von der L{\"u}he}, O.,
  {Valtier}, J.-C., {Vannier}, M., \& {Ventura}, N. 2007, \aap, 464, 55

\bibitem[{{Tatulli} {et~al.}(2008){Tatulli}, {Malbet}, {M{\'e}nard}, {Gil},
  {Testi}, {Natta}, {Kraus}, {Stee}, \& {Robbe-Dubois}}]{TATULLI+08}
{Tatulli}, E., {Malbet}, F., {M{\'e}nard}, F., {Gil}, C., {Testi}, L., {Natta},
  A., {Kraus}, S., {Stee}, P., \& {Robbe-Dubois}, S. 2008, \aap, 489, 1151

\bibitem[{{Thi} {et~al.}(2005){Thi}, {van Dalen}, {Bik}, \& {Waters}}]{THI+05}
{Thi}, W.-F., {van Dalen}, B., {Bik}, A., \& {Waters}, L.~B.~F.~M. 2005, \aap,
  430, L61

\bibitem[{{Weidenschilling}(1977)}]{WEID+77}
{Weidenschilling}, S.~J. 1977, \apss, 51, 153

\bibitem[{{Weigelt} {et~al.}(2011){Weigelt}, {Grinin}, {Groh}, {Hofmann},
  {Kraus}, {Miroshnichenko}, {Schertl}, {Tambovtseva}, {Benisty}, {Driebe},
  {Lagarde}, {Malbet}, {Meilland}, {Petrov}, \& {Tatulli}}]{WEIGELT+11}
{Weigelt}, G., {Grinin}, V.~P., {Groh}, J.~H., {Hofmann}, K.-H., {Kraus}, S.,
  {Miroshnichenko}, A.~S., {Schertl}, D., {Tambovtseva}, L.~V., {Benisty}, M.,
  {Driebe}, T., {Lagarde}, S., {Malbet}, F., {Meilland}, A., {Petrov}, R., \&
  {Tatulli}, E. 2011, \aap, 527, A103

\bibitem[{{White} \& {Ghez}(2001)}]{WG01}
{White}, R.~J. \& {Ghez}, A.~M. 2001, \apj, 556, 265

\bibitem[{{Woillez} {et~al.}(2012){Woillez}, {Akeson}, {Colavita}, {Eisner},
  {Millan-Gabet}, {Monnier}, {Pott}, {Ragland}, {Wizinowich}, {Abajian},
  {Appleby}, {Berkey}, {Cooper}, {Felizardo}, {Herstein}, {Hrynevych},
  {Medeiros}, {Morrison}, {Panteleeva}, {Smith}, {Summers}, {Tsubota}, {Tyau},
  \& {Wetherell}}]{WOILLEZ+12}
{Woillez}, J., {Akeson}, R., {Colavita}, M., {Eisner}, J., {Millan-Gabet}, R.,
  {Monnier}, J., {Pott}, J.-U., {Ragland}, S., {Wizinowich}, P., {Abajian}, M.,
  {Appleby}, E., {Berkey}, B., {Cooper}, A., {Felizardo}, C., {Herstein}, J.,
  {Hrynevych}, M., {Medeiros}, D., {Morrison}, D., {Panteleeva}, T., {Smith},
  B., {Summers}, K., {Tsubota}, K., {Tyau}, C., \& {Wetherell}, E. 2012, \pasp,
  124, 51

\bibitem[{{Woillez} {et~al.}(2014){Woillez}, {Wizinowich}, {Akeson},
  {Colavita}, {Eisner}, {Millan-Gabet}, {Monnier}, {Pott}, \&
  {Ragland}}]{WOILLEZ+14}
{Woillez}, J., {Wizinowich}, P., {Akeson}, R., {Colavita}, M., {Eisner}, J.,
  {Millan-Gabet}, R., {Monnier}, J.~D., {Pott}, J.-U., \& {Ragland}, S. 2014,
  \apj, 783, 104

\end{thebibliography}
\end{document}